\documentclass[aps,pra,twocolumn,superscriptaddress]{revtex4-2}  

\usepackage[dvipsnames]{xcolor}

\definecolor{darkred}{rgb}{0.6,0.05,0.05}
\definecolor{darkgreen}{rgb}{0.05,0.6,0.05}
\definecolor{darkblue}{rgb}{0.05,0.05,0.6}
\usepackage[colorlinks]{hyperref}
\hypersetup{colorlinks,
linkcolor=darkred,
filecolor=darkgreen,
urlcolor=darkblue,
citecolor=darkgreen}

\usepackage{mathtools}
\usepackage{amsmath}
\allowdisplaybreaks
\usepackage{amssymb}
\usepackage[normalem]{ulem}
\usepackage{amsthm}
\usepackage{xfrac}
\usepackage{physics}
\usepackage{graphicx}
\usepackage{hyperref}
\usepackage[capitalise]{cleveref}
\usepackage{dcolumn}% Align table columns on decimal point
\usepackage{bm}% bold math
\usepackage{comment}

\newcommand{\cami}[1]{#1}
\newcommand{\fab}[1]{#1}
\newcommand{\fil}[1]{#1}

\newcommand{\rme}{{\rm e}}
\newcommand{\rmd}{{\rm d}}
\newcommand{\rmi}{{\rm i}}

\newcommand{\cev}[1]{\reflectbox{\ensuremath{\vec{\reflectbox{\ensuremath{#1}}}}}}
\newcommand{\vecdouble}[1]{\overset{\leftrightarrow}{#1}}

\begin{document}
%\linenumbers

\author{Filippo Ferrari}
\affiliation{Institute of Physics, \'{E}cole Polytechnique F\'{e}d\'{e}rale de Lausanne (EPFL), CH-1015 Lausanne, Switzerland}
\affiliation{Center for Quantum Science and Engineering, \\ \'{E}cole Polytechnique F\'{e}d\'{e}rale de Lausanne (EPFL), CH-1015 Lausanne, Switzerland}
\author{Fabrizio Minganti}

\altaffiliation{Currently at Alice \& Bob}
\affiliation{Institute of Physics, \'{E}cole Polytechnique F\'{e}d\'{e}rale de Lausanne (EPFL), CH-1015 Lausanne, Switzerland}
\affiliation{Center for Quantum Science and Engineering, \\ \'{E}cole Polytechnique F\'{e}d\'{e}rale de Lausanne (EPFL), CH-1015 Lausanne, Switzerland}
\author{Camille Aron}
\affiliation{Institute of Physics, \'{E}cole Polytechnique F\'{e}d\'{e}rale de Lausanne (EPFL), CH-1015 Lausanne, Switzerland}
\affiliation{Laboratoire de Physique de l’\'{E}cole Normale Sup\'{e}rieure, ENS, Universit\'{e} PSL,
CNRS, Sorbonne Universit\'{e}, Universit\'{e} Paris Cit\'{e}, F-75005 Paris, France}
\author{Vincenzo Savona}
\affiliation{Institute of Physics, \'{E}cole Polytechnique F\'{e}d\'{e}rale de Lausanne (EPFL), CH-1015 Lausanne, Switzerland}
\affiliation{Center for Quantum Science and Engineering, \\ \'{E}cole Polytechnique F\'{e}d\'{e}rale de Lausanne (EPFL), CH-1015 Lausanne, Switzerland}

%\title{Chaos and spatial prethermalization in driven-dissipative bosonic chains}

\title{\fil{Chaotic and quantum dynamics in driven-dissipative bosonic chains}}

\date{\today}
             
\begin{abstract}
Thermalization in quantum many-body systems typically unfolds over timescales governed by intrinsic relaxation mechanisms. Yet, its spatial aspect is less understood.
We investigate this phenomenon in the nonequilibrium steady state (NESS) of a Bose-Hubbard chain subject to coherent driving and dissipation at its boundaries, a setup inspired by current designs in circuit quantum electrodynamics.  The dynamical fingerprints of chaos in this NESS are probed using semiclassical out-of-time-order correlators (OTOCs) within the truncated Wigner approximation (TWA).
At intermediate drive strengths, we uncover a two-stage thermalization along the spatial dimension: phase coherence is rapidly lost near the drive, while amplitude relaxation occurs over much longer distances. \cami{This separation of scales gives rise to an extended hydrodynamic regime exhibiting anomalous temperature profiles, which we designate as a ``prethermal'' domain.
At stronger drives, the system enters a nonthermal, non-chaotic finite-momentum condensate characterized by sub-Poissonian photon statistics and a spatially modulated phase profile, whose stability is undermined by quantum fluctuations.}
We explore the conditions underlying this protracted thermalization in space and argue that similar mechanisms are likely to emerge in a broad class of extended driven-dissipative systems.
\end{abstract}

\maketitle

\section{Introduction}
\fil{Chaos and thermalization are ubiquitous features in classical and quantum many-body systems.
Classical chaotic dynamics are well understood in terms of Lyapunov instability: \cami{two initially nearby trajectories in phase space diverge exponentially in time \cite{strogatz_nonlinear_2018}.
In quantum mechanics, however, the notion of phase-space trajectory is lost, and the characterization of chaos becomes more subtle. Quantum chaotic dynamics are typically diagnosed through the statistical properties of the energy spectrum \cite{dalessio_quantum_2016} as well as through the temporal behavior of out-of-time-ordered correlation functions}~\cite{hashimoto_out--time-order_2017}.
With the recent advances of quantum computing platforms, especially within circuit quantum electrodynamics (QED), the study of chaos in quantum devices has gained significant relevance~\cite{berke_transmon_2022, shillito_dynamics_2022, cohen_reminiscence_2023, dumas_measurement-induced_2024, garcia-mata_impact_2025, chavez-carlos_driving_2025}.
State-of-the art driven-dissipative circuit QED architectures are nowadays able to control tens of coupled nonlinear oscillators, enabling the experimental realization of atom-photon bound states~\cite{liu_quantum_2017, sundaresan_interacting_2019, scigliuzzo_controlling_2022}, two-dimensional Bose-Hubbard lattices~\cite{karamlou_probing_2024}, quantum chaotic and thermalized models~\cite{braumuller_probing_2022, zhang_superconducting_2023, andersen_thermalization_2025} and concatenated bosonic qubits~\cite{putterman_hardware-efficient_2025}. 
\cami{In the ongoing effort to fabricate and operate increasingly complex systems—featuring a growing number of individual degrees of freedom and enhanced nonlinearities,} chaotic dynamics and thermalization are both an opportunity for quantum simulations and a threat for the coherent manipulation of quantum information.}

Large, isolated and nonintegrable many-body systems are generically expected to thermalize at sufficiently long times.
Classically, the microscopic understanding of thermalization is addressed by Boltzmann's H-theorem, which relies on the assumption of molecular chaos~\cite{boltzmann_uber_1872, brown_boltzmanns_2009}. 
Quantum mechanically, the eigenstate thermalization hypothesis provides a theoretical framework to explain how closed Hamiltonian systems can achieve thermalization under unitary dynamics~\cite{deutsch_quantum_1991, srednicki_chaos_1994, srednicki_approach_1999, dalessio_quantum_2016, rigol_thermalization_2008}.
Typically, thermalization occurs in two stages. 
First, non-conserved quantities rapidly relax to local equilibrium values.
Then, the hydrodynamic modes---long-wavelength excitations associated with conserved quantities such as energy or particle density---relax through a slower, often diffusive, process.
This prolonged transient before complete thermalization is particularly pronounced in the proximity of integrability, where it has sometimes been dubbed prethermal regime. This is notably the case when an integrability-breaking perturbation is turned on suddenly~\cite{bertini_prethermalization_2015, babadi_far--equilibrium_2015, birnkammer_prethermalization_2022} or in a periodic fashion~\cite{abanin_rigorous_2017, abanin_effective_2017, else_prethermal_2017, machado_long-range_2020, pizzi_classical_2021}. The system initially relaxes to a quasi-equilibrium state~\cite{vidmar_generalized_2016, mori_thermalization_2018}, and subsequently converges to a true thermal state on an exceedingly long timescale~\cite{berges_prethermalization_2004}.

Quantum devices are inherently prone to intrinsic losses as well as extrinsic dissipation channels introduced by measurement, and their open, nonequilibrium nature poses significant conceptual challenges. 
While dissipative quantum chaos has gained considerable attention~\cite{akemann_universal_2019, hamazaki_universality_2020, sa_complex_2020, dahan_classical_2022, prasad_dissipative_2022, kawabata_symmetry_2023, villasenor_breakdown_2024, ferrari_dissipative_2025}, thermalization in chaotic dissipative systems remains only partially explored.
In this work, we investigate the route to thermalization in driven-dissipative chains of nonlinear bosonic modes, where photons are coherently injected at one end of the chain, and the coupling to the input and output channels induces incoherent photon losses at both ends of the chain.
This is a prototypical and versatile model for quantum hardware based on superconducting circuits, within reach of current experimental setups~\cite{FitzpatrickPRX17, FedorovPRL21, jouanny_high_2025}.
Similar chains have been realized in other quantum architectures, including trapped ions~\cite{MeiPRL22}, semiconductor micropillars~\cite{rodriguez_interaction-induced_2016}, and ultracold gases in optical lattices~\cite{Benary2022}.

\cami{At the classical level, \textit{i.e.}, neglecting the quantum fluctuations, recent studies of boundary-driven dissipative Klein-Gordon chains have revealed a rich phenomenology, including unconventional transport behaviors~\cite{debnath_nonequilibrium_2017, prem_dynamics_2023, Abhishek2024}, which are absent in their isolated counterparts~\cite{parisi_approach_1997}. 
As the drive strength is increased, three distinct nonequilibrium steady-state (NESS) regimes have been identified:
First, a quasi-linear regime in which nonlinearities are essentially irrelevant, the dynamics remain regular, and transport is ballistic. Second, a hydrodynamic regime characterized by local thermal equilibria and superdiffusive transport. At even stronger drives, this regime gives way to a resonant nonlinear wave (RNW) regime, characterized by a spatially periodic phase profile and ballistic transport. The RNW regime has been shown to remain stable under weak but finite thermal fluctuations induced by the environments at the two boundaries of the chain.
}

\cami{In this article, we explore the structure of the quantum many-body NESS and we investigate how quantum fluctuations impact this rich dynamical landscape.}
We leverage the fact that these systems are experimentally operated in regimes well described by semiclassical approaches to promote well-established concepts and tools from classical chaos theory to analyze the ergodic properties of the nonequilibrium steady state (NESS) of the open quantum chain.
Specifically, we use the truncated Wigner approximation (TWA) and out-of-time-order correlators (OTOCs) to follow the transition from regular to chaotic regimes beyond the classical descriptions of Refs.~\cite{debnath_nonequilibrium_2017, prem_dynamics_2023, Abhishek2024}.
In the chaotic regime, we spatially resolve the transition from the strongly nonequilibrium state near the coherent drive to the low-temperature state expected at the opposite end of the chain.
Between these extremes, we identify an extended phase, referred to as the prethermal phase, in which the $\mathbb{U}(1)$ symmetry, initially broken by the boundary drive, is restored, and the chain locally equilibrates to high-density states.
\cami{Notably, this phase exhibits anomalous heating: the temperature increases with the distance from the boundary where energy is injected, resulting in a temperature gradient that opposes both the photon and energy currents.}
We attribute this phenomenon to a significant mismatch between the short relaxation scale of the phase degree of freedom and the longer hydrodynamic relaxation of the amplitude sector of the photonic field.
\cami{Away from the hydrodynamic regime, we find that quantum effects significantly impact the RNW regime.  In particular, quantum fluctuations demote the long-range coherence of the phase modulation and can even destabilize the RNW regime in sufficiently long chains.}
Our findings are directly relevant to current experimental platforms, and we propose diagnostics based on routinely measured quantities, which can be determined through quantum-state tomography via, \textit{e.g.}, heterodyne detection.

The paper is structured as follows. 
In Sec.~\ref{sec:model}, we introduce our model of boundary driven-dissipative bosonic chains and the TWA approach. 
In Sec.~\ref{sec:NESS}, we describe the steady-state phase diagram as a function of the strength of the boundary drive.
In Sec.~\ref{sec:chaotic_regime}, we detail the hydrodynamic regime and its prethermal phase.
\cami{In Sec.~\ref{sec:RNW_regime}, we analyze the RNW regime along with its distinct quantum features.}
We conclude in Sec.~\ref{sec:discussion} by outlining potential directions for future work.

\begin{figure*}[t!]
\centering
\includegraphics[width=1 \textwidth]{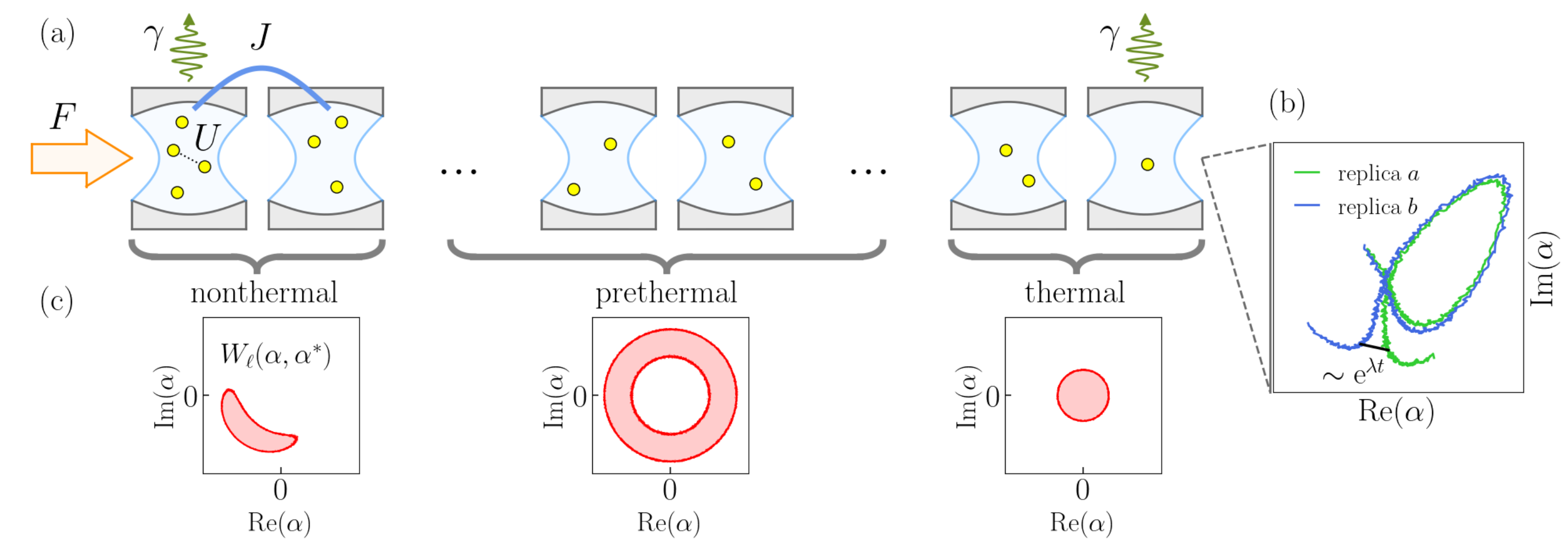}\vspace{-0.2em}
\caption{
\textbf{Boundary-driven dissipative Bose-Hubbard chain: schematics and key results}. 
(a) Tight-binding array of $L$ nonlinear resonators described by the Hamiltonian in Eq.~(\ref{eqs:hamiltonian}) subject to a drive of amplitude $F$ coherently injecting photons at the leftmost site and to single-photon losses at both ends of the chain. The interplay of interaction, drive, and dissipation leads to a nonequilibrium steady state (NESS).
(b) Lyapunov growth between neighboring Wigner trajectories is used to identify chaotic dynamics.
(c) In the chaotic regime, the chain hosts three distinct domains illustrated by their steady-state local Wigner functions $W_\ell(\alpha,\alpha^*)$: a nonsymmetric nonthermal domain near the left boundary, an extensive prethermal phase where the $\mathbb{U}(1)$ symmetry of the phase is restored and which hosts a high density of photons only saturated by the Kerr non-linearity,
and a $\mathbb{U}(1)$-symmetric thermal domain near the right boundary which is characterized by fluctuations over the vacuum state.
}
\label{fig:sketch}
\end{figure*}

\section{Results}

\subsection{Boundary-driven dissipative Bose-Hubbard chain}\label{sec:model}

We consider a one-dimensional chain of $L$ single-mode photonic resonators with nearest-neighbor coupling, modeled by the Bose-Hubbard model. The leftmost resonator of the chain is driven by a continuous wave drive, and the resonators at both ends of the chain experience single-photon losses, as depicted in Fig.~\ref{fig:sketch}.
The intrinsic losses of the resonators within the bulk of the chain are assumed to be negligible.
The dynamics are modeled by a Lindblad master equation for the system's density matrix $\hat \rho(t)$ reading (we set $\hbar = 1$)
\begin{equation}\label{eqs:lindblad}
    \frac{\partial\hat{\rho}}{\partial t}  = -\rmi[\hat{H}, \hat{\rho}] +
       \mathcal{D}[\hat L_1] \hat \rho +    \mathcal{D}[\hat L_L]  \hat \rho\,.
\end{equation}
The system Hamiltonian is expressed in the frame rotating at the drive frequency $\omega_\rmd$ as
\begin{align}\label{eqs:hamiltonian}
    \hat{H} =& \sum_{\ell=1}^L\left(-\Delta \, \hat{a}_\ell^{\dagger}\hat{a}_\ell + \frac{1}{2}\, U\,\hat{a}_\ell^{\dagger}\hat{a}_\ell^{\dagger}\hat{a}_\ell\hat{a}_\ell\right)\\* &- J\sum_{\ell=1}^{L-1}\big{(}\hat{a}_{\ell+1}^{\dagger}\hat{a}_\ell + \hat{a}_\ell^{\dagger}\hat{a}_{\ell+1}\big{)} + F(\hat{a}_1^{\dagger} + \hat{a}_1)\,. \nonumber
\end{align}
Here, $\hat{a}_\ell^\dagger$ ($\hat{a}_\ell$) are the bosonic creation (annihilation) operators for the photons in the $\ell$-th resonator mode of frequency $\omega_0$. $\Delta := \omega_\rmd - \omega_0$ is the pump-to-resonator detuning, $J$ is the hopping amplitude between neighboring resonators, and $U$ is the strength of the onsite Kerr nonlinearity. 
Our model is motivated by state-of-the-art experimental cavity or circuit QED setups~\cite{FitzpatrickPRX17, FedorovPRL21, jouanny_high_2025} with a weak but finite interaction, $|U| \ll |\Delta|, |J|$. While $U=0$ makes the model trivially integrable, a finite $U$ ensures the non-integrability of the Bose-Hubbard Hamiltonian on sizable chains~\cite{kolovsky_quantum_2004}.
\cami{Unlike in the classical Klein-Gordon chains studied in Refs.~\cite{prem_dynamics_2023, Abhishek2024}, the Kerr nonlinearity features a $\mathbb{U}(1)$ symmetry associated with photon number conservation in the bulk of the chain.}
$F$ is the amplitude of the driving field, which explicitly breaks this symmetry at one end of the chain. We take $U>0$, $\Delta > 0$, $J>0$, and $F\geq 0$ ($U<0$, $\Delta < 0$, $J<0$, and $F\leq 0$ yields identical results).
The Lindblad dissipators at sites $\ell=1$ and $\ell=L$ are defined as
\begin{align}
    \mathcal{D}[\hat L_\ell] \hat \rho := \hat L_\ell \hat{\rho} \hat L_\ell^\dagger - \frac{1}{2}\left\{\hat L_\ell^{\dagger}\hat L_\ell, \hat\rho \right\},
\end{align}
where $\hat L_\ell = \sqrt{\gamma} \hat a_\ell$ are the local jump operators modeling incoherent one-photon losses to the zero-temperature environment at a rate $\gamma > 0$. 
When not driven and isolated, \textit{i.e.}, $F=\gamma=0$, the Bose-Hubbard chain is far from the Mott insulating regime, and it is naturally expected to thermalize (see, \textit{e.g.}, Refs.~\cite{KollathPRL07,SchlagheckPRE16}). 
In the presence of drive and dissipation, it is expected to reach a unique NESS, $\lim\limits_{t\to\infty} \hat \rho(t)$, carrying uniform DC currents of photons and energy flowing from the left drive to the right drain. 
As initial conditions for Eq.~\eqref{eqs:lindblad}, we simply choose the vacuum state, $\hat \rho(0) = \bigotimes_{\ell=1}^L |0 \rangle_\ell \langle 0 |_\ell$, where $|0 \rangle_\ell$ denotes the Fock state with zero excitation in the $\ell$-th resonator.
We take $\gamma$ as the unit of energy and set $\Delta=2.5$, $J=2$, and $U=0.1$ for the remainder of this study.

To access the NESS of Eq.~\eqref{eqs:lindblad}, we use the truncated Wigner approximation (TWA), an approach based on a semiclassical treatment of the bosonic fields that accounts for leading-order quantum fluctuations~\cite{carmichael_statistical_1999, polkovnikov_phase_2010}.
In the TWA, Eq.~\eqref{eqs:lindblad} is mapped to a set of $L$ coupled stochastic differential equations for the complex field amplitudes $\alpha_{\ell}$, reading
\begin{align}\label{eqs:stochastic_differential_equations}
    \rmi \frac{\partial  \alpha_1}{\partial t} &=   -f(\alpha_1)  - J  \alpha_{2} + F - \frac{\rmi\gamma}{2}\alpha_1 + \sqrt{\frac{\gamma}{2}} \xi_1(t)
    \,, \nonumber \\
    \rmi \frac{\partial  \alpha_\ell}{\partial t} &= -f(\alpha_\ell)  - J (\alpha_{\ell-1} + \alpha_{\ell+1})
    \,,\ \ell=2,...,L-1 \nonumber  \\
    \rmi \frac{\partial  \alpha_L}{\partial t} &=  -f(\alpha_L)  - J  \alpha_{L-1}  - \frac{\rmi\gamma}{2}\alpha_L + \sqrt{\frac{\gamma}{2}} \xi_L(t)
    \,,
\end{align}
where $f(\alpha) := \Delta \, \alpha - U\,(|\alpha|^2-1)\alpha$, $\xi_1$ and $\xi_L$ are complex Gaussian white noises such that $\langle\xi_1(t)\rangle =\langle \xi_L(t)\rangle=0$ and $\langle\xi_1(t)\xi_1^*(t')\rangle = \langle\xi_L(t)\xi_L^*(t')\rangle =\delta(t-t')$. 
\cami{In the absence of drive and dissipation, \textit{i.e.} $F = \gamma = 0$, this set of equations corresponds to a discrete version of nonlinear Schr\"odinger equations~\cite{Politi, iubini_coupled_2016, iubini_coupled_2019, manas_spatiotemporal_2020, komorowski_heat_2023}.}
Quantum mechanical effects are encoded both in the vacuum initial conditions $\alpha_\ell(0)$ for $\ell = 1,..., L$ drawn from a complex Gaussian distribution with zero mean and variance $1/2$, and in the stochastic coupling to the baths.
\cami{The limiting equations in the classical regime are discussed in the Supplementary Information.}
A solution of Eqs.~\eqref{eqs:stochastic_differential_equations} is called a Wigner trajectory. Individual Wigner trajectories capture the stochastic nature of the interactions between the quantum system and its environment. 
In this framework, observables are calculated by sampling the Wigner trajectories over many realizations of the quantum noise. The quantum state at site $\ell$ can be conveniently visualized using the local Wigner function $W_\ell(t;\alpha, \alpha^*)$ which can be reconstructed from the statistical distribution of $\alpha_\ell(t)$ in the complex $\Re(\alpha_\ell)-\Im(\alpha_\ell)$ plane (see Methods).

Although the TWA is exact only for quadratic models ($U=0$), it has been successfully applied to describe dissipative phase transitions, disordered systems, time crystals, and quantum chaos in a variety of weakly nonlinear driven-dissipative systems~\cite{dagvadorj_nonequilibrium_2015, dujardin_elastic_2015, dujardin_breakdown_2016, vicentini_critical_2018, vicentini_optimal_2019, schlagheck_enhancement_2019, seibold_dissipative_2020, deuar_fully_2021}.
We justify the use of the TWA in our model by the weakness of the Kerr nonlinearity---a condition easily achievable in current circuit QED architectures~\cite{blais_circuit_2021}; under this condition, the TWA faithfully describes the NESS of single or multiple nonlinear driven resonators~\cite{vicentini_critical_2018} and boundary-driven dimers~\cite{seibold_dissipative_2020}.
We further validate this approach by benchmarking it against exact results for small system sizes and comparing it with other approximation schemes (see the Supplementary Information).
The choice of the TWA is motivated by the considerable reduction of the computational complexity it offers: the exponential growth of the Hilbert space in Eq.~\eqref{eqs:lindblad} is reduced to a linear growth in the number of stochastic equations~\eqref{eqs:stochastic_differential_equations}, thus enabling direct numerical integration of long chains of resonators up to times past the transient dynamics and deep in the steady-state regime.
Furthermore, Wigner trajectories extend the classical notion of phase-space trajectories to the quantum regime, thereby offering a practical means of investigating the exponential sensitivity to initial conditions (or lack thereof), which is a constitutive hallmark of chaotic dynamics~\cite{strogatz_nonlinear_2018}.
Notably, while the density matrix and local Wigner functions remain constant in the NESS, individual Wigner trajectories keep fluctuating and can thus be used to analyze the integrable versus chaotic character of the dynamics in the NESS~\cite{dahan_classical_2022, ferrari_dissipative_2025}.

\subsection{\cami{Dynamical regimes in the NESS}}\label{sec:NESS}
We begin by investigating the NESS phase diagram as a function of the drive amplitude $F$ and the chain length $L$. Unless specified otherwise, all expectation values reported below are computed in the NESS. To characterize the different steady-state regimes, we extract observables at the last site of the chain.
We monitor two standard quantities in quantum optics that can easily be probed within the TWA framework: the steady-state photon number $n_\ell :=  \langle \hat{a}_\ell^\dagger \hat{a}_\ell \rangle$ and the variance of its fluctuations $(\Delta n_\ell)^2  :=  \langle(\hat{a}_\ell^{\dagger}\hat{a}_\ell)^2\rangle - \langle\hat{a}_\ell^{\dagger}\hat{a}_\ell\rangle^2$ by means of the quantity
\begin{align}\label{eqs:delta_n}
    &\delta n_\ell := (\Delta n_\ell)^2 - n_\ell 
    \end{align}
which quantifies the relative distance to Poissonian statistics~\footnote{Compared to the conventional indicator $g^{(2)}$, $\delta n_\ell$ defined in Eq.~(\ref{eqs:delta_n}) is less prone to numerical errors caused by a finite sampling of Wigner trajectories.}. $\langle ... \rangle$ indicates the average over Wigner trajectories once the NESS is reached.
$\delta n_\ell = 0 $ corresponds to Poissonian statistics typical of coherent states, $\delta n_\ell > 0$ corresponds to super-Poissonian statistics, and sub-Poissonian statistics with $\delta n_\ell < 0$ are incompatible with a classical description of the state~\cite{scully_quantum_1997}.
\cami{The quantity $\delta n_\ell$ is directly related to the second-order coherence function $g^{(2)}_\ell$, a standard observable in quantum optics, via $\delta n_\ell  = n_\ell^2( g^{(2)}_\ell -1) $.}

In addition to the above static diagnostics, we probe the chaotic character of the dynamics, or lack thereof, by means of an out-of-time correlator (OTOC) in the NESS. OTOCs are commonly used as a diagnostic of the spatiotemporal spread of chaos in both quantum mechanical and classical systems.
We focus on the OTOC between the number and the phase degrees of freedom of the resonators.
In the Methods, we show that the semiclassical formulation of this phase OTOC reads~\cite{das_light-cone_2018, bilitewski_temperature2018, schuckert_thermal2019, manas_spatiotemporal_2020, bilitewski_classical2021, sibaram_thermal2021, deger_arresting2022}
\begin{equation}\label{eqs:semiclassical_OTOC}
    D_{k,\ell}(\tau) := 1 - \lim_{t\to\infty}\left\langle \cos\left[ \varphi_\ell^{(a)}(t+\tau) - \varphi_\ell^{(b)}(t+\tau)\right]\right\rangle.
\end{equation}
\fil{Here, $\varphi_\ell(t) := \textrm{arg}[\alpha_\ell(t)]$ is the phase of the complex field in the $\ell$-th resonator along an individual Wigner trajectory. The superscripts $(a)$ and $(b)$ refer to two replicas of the system which are identical until time $t$, at which an infinitesimal perturbation is applied to the resonator at site $k$ in replica $b$ according to $\varphi_k^{(b)}(t) = \varphi_k^{(a)}(t) + \varepsilon$,} \fab{with $\varepsilon\ll1$}.
\fil{The subsequent evolution is computed using the same quantum noise realization for the two replicas.}
\fab{We stress that in Eq.~\eqref{eqs:semiclassical_OTOC}, the replica (b) is the one perturbed at site $k$ and, therefore, $\varphi_\ell^{(b)}$ implicitly depends on the $k$ index.}
%\cami{and the limit $\varepsilon\to0$ is implied}.

\begin{figure*}[t!]
\centering
\includegraphics[width=0.95 \textwidth]{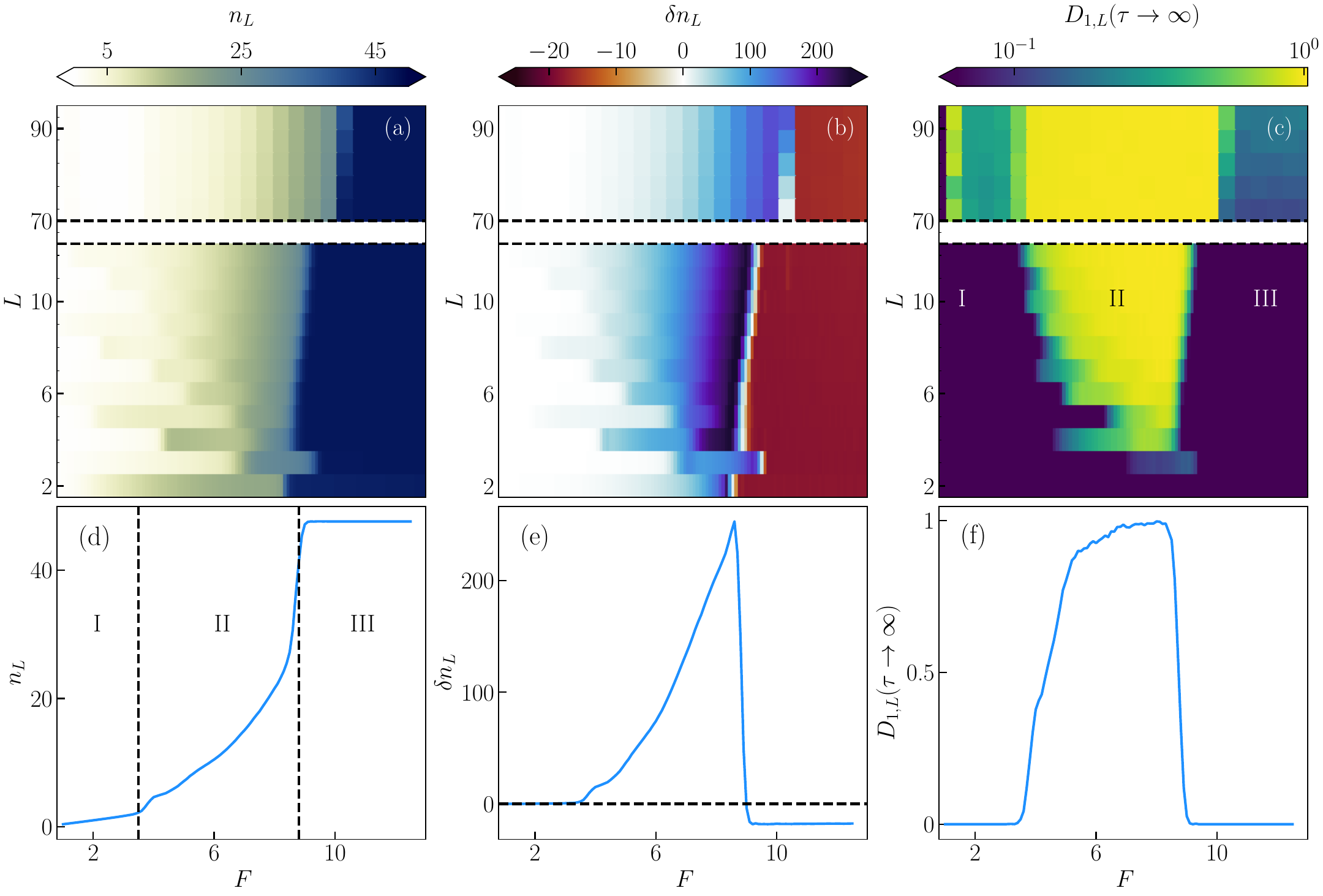}\vspace{0.8em}
\caption{
\textbf{Nonequilibrium steady-state phase diagram}.
Steady-state properties of the last resonator in the chain, $\ell = L$, as functions of the chain length $L$ and drive strength $F$.  
(a) Photon number $n_L$.  
(b) Photon-number fluctuation $\delta n_L$, defined in Eq.~\eqref{eqs:delta_n}.  
(c) Saturation value of the steady-state phase OTOC, $D_{1,L}(\tau \to \infty)$, defined in Eq.~\eqref{eqs:semiclassical_OTOC}.  
The three distinct regimes labeled I, II, and III are discussed in the text.  
\cami{
Panels (d), (e), and (f) show cuts of panels (a), (b), and (c), respectively, at fixed chain length $L = 10$. }
Results are computed by averaging over $N_{\rm traj} = 10^3$ independent Wigner trajectories. 
Statistics are further improved by averaging over a time window $\Delta \tau$ after reaching the NESS: $\Delta \tau = 25$ for $D_{1,L}(\tau \to \infty)$, and $\Delta \tau = 2 \times 10^3$ for $n_L$ and $\delta n_L$.  
Throughout the manuscript, the dissipation rate $\gamma$ sets the unit of energy, and the other parameters are fixed to $\Delta = 2.5$, $J = 2$, and $U = 0.1$.
}
\label{fig:phase_diagram}
\end{figure*}

Both the initial growth and the late regimes of $D_{k,\ell}(\tau)$ shed light on the chaotic versus regular nature of the dynamics in the NESS.
In chaotic dynamics, the typical distance between two trajectories with nearly identical initial conditions is expected to grow exponentially -- a hallmark of Lyapunov instability, which is often taken as a defining feature of classical chaos. Quantum mechanically, this exponential growth is expected to be captured by OTOCs in systems that have a well-defined semiclassical limit or large-$N$ limit. 
%The saturation value $D_{k,\ell}( \tau \to \infty)$ is related to the square of the available local phase space volume. 
$D_{k,\ell}( \tau \to \infty) \ll 1$ corresponds to situations where the phases in replica $a$ and $b$ remain strongly correlated, suggestive of regular dynamics.
In contrast, in chaotic regimes where the trajectories decorrelate rapidly, $\varphi^{(a)}$ and $\varphi^{(b)}$ can be seen as uniformly distributed random phases, resulting in a saturation value at its ergodic bound $D_{k,\ell}(\tau \to \infty) \approx 1$.
\fil{We refer the Reader to the Methods section for additional details on the OTOC's dynamics.}
Here, we use the saturation value of the steady-state phase OTOC, $D_{1,\ell}(\tau\to\infty)$, as a proxy to map the chaotic versus regular character of the nonequilibrium steady-state dynamics as the drive strength $F$ and the chain length $L$ are varied. 
\fab{We notice that, while $n_\ell$ and $\delta n_\ell$ (or $g^{(2)}_\ell$) can be monitored in the laboratory, accessing the behavior of $D_{k\ell}$ may be more difficult, as it requires an accurate cloning of the time evolution.}

\begin{figure*}[t!]
\includegraphics[width=1 \textwidth]{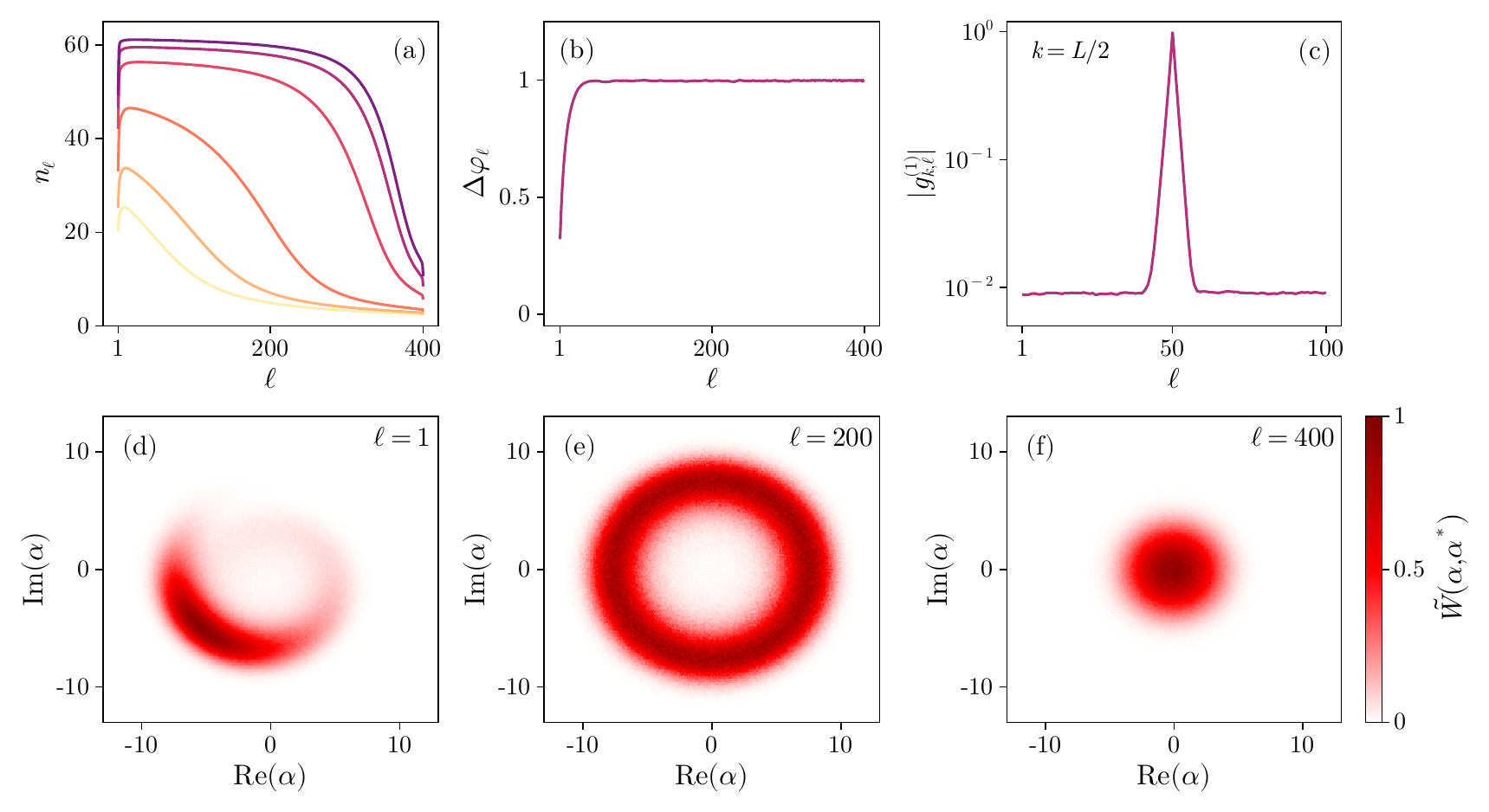}\vspace{0.8em}
\caption{
\textbf{Chaotic regime: Two-stage relaxation in space.}
Spatial profiles of equal-time photon statistics in the chaotic NESS.
(a) Photon density $n_\ell  = \langle  |\alpha_\ell|^2 \rangle - 1/2$,  showing a growing relaxation length scale with increasing drive strength $F = 5.5,\, 6,\, 6.5,\, 7,\, 7.5,\, 8$ (from yellow to purple), and a plateau across most of the chain at stronger drives.
\cami{(b) Circular phase variance $\Delta \varphi_\ell := 1 - |\langle \rme^{\rmi\varphi_\ell}\rangle |$, which rapidly saturates to unity, indicating a uniform phase distribution.}
(c) First-order coherence function $|g_{k, \ell}^{(1)}|$ defined in Eq.~(\ref{eqs:g1}) showing exponential decay of phase correlations on microscopic length scales away from $\ell = k$.
\cami{(d-f) Normalized local Wigner functions, $\tilde{W}_\ell(\alpha, \alpha^*) := W_\ell(\alpha, \alpha^*)/\textrm{max}[W_\ell(\alpha, \alpha^*)]$ for representative sites in a chain of length $L = 400$.}
Results are computed by averaging over $N_{\rm traj} = 10^2$ independent Wigner trajectories and over a time window $\Delta\tau=10^4$ once the steady state is reached.
The drive strength is set to $F=7.5$ in panels (b-f), and the other parameters are set as in Fig.~\ref{fig:phase_diagram}.
}
\label{fig:correlations}
\end{figure*}

Focusing on the last site, $\ell = L$, the results are collected in Fig.~\ref{fig:phase_diagram}. Qualitatively similar results are obtained throughout the chain, and we refer the reader to the Supplementary Information for results at $\ell=1$ and $\ell = L/2$.
We identify three distinct regimes:
\begin{enumerate}
    \item[(I)] \textit{Regular quasilinear regime}; at weak drive, only a few photons populate the last site. The photon statistics are Poissonian ($\delta n_L \approx 0$) and the small saturation value of the phase OTOC,  $D_{1,L}(\tau \to \infty) \approx 10^{-2}-10^{-1}$, indicates regular dynamics. 
    In this regime, single-particle excitations are dilute, rendering nonlinearities negligible and preventing the onset of chaos.
    
    \item[(II)] \textit{Chaotic regime}; at stronger drives, the photon number markedly increases with respect to the quasilinear regime and nonlinearities are now relevant. The large value of $\delta n_L$ indicates strong super-Poissonian fluctuations, \cami{compatible with a thermal description of light.}
    Concurrently, the phase OTOC saturates to $D_{1,L}(\tau \to \infty) \approx 1$, signaling significant dephasing between the fields in the two replicas.
    These observations, along with the exponential growth of $D_{1,\,\ell}(\tau)$ shown in the Methods section, confirm the chaotic nature of the quantum dynamics. This regime will be discussed in detail in Sect.~\ref{sec:chaotic_regime}.

    \item[(III)] 
    \textit{Resonant nonlinear wave (RNW) regime}; 
    at even stronger drives, the dynamics reach an incompressible regime where the large photon number marginally increases with $F$~\cite{biondi_nonequilibrium2017}.
    The small saturation value $D_{1,\, L}(\tau\to\infty) \approx  10^{-2}-10^{-1}$ indicates persistent phase correlations between the two replicas. These observations, along with the sub-exponential growth of $D_{1,\,\ell}(\tau)$ shown in the Methods section, are indicative of regular dynamics.
    \cami{This regime exhibits distinct quantum signatures, most notably the emergence of sub-Poissonian photon statistics, as indicated by $\delta n_L < 0$. These features will be discussed in detail in Sect.~\ref{sec:RNW_regime}.}
\end{enumerate}

Notably, the crossover from the regular quasilinear regime (I) to the chaotic regime (II) is smooth in both $n_L$ and $\delta n_L$ while the transition between the chaotic regime (II) and the RNW regime (III) is characterized by abrupt variations [see Figs.~\ref{fig:phase_diagram} (d) and (e)].
In particular, the drive strength $F$ at which the photon number leaps coincides with that of the maximum of $\delta n_L$.

\subsection{Chaotic regime} \label{sec:chaotic_regime}
\subsubsection{Two-stage relaxation in space}
In this section, we focus on the chaotic regime (II). 
We decompose the complex field in terms of amplitude and phase degrees of freedom, $\alpha_\ell = |\alpha_\ell| \, \rme^{\rmi \varphi_\ell}$,
and separately analyze their spatial relaxation along the chain, from $\ell =1$ to $\ell = L$.
Figures~\ref{fig:correlations} (a) and (b) shows the steady-state profile of photon density $n_\ell = \langle |\alpha_\ell|^2 \rangle - 1/2$ and \cami{the circular phase variance $\Delta \varphi_\ell := 1- |\langle \rme^{\rmi\varphi_\ell} \rangle|$ across the chain. The latter quantifies the spread of the phase distribution: $\Delta\varphi_\ell \to 0$ if the phase is well-defined, whereas $\Delta\varphi_\ell \to 1$ for a uniformly distributed phase.}
\cami{The field amplitude decays slowly along the chain over a characteristic length scale $\xi_n$, which increases monotonically with both the chain length $L$ and the drive strength $F$.
We distinguish two regimes: for weak drives, $F \lesssim 6.5$, the photon density relaxes just beyond the driven site; for intermediate drives, $F \gtrsim 6.5$, the density first saturates to a large value over an extended region, with its decay occurring only near the end of the chain.}
These large spatial scales can be explained by the local conservation of the photon number within the bulk of the chain, which hinders the rapid relaxation of the amplitude degree of freedom. Instead, this relaxation occurs through much slower hydrodynamic processes involving, notably, the driven-dissipative conditions at the two boundaries of the chain.
In contrast, the phase degrees of freedom relax over a microscopic length scale $\xi_\varphi$, typically spanning only a few lattice sites.
\cami{The saturation of the phase variance, $\Delta\varphi_\ell \to 1$, signals the restoration of the $\mathbb{U}(1)$ symmetry of the underlying bulk Hamiltonian. 
This mechanism is illustrated in Figs.~\ref{fig:correlations} (d-f), where the local Wigner functions $W_\ell(\alpha, \alpha^*)$ rapidly lose the anisotropy imprinted by the coherent boundary drive and become invariant under rotations in the complex plane throughout the rest of the chain.
}
\fil{The overall change of the topology of $W_\ell(\alpha, \alpha^*)$, from ring shape to bell shape, will be discussed below.}

To corroborate the short relaxation of the phase sector, we compute the spatial correlations of the phases in the chain by means of the first-order coherence function,
\begin{equation} \label{eqs:g1}
    g^{(1)}_{k, \ell} := \frac{\langle\hat{a}_{k}^{\dagger}\hat{a}_\ell\rangle}{\sqrt{\langle\hat{a}_{k}^{\dagger}\hat{a}_k\rangle\langle\hat{a}_{\ell}^{\dagger}\hat{a}_\ell\rangle}}\,.
\end{equation}
The results are presented in Fig.~\ref{fig:correlations} (c) as a function of $\ell$ for a center site $k=L/2$. 
The profile of the spatial correlations shows exponential decay, $|g^{(1)}_{k, \ell}|\sim  \rme^{-|k-\ell|/\xi_\varphi}$, typical of a disordered phase with a correlation length $\xi_\varphi$ on the order of a few sites, before saturating to values below $10^{-2}$.

Altogether, these results point to a two-stage thermalization process in space: the phase sector relaxes to a disordered state over a short, microscopic length scale, while the amplitude sector undergoes a slower relaxation that depends on the boundary conditions and unfolds over the entire system.
As we shall demonstrate below, this opens up a significant region of the chain, beyond the few sites directly influenced by the boundary drive, but before the complete relaxation to near-vacuum states subject to thermal fluctuations, where an effective $\mathbb{U}(1)$ symmetry emerges. In this intermediate region, the highly populated sites along the chain can be interpreted as being in local equilibrium, characterized by a large and slowly varying chemical potential.

\begin{figure*}[t!]
\includegraphics[width=1 \textwidth]{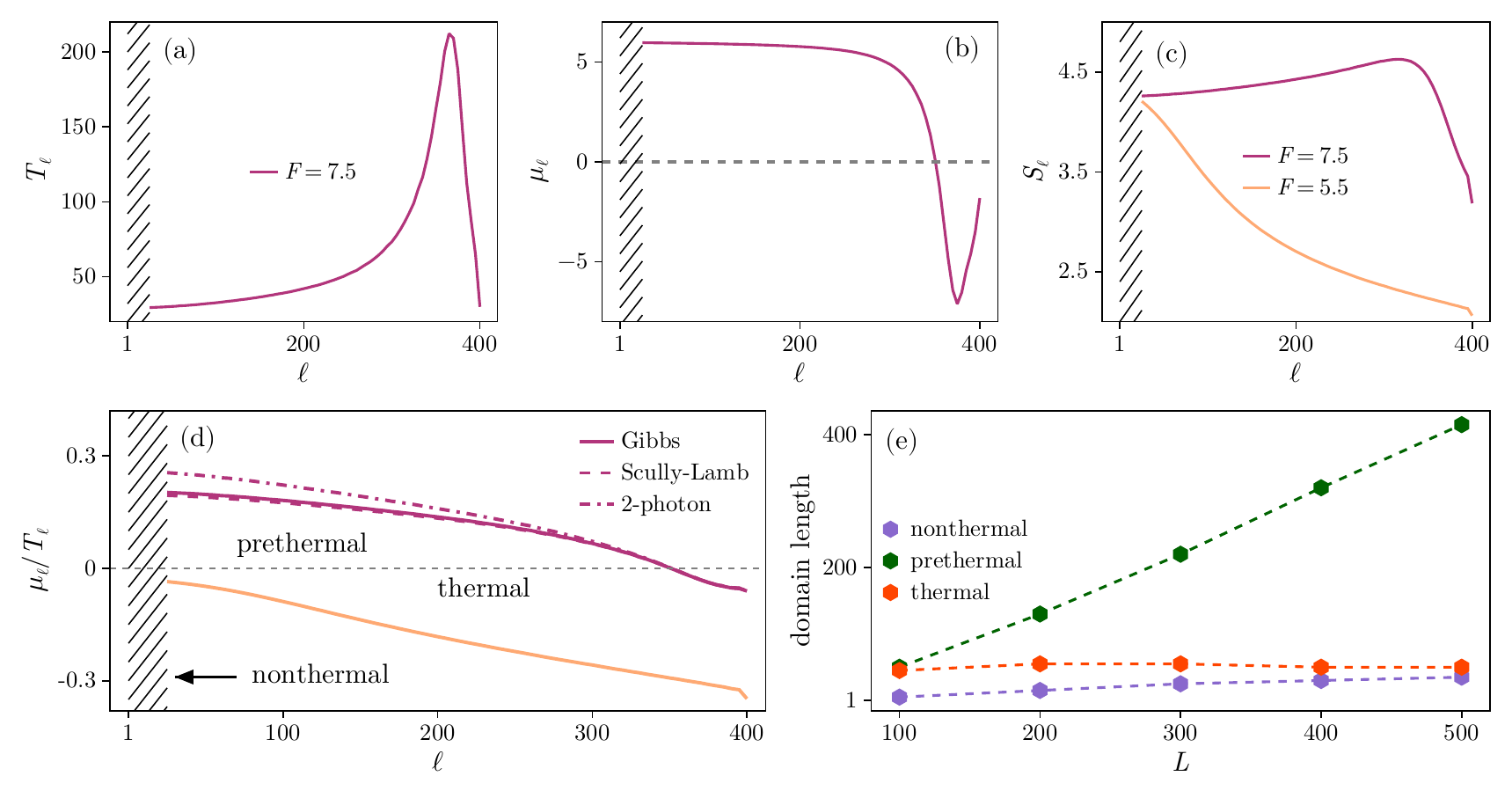}\vspace{0.8em}
\caption{
\textbf{Chaotic regime: hydrodynamic description.}
\cami{Spatial profile of the local (a) effective temperature $T_{\ell}$, (b) effective chemical potential $\mu_\ell$, and (c)  entropy density $S_\ell := -\textrm{Tr}[ \hat \rho_\ell \log \hat \rho_\ell]$ across a chain of length  $L=400$. The results are obtained by mapping the local steady-state physics at site $\ell$ to the Gibbs ansatz defined in Eq.~(\ref{eq:Gibbs}).}
(d)  Spatial profile of the dimensionless ratio $\mu_{\ell}/T_\ell$ for the same system. The results are obtained by mapping the local physics at site $\ell$ to the Gibbs ansatz, to the 2-photon impurity ansatz defined in Eqs.~(\ref{eqs:SC_hamiltonian}-\ref{eqs:SC_dissipator}), and to the generalized Scully-Lamb ansatz. 
Three distinct spatial domains are discussed in the text: 1) nonsymmetric nonthermal domain where all the ans\"atze fail; 2) prethermal domain where $\mu_{\ell}/T_\ell > 0$; 3) thermal domain with $\mu_{\ell}/T_\ell \leq 0$.
(e) Respective sizes of the three domains versus the total chain length $L$. 
The drive strength is set to $F=7.5$ and the other parameters are set as in Fig.~\ref{fig:phase_diagram}.
\label{fig:pre_thermalization}}
\end{figure*}

\subsubsection{Hydrodynamic description}

We test this hydrodynamic scenario by probing the local symmetries, the local thermodynamics, and the temporal fluctuations across the chain.
On general grounds, the local steady-state density matrix $\hat \rho_\ell := \lim\limits_{t\to\infty} \textrm{Tr}_{k\neq \ell} \hat \rho(t)$ can be cast as the solution of an effective driven-dissipative impurity model constructed by singling out one site of the chain, and tracing out its neighbors such as to treat them as incoherent sources and drains. In a generic NESS, however, the rigorous derivation of an effective driven-dissipative impurity model is a daunting task, and we simply rely on symmetries and heuristics to identify minimal impurity models that successfully locally capture both the statics as well as temporal fluctuations.

Let us first address the statics.
Guided by hydrodynamic principles, we assume that, away from the boundary drive, the local steady-state density matrix $\hat \rho_\ell$ is close to an equilibrium $\mathbb{U}(1)$-symmetric Gibbs state of the form (we set $k_{\rm B} = 1$) 
 \begin{equation} \label{eq:Gibbs}
    \hat{\rho}^{\rm eq}_\ell = \exp[- (\hat h - \mu_\ell \, \hat a^\dagger \hat a)/T_\ell]/ \mathcal{Z}_\ell\,,
\end{equation}
where $\mathcal{Z}_\ell$ is such that $\mathrm{Tr}\left(\hat{\rho}^{\rm eq}_\ell\right)  =1$, and the local Hamiltonian 
\begin{equation}\label{eqs:SC_hamiltonian}
    \hat{h} = \frac{1}{2}U\,  (\hat{a}^{\dagger}\hat{a})^2\,
\end{equation}
corresponds to a single site of the chain Hamiltonian in Eq.~(\ref{eqs:hamiltonian}). $T_{\ell}$ and $ \mu_\ell$ are, respectively, the effective temperature and chemical potential at site $\ell$, to be determined.
Note that the quadratic onsite contributions to $\hat H$, namely $-\Delta - U/2$, have been absorbed into a redefinition of $\mu_\ell$.

We assess the validity of this effective description and the underlying restoration of the $\mathbb{U}(1)$ symmetry by fitting the local steady-state Wigner function $W_\ell(\alpha, \alpha^*)$ along the chain to those predicted by the above Gibbs ansatz. The two fitting parameters are $T_\ell$ and $\mu_\ell$.
We repeat this procedure at all sites from $\ell = 1$ to $\ell = L$. Excellent matches are obtained at all sites except near the driven boundary, thereby confirming the rapid restoration of the $\mathbb{U}(1)$ symmetry away from the drive and validating the use of the Gibbs state as a local thermometer.
\cami{In regimes where the photon density is very low, the Kerr nonlinearity is effectively inactive, and it becomes numerically challenging to extract $T_\ell$ and $\mu_\ell$ independently. However, the ratio $\mu_\ell / T_\ell$, which governs the shape of the Wigner function in the dilute limit, can still be determined with high accuracy.}

Let us now turn to the description of the dynamical fluctuations in the steady state.
We start from the local $\mathbb{U}(1)$-symmetric Hamiltonian $ \hat{h}$ introduced in Eq.~(\ref{eqs:SC_hamiltonian}) and promote it to Lindblad dynamics by supplementing it with the first non-trivial jump operators allowed under weak $\mathbb{U}(1)$ symmetry~\cite{minganti_correspondance_2020},
\begin{equation}\label{eqs:SC_dissipator}
    \hat{L}_\uparrow \!=\! \sqrt{\gamma^\uparrow_\ell} \hat a^\dagger\!,
    \hat{L}_\downarrow \!=\! \sqrt{\gamma^\downarrow_\ell}\hat a,
    \hat{L}_\phi \!=\! \sqrt{\gamma^\phi_\ell}\hat{a}^{\dagger} \hat a,
    \hat{L}_{\rm s} \!=\! \sqrt{\gamma^{\rm s}_\ell} \hat a^2.
\end{equation}
The non-negative parameters  $\gamma^\uparrow_\ell$, $\gamma^\downarrow_\ell$, $\gamma^\phi_\ell$, and $\gamma^{\rm s}_\ell$ are effective rates of incoherent pumping, decay, dephasing, and 2-photon decay at site $\ell$. The inclusion of the latter is essential to ensure the saturation of the photon number when $\gamma^\uparrow_\ell > \gamma^\downarrow_\ell$.
Equations~(\ref{eqs:SC_hamiltonian}) and (\ref{eqs:SC_dissipator}) define a driven-dissipative impurity ansatz that supports a NESS. The parameters $\gamma^{\uparrow}_\ell$, $\gamma^{\downarrow}_\ell$ and $\gamma^{\rm s}_\ell$ are determined by fitting the local Wigner functions $W_\ell(\alpha, \alpha^*)$ to those predicted by the impurity ansatz. The latter are independent of the dephasing rate $\gamma_\ell^\phi$ that can be extracted by fitting two-time correlation functions. The overall fitting procedure is detailed in the Supplementary Information.
Excellent matches are obtained everywhere away from the boundary drive.

Although this impurity ansatz does not rely on assuming local equilibria, we have verified that both the previous Gibbs ansatz and this impurity ansatz consistently capture and reproduce the same static properties when related through the detailed-balance condition
\begin{equation}\label{eqs:effective_temperature}
    \mu_\ell/T_\ell =   \log(\gamma^{\uparrow}_{\ell} \big{/}\gamma^{\downarrow}_{\ell})\,.
\end{equation}

We stress that the above steady-state impurity modeling is not unique. In particular, we also tested a generalized version of the Scully-Lamb model~\cite{minganti_continuous_2021, minganti_liouvillian_2021} defined by $ \hat{L}_{\rm s} =0$ and a modified $\hat{L}_\uparrow = {\hat{a}^{\dagger}(\gamma^{\uparrow} - \mathcal{S} \, \hat{a}\hat{a}^{\dagger})}/{\sqrt{\gamma^{\uparrow}}}$, where $\mathcal{S} \geq 0$ is a photon saturation rate. In the context of lasing, this model can be derived as an effective theory for the optical degree of freedom when the (inverted) atomic population modeled by two-level systems has been integrated out~\cite{scully_quantum_1967, yamamoto_mesoscopic_1999}. This ansatz proved to be equally successful in capturing both the statics and the dynamical fluctuations yielding, notably, comparable values of the effective parameters $\gamma^\uparrow_\ell$ and $\gamma^\downarrow_\ell$.

A detailed comparison between these three thermometers, \fil{and temperature computed from the equipartition theorem whenever the latter is applicable,} can be found in the Supplementary Information. As reported in Fig.~\ref{fig:pre_thermalization} (d), we obtained an excellent agreement between these multiple approaches.

\cami{
\subsubsection{Nonthermal, prethermal, and thermal domains}
The above generalized thermometers provide the spatial profiles of the local temperature and chemical potential, $T_\ell$ and $\mu_\ell$, respectively, which we use to analyze the thermodynamics along the chain, see Figs.~\ref{fig:pre_thermalization} (a) and (b). In Fig.~\ref{fig:pre_thermalization} (c), we present the dimensionless ratio $T_\ell/\mu_\ell$.
As a complementary observable, we also monitor the entropy density $S_\ell := -\textrm{Tr}[ \hat \rho_\ell \log \hat \rho_\ell]$, see Fig.~\ref{fig:pre_thermalization} (c), that can be computed independently, \textit{i.e.} without relying on the thermodynamic ans\"atze.}

\cami{Depending on the strength of the drive, we distinguish two regimes. At relatively weak drives ($F \lesssim 6.5$), $\mu_\ell$ remains negative throughout the chain and $S_\ell$ monotonously decreases along the chain. This is consistent with the monotonously decaying photon density profiles shown in Fig.~\ref{fig:correlations} (a): negative chemical potentials lead to photon depletion toward a vacuum state dressed by thermal fluctuations.
At intermediate drives ($F \gtrsim 6.5$), $\mu_\ell$ becomes positive and approximately constant across most of the chain, before dropping abruptly to negative values near the non-driven right boundary. This correlates with the photon density profiles shown in Fig.~\ref{fig:correlations} (a): large positive chemical potentials sustain a significant photon population limited only by the Kerr nonlinearity.
Concurrently, $T_\ell$ increases across most of the chain, reaching very high values before dropping sharply near the right end. Although the low temperatures observed at both boundaries are consistent with the proximity of the zero-temperature dissipative baths, the magnitude and position of the temperature maximum may at first seem to challenge the Clausius principle. This anomalous profile is corroborated by the entropy density profile, $S_\ell$, which exhibits an accumulation of entropy precisely where $T_\ell$ attains its peak. This counterintuitive behavior motivates our designation of this anomalous regime as ``prethermal''.
Similar anomalous heating profiles have previously been reported in the context of the discrete nonlinear Schr\"odinger equation subjected to incoherent thermal reservoirs at the two ends of a one-dimensional chain~\cite{Politi}. This effect has been attributed to the coupling between heat and particle transport. Specifically, it arises from the coexistence of two conserved quantities in the bulk, photon number and energy, whose associated Noether currents, $\boldsymbol{j}_n$ and $\boldsymbol{j}_\epsilon$, are coupled at the level of linear response theory.
}

\begin{figure}[t!]
\centering
\includegraphics[width=0.5\textwidth]{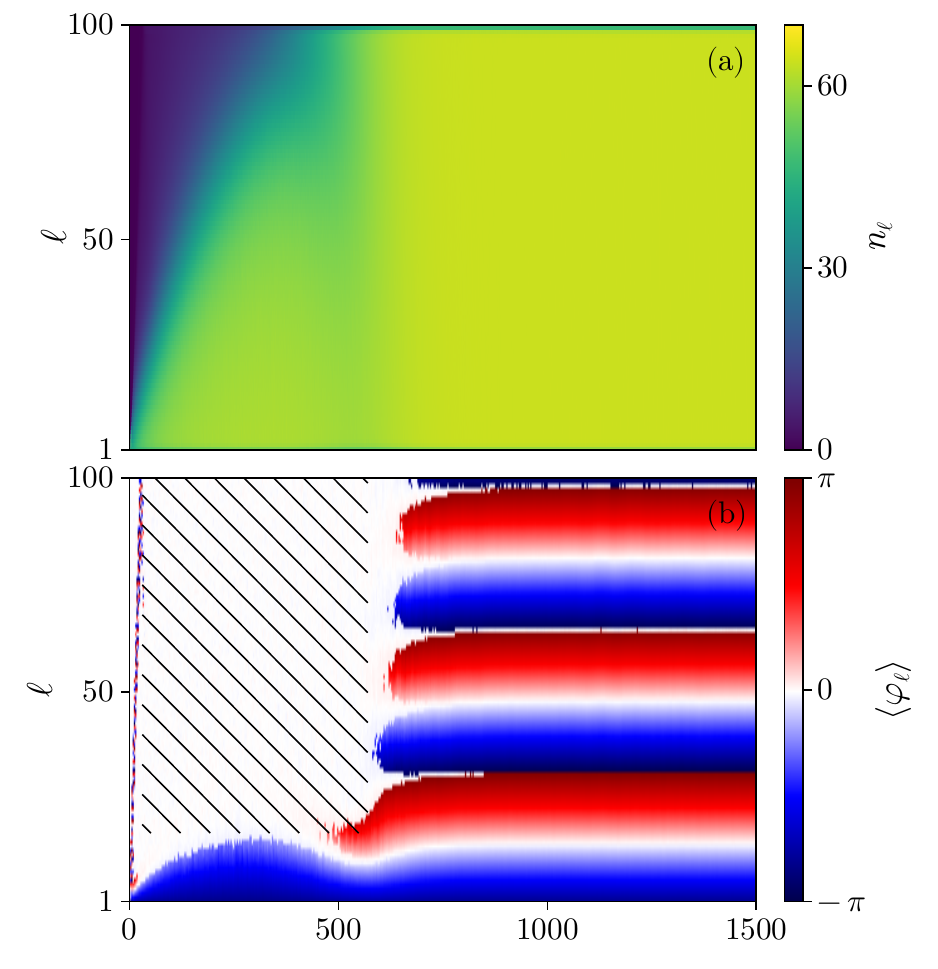}\vspace{0.8em}
\caption{
\cami{\textbf{Onset of the RNW Regime.} Spatiotemporal evolution of (a) the photon density $n_\ell$, and
(b) the average phase $\langle \varphi_\ell \rangle$
as functions of time $t$ and site index $\ell$ in a $L=100$ chain in the RNW regime.
The hatched region in (b) indicates the transient chaotic regime in which the $\mathbb{U}(1)$ symmetry is restored and the average phase becomes ill-defined.}
\fil{Results are obtained upon averaging over $N_{\rm traj} = 5\times 10^3$ independent Wigner trajectories.
The drive amplitude is fixed to $F=12.5$.
The other parameters are set as in Fig.~\ref{fig:phase_diagram}.}
}
\label{fig:regular_array}
\end{figure}

\begin{figure*}[t!]
\centering
\includegraphics[width=1 \textwidth]{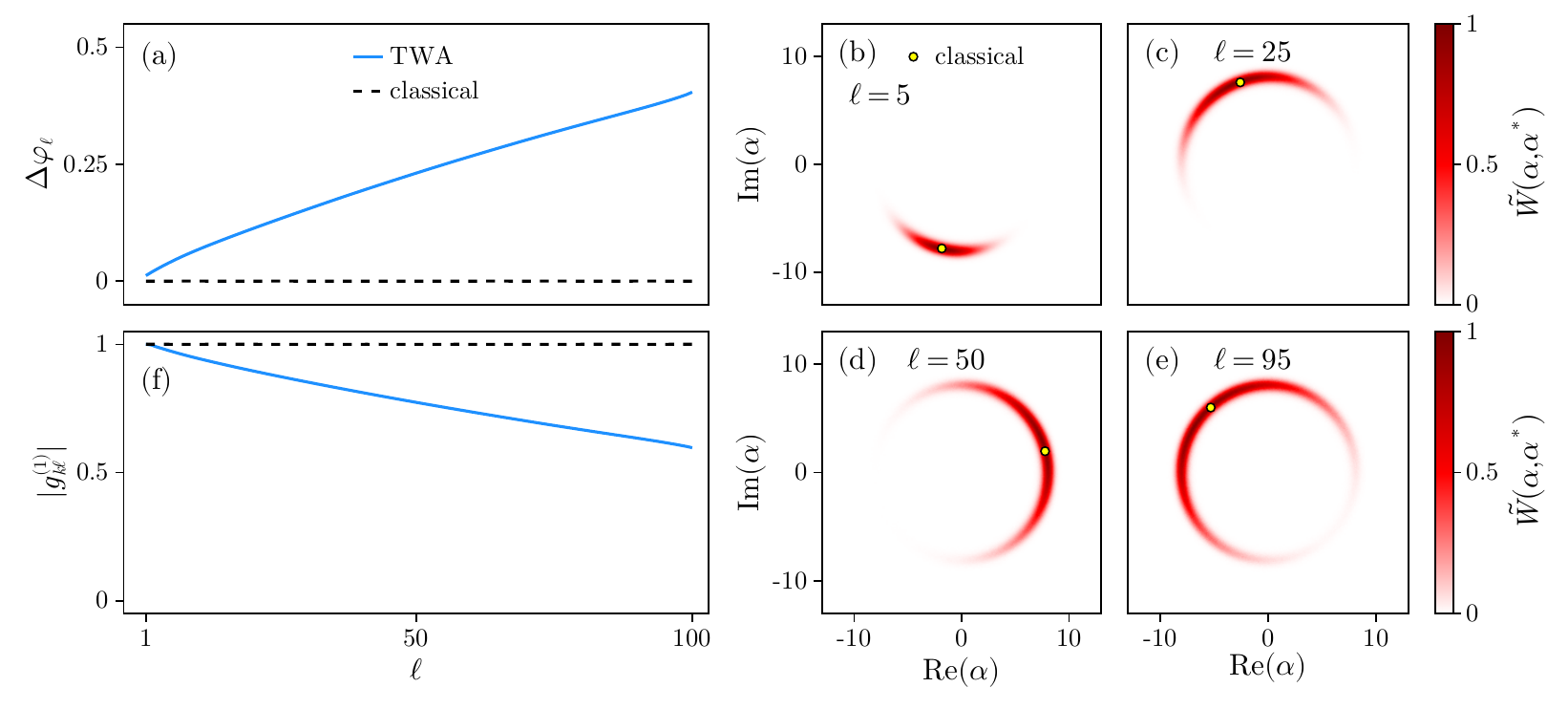}
\caption{
\cami{\textbf{RNW regime: phase decoherence from quantum fluctuations.}}
\cami{
(a) Phase fluctuations along the chain, quantified by the circular variance $\Delta\varphi_\ell := 1 - |\langle \rme^{\rmi\varphi_\ell} \rangle|$, plotted as a function of site index $\ell$ for a chain of length $L = 100$ in the RNW regime. The solid blue line corresponds to the TWA results, while the dashed black line shows the classical Gross-Pitaevskii solution.   
(b–e) Local Wigner functions $W_\ell(\alpha, \alpha^*)$ at representative sites throughout the chain. We show the normalized distribution $\tilde{W}(\alpha, \alpha^*) := W_\ell(\alpha, \alpha^*) / \max[W_\ell(\alpha, \alpha^*)]$. Yellow markers denote the classical Gross-Pitaevskii solutions.  
(f) Same as in panel (a), but for the first-order coherence function $|g^{(1)}_{k\ell}|$ with $k = 1$.
Results are computed by averaging over $N_{\rm traj} = 5\times10^3$ independent Wigner trajectories and over a time window $\Delta\tau=10^3$ once the steady state is reached.
The drive amplitude is fixed to $F=12.5$.
The other parameters are set as in Fig.~\ref{fig:phase_diagram}.
}
}
\label{fig:RNW}
\end{figure*}

Altogether, we identify three distinct spatial domains in the chaotic chain, from the left to the right:

\begin{enumerate}
    \item \textit{Nonsymmetric nonthermal domain}; The first few sites are not captured by the different hydrodynamic ans\"atze because of the proximity to the $\mathbb{U}(1)$-breaking drive. This breaking of $\mathbb{U}(1)$ symmetry is visible in the asymmetry of local Wigner functions represented in Fig.~\ref{fig:correlations} (d). Here, the quantities $T_\ell$ and $\mu_\ell$ are ill-defined, as indicated by the hatched region in the figure.
    
    \item \textit{$\mathbb{U}(1)$-symmetric prethermal domain};
    In this intermediate region of the chain, \cami{only realized under strong enough driving,}
    local quantities are well captured by hydrodynamics with a \emph{positive} effective chemical potential, $\mu_{\ell} > 0$, and a finite 2-photon decay rate $\gamma^{\rm s}$. The relatively constant number of photons reported in Fig.~\ref{fig:correlations} (a), along with the large ring-shaped Wigner functions in Fig.~\ref{fig:correlations} (e), can be interpreted as the result of the competition between a large chemical potential that strives to add photons and the Kerr non-linearity that acts as a saturation mechanism.
This state should not be interpreted as lasing: unlike standard lasers, the temporal phase coherence here is short-lived, and the $\mathbb{U}(1)$ symmetry remains unbroken (see Supplementary Information). Instead, the phase undergoes diffusive dynamics~\cite{minganti_spectral_2018}, which restores the symmetry of the chain Hamiltonian. 
\cami{From a thermodynamic perspective, this domain exhibits anomalous heating, characterized by temperature $T_\ell$ and entropy $S_\ell$ that increase with distance from the coherent drive. This originates from the fact that the energy current is predominantly transported by a flux of photons with large chemical potential, rather than through conventional heat flow.}

    \item \textit{$\mathbb{U}(1)$-symmetric thermal domain};  The right side of the chain is captured by the impurity ansatz with a negative chemical potential, $\mu_{\ell} \leq  0$.
    The corresponding Wigner functions, illustrated in Fig.~\ref{fig:correlations} (f), display bell-shaped envelopes that can be interpreted as the outcome of a competition between the chemical potential depleting the photon population towards the vacuum, and the thermal fluctuations that sustain a residual population of weakly-interacting photons. \cami{Thermodynamically, the energy current is now mostly driven by the strong thermal gradient that has built up in this region of the chain.}
\end{enumerate}

\begin{figure}[t!]
\centering
\includegraphics[width=0.47 \textwidth]{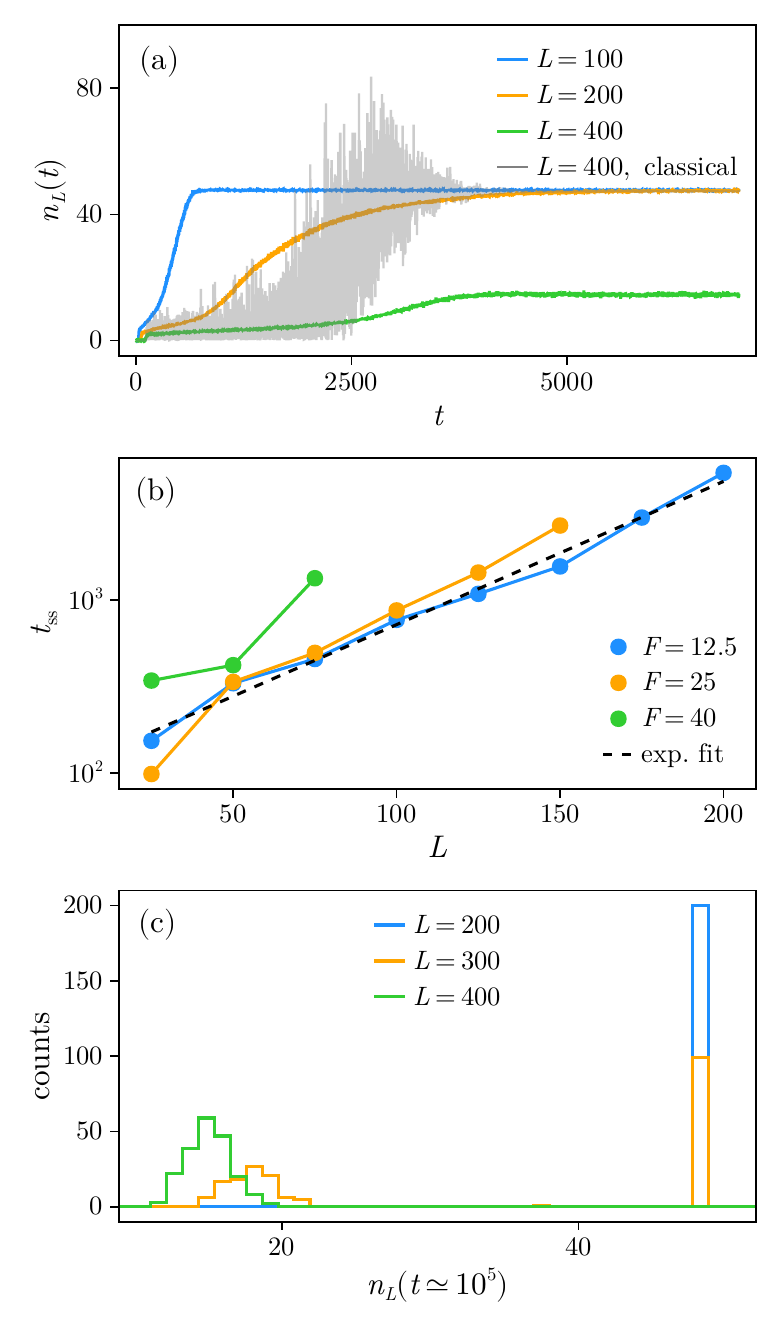}\vspace{0.8em}
\caption{
\textbf{\cami{RNW regime: destabilization by quantum fluctuations.}}
\cami{
(a) Transient dynamics of the photon number at the last site of the chain, $n_L(t)$, plotted as a function of time $t$ for system sizes $L = 100$, $200$, and $400$. The gray line is the classical Gross-Pitaevskii solution for $L = 400$.  
(b) Characteristic timescale $t_{\rm ss}$ to reach the RNW steady state as a function of system size $L$, for three different drive strengths. Each data series is truncated once the RNW regime fails to emerge within the accessible simulation time. The black dashed line is an exponential fit to the data for $F = 12.5$.  
(c) Evidence of metastability near the transition between the RNW and chaotic regimes. The photon number at the last site, $n_L(t_{\rm max} \simeq 10^5)$, is computed from $N_{\rm traj} = 200$ Wigner trajectories and time-averaged over a window $\Delta t = 10^3$. The resulting histogram is shown for $L = 200$, $300$, and $400$. 
In panels (a) and (c), the drive amplitude is fixed at $F = 12.5$.
The other parameters are set as in Fig.~\ref{fig:phase_diagram}.} 
}
\label{fig:RNW_2}
\end{figure}

Remarkably, we argue in Fig.~\ref{fig:pre_thermalization} (e) that the sizes of the non-symmetric and thermal domains are limited to the subdominant portions of the chain, while the size of the prethermal domain increases linearly with the total length of the chain $L$. This suggests that the latter may be an extensive thermodynamic phase that occurs ahead of the complete thermalization at the rightmost portion of the chain. 

\subsection{Regular RNW regime}\label{sec:RNW_regime}

\fil{In this Section, we focus on the RNW regime that was first observed in Ref.~\cite{prem_dynamics_2023} in the context of driven-dissipative Klein-Gordon chains coupled to \cami{zero-temperature} boundary baths, and further studied in Ref.~\cite{Abhishek2024}.}
\cami{We discuss how the presence of quantum fluctuations modifies the classical picture.}

\cami{
Performing a quench from the vacuum and evolving under strong coherent drive, the dynamics exhibit two successive temporal regimes, illustrated in Fig.~\ref{fig:regular_array} in terms of the average phase $\langle \varphi_\ell \rangle$ and photon density $n_\ell$. 
Initially, the system undergoes a prolonged transient during which $\varphi_\ell$ is uniformly distributed, rendering $\langle \varphi_\ell \rangle$ ill-defined away from the boundary drive. In this stage, $n_\ell$ gradually builds up along the chain, starting from the driven boundary site and progressively reaching large values throughout the entire system.
This transient corresponds to the chaotic regime discussed in the previous section.
It is followed at late times by the onset of the RNW steady state, where  $\varphi_{\ell}$  acquires a finite expectation value that slowly rotates along the chain. The dominant wavelength of these oscillations, approximately 30 sites here, as well as the large and uniform photon density, are expected to be dynamically set and independent of the total chain length $L$~\cite{Abhishek2024}.
As we show in the Supplementary Information, both $\langle \varphi_{\ell} \rangle$ and $n_\ell$ in the RNW steady state closely match their classical counterparts, indicating that these observables do not capture distinct quantum features of the RNW regime. The onset of the RNW steady state is governed by a characteristic timescale $t_{\rm ss}$ that will be discussed below.}

\subsubsection{Quantum signatures in the steady state}
\cami{
Let us now characterize the distinct quantum signatures of the RNW regime once it is established.
In Fig.~\ref{fig:RNW} (a), we monitor the phase fluctuations by means of the circular variance $\Delta\varphi_\ell := 1-|\langle \rme^{\rmi\varphi_\ell}\rangle|$
as a function of the site index $\ell$. 
In contrast to the driven-dissipative Klein-Gordon chain with zero-temperature boundary baths, where phase fluctuations are absent~\cite{prem_dynamics_2023, Abhishek2024}, we find that $\Delta \varphi_\ell$ increases monotonically along the chain.
This behavior is further illustrated in Fig.~\ref{fig:RNW} (b-e) through the Wigner functions $W_\ell\left(\alpha, \alpha^*\right)$ at four representative sites in the bulk of the chain; (b) $\ell=5$, (c) $\ell=25$, (d) $\ell=50$, (e) $\ell=95$.
The amplitude degree of freedom remains effectively frozen, while the phase undergoes diffusion. 
As a result, $W_\ell\left(\alpha, \alpha^*\right)$  is concentrated along an arc in phase space, with its angular extent gradually broadening along the chain.
The suppressed fluctuations in the radial direction underpin the sub-Poissonian photon statistics characteristic of the RNW regime.

In Fig.~\ref{fig:RNW} (f), the phase coherence $|g_{1,\ell}^{(1)}|$ exhibits long-range correlations extending across the entire chain. However, contrary to the classical prediction of a uniform $|g_{1,\ell}^{(1)}|$, we observe a slow spatial decay of phase coherence along the chain. This decay arises from quantum fluctuations, which demote the true long-range order predicted in the classical limit. Determining the precise functional form of this decay is challenging due to numerical limitations on accessible system sizes, and because large systems tend to lose the RNW regime, as we discuss below.
We also demonstrate that the RNW regime is stable against the presence of weak but finite intrinsic photon losses at all sites, up to decay rates $\gamma_{\rm int} = 10^{-3}$.
For larger values of $\gamma_{\rm int}$, the RNW first collapses into the chaotic regime and ultimately collapses to a collection of local vacuum states.}

\subsubsection{Metastability and transition to chaos}

\cami{We found compelling numerical evidence that the RNW steady state does not persist in the thermodynamic limit in the presence of quantum fluctuations. Following a quench from the vacuum under strong coherent drive, we observe that the emergence of the RNW steady state is governed by a characteristic timescale $t_{\rm ss}$ which increases exponentially with the system size $L$. Moreover, even when evolving the dynamics beyond $t_{\rm ss}$, the system fails to reach the RNW steady state for chains larger than a critical length $L > L^*$, instead remaining confined to the chaotic regime described in the previous section.
}

\cami{
This is illustrated in Fig.~\ref{fig:RNW_2} (a), where the photon density at the end of the chain, $n_L(t)$, is monitored for various $L$. While $n_L(t)$ converges to its RNW value for chains of length $L \leq 200$, it saturates to system-size-dependent values consistent with the chaotic regime for $L \gtrsim 200$. We verified the chaotic nature of this NESS using the diagnostics introduced earlier (data not shown). To highlight the quantum origin of this dynamical phenomenon, we compare with the classical case: the convergence to the RNW steady state is restored when setting the quantum fluctuations to zero in Eqs.~(\ref{eqs:stochastic_differential_equations}).
\fil{Moreover, the classical solution exhibits transient chaos before approaching the regular RNW steady state.}

Remarkably, within the quantum picture, the size $L^*$ at which the breakdown of the RNW regime occurs decreases with increasing drive strength $F$, ruling out a reentrant chaotic phase in the NESS phase diagram as a function of $L$. 
In Fig.~\ref{fig:RNW_2} (b), we demonstrate the exponential scaling of $t_{\rm ss}$ that is operationally defined via the convergence criterion $|n_L(t > t_{\rm ss}) - n_L(t_{\rm max})| < \varepsilon$, with $\varepsilon = 0.01$, whenever the RNW regime has been reached at the final time $t_{\rm max}$.

We attribute this phenomenon to the emergence of metastability: in the absence of fluctuations, multiple stationary states coexist, namely the chaotic and the RNW steady states~\citep{prasad_dissipative_2022, Abhishek2024}. The addition of quantum fluctuations allow trajectories to stochastically transition between these, and the RNW regime becomes metastable for $L>L^*$.
To probe this, we evolve an ensemble of Wigner trajectories from the vacuum up to a late time $t_{\rm max}$, and subsequently analyze the time-averaged photon number at the chain’s end over a window $\Delta t$. The latter is chosen to suppress high-frequency noise while preserving distinctions between chaotic and regular dynamics. Histograms of this observable, shown in Fig.~\ref{fig:RNW_2} (c) for a fixed $t_{\rm max}=10^5$, $\Delta t = 10^3$, reveal a clear dynamical crossover when increasing the system size: unimodal at $L=200$ (fully RNW), bimodal at $L=300$ (coexistence), and unimodal (fully chaotic) at $L=400$. 
Notably, at $L=300$, a significant fraction of the trajectories remains in the chaotic manifold although $t_{\rm max} \gg t_{\rm ss}$.
}

\section{Discussion}\label{sec:discussion}
\cami{The NESS of the boundary-driven dissipative chain of quantum nonlinear oscillators revealed an extensive spatial domain, where the chaotic dynamics drive the system to a hydrodynamic regime describable via local equilibria with a large, emergent chemical potential. While consistently with its hydrodynamic nature, this regime displays no distinct quantum signatures, it remarkably exhibits anomalous heating, with temperature increasing away from the coherent drive.}
More broadly, the emergence of a such a prethermal chaotic phase hinges on three key ingredients: first, an interacting bulk with two conserved charges; second, a local nonequilibrium drive that explicitly breaks the associated symmetries; and finally, weakly-symmetric dissipation channels positioned far from the drive, enabling nontrivial steady states and slow hydrodynamic relaxation. Together with the semiclassical framework developed here, these elements provide a minimal blueprint for realizing prethermal chaotic phases in a broad class of boundary-driven systems.

\cami{Quantum fluctuations were found to be relevant to the RNW regime: (i) inducing phase fluctuations along the chain that degrade the long-range coherence of the finite-momentum condensate, and (ii) ultimately destabilizing the condensate in sufficiently long chains.
These findings naturally raise the question of how anomalous heating in the prethermal domain and phase fluctuations in the RNW regime impact transport properties, which were reported to be superdiffusive and ballistic, respectively, in classical Klein-Gordon chains~\cite{debnath_nonequilibrium_2017, prem_dynamics_2023, garbe_bosonic_2024, muraev_signatures_2024, markovic_chaos_2024}. In particular, it is an intriguing possibility that phase fluctuations in the RNW regime might hold fingerprints of Kardar-Parisi-Zhang (KPZ) universality.
Also, it remains an open question to assess the role of quantum fluctuations, beyond semiclassics, on both the existence of the prethermal domain and the instability of the RNW regime in the thermodynamic limit.}
Resolving these challenges will demand either significantly larger numerical simulations or novel approaches beyond TWA, capable of capturing intermediate Kerr nonlinearities while incorporating quantum fluctuations.

\begin{widetext}

\section{Methods}\label{sec:methods}

\subsection{Stochastic semiclassical description}
The TWA is a semiclassical approximation of the quantum many-body dynamics that accounts for leading-order quantum fluctuations. It relies on a phase-space representation of the system's density matrix $\hat \rho$ in terms of the Wigner function $W(\alpha_1, \alpha_1^*, ..., \alpha_L, \alpha_L^*)$, where $\alpha_\ell$ and $\alpha_\ell^*$ for $\ell = 1,..., L$ are the complex amplitudes associated with the local coherent states.
In this framework, the Lindblad master equation on the operator $\hat \rho$ in Eq.~\eqref{eqs:lindblad} is mapped to a partial differential equation on $W$. 
Notably, two-body interactions in the Hamiltonian yield contributions up to third order derivatives of the type $\alpha_\ell \, \partial^3 W/ \partial \alpha^*_\ell \partial^2 \alpha_\ell$.
The approximation consists of discarding third and higher-order derivatives, reducing the dynamics to a Fokker-Planck equation where $W$ can be interpreted as a probability distribution of the phase-space variables. The latter equation can then be unraveled into a set of $L$ coupled stochastic differential equations on the complex amplitudes $\alpha_{\ell}$ given by Eq.~\eqref{eqs:stochastic_differential_equations}. 
In practice, we compute the solutions of these Langevin-like equations, the so-called Wigner trajectories, by means of numerical solvers specific to stochastic differential equations. Observables are computed by averaging over a large number of trajectories generated by different realizations of the quantum noise.

\begin{figure*}[t!]
\centering
\includegraphics[width=1 \textwidth]{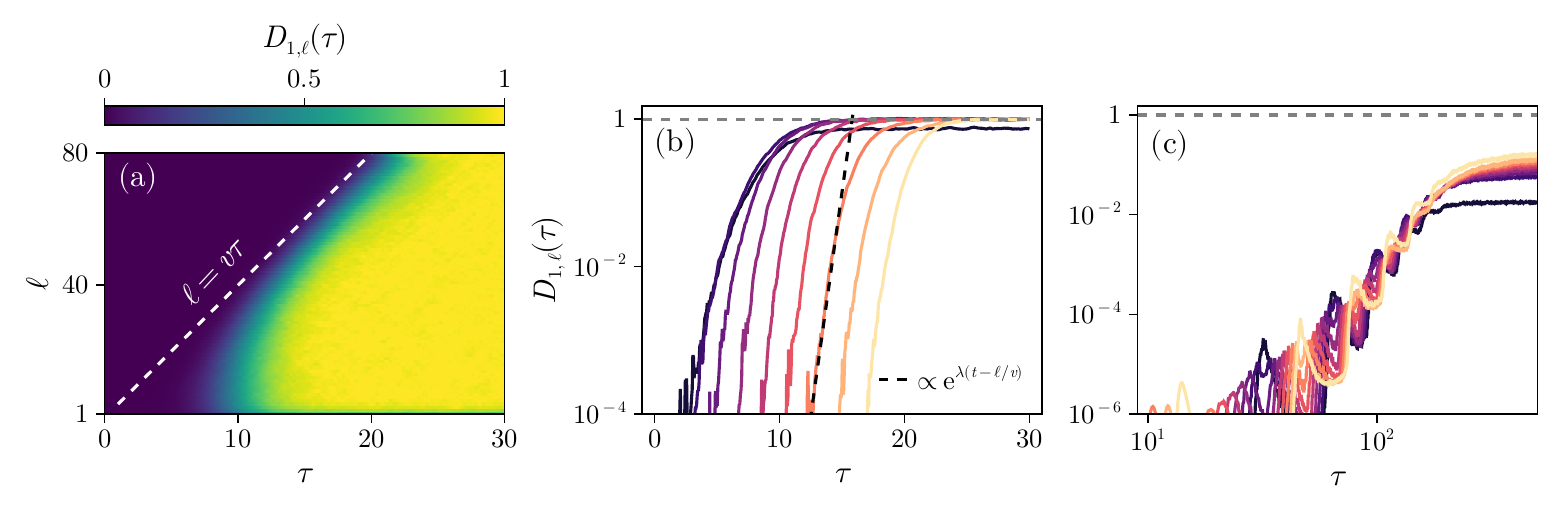}\vspace{0.8em}
\caption{
\textbf{Semiclassical OTOC dynamics}.
\cami{
(a) Spatio-temporal scrambling of information in the NESS probed by the semiclassical OTOC $D_{1, \ell}(\tau)$, defined in Eq.~(\ref{eqs:OTOC_definition2}) and computed via Eq.~(\ref{eqs:semiclassical_OTOC}), across a chain of length $L=80$ in the chaotic regime.
The dashed line indicates ballistic spreading of information at the butterfly velocity $v=2J$.
(b) Dynamics of $D_{1, \ell}(\tau)$ in the chaotic regime computed at sites $\ell=1, 10, 20, ..., 80$ (from dark purple to light pink) of an $L=80$ chain. 
The exponential growth in $\tau$ at a Lyapunov rate $\lambda \simeq 2.8$ is a hallmark of chaotic dynamics.}
(c) Dynamics of $D_{1, \ell}(\tau)$ in the RNW regime at the same sites as in (b).
The sub-exponential growth in time $\tau$ signals non-chaotic dynamics.
The dashed line indicates the ergodic bound on the saturation value of the semiclassical OTOC.
The results are computed upon averaging over $N_{\rm traj} = 5\times10^3$ independent Wigner trajectories.
In panels (a) and (b), the drive amplitude is fixed to $F=7.5$, in panel (c) to $F=12.5$.
The other parameters are set as in Fig.~\ref{fig:phase_diagram}.
}
\label{fig:OTOC}
\end{figure*}

The dictionary between the original Lindblad master equation framework and the Wigner-trajectory implementation of the TWA framework reads, notably,
\begin{center}
\begin{tabular}{|c|c|c|c|}
    \hline
    Observable & Lindblad & Wigner trajectories \\
    \hline
    Field $\langle \hat{a}_{\ell} \rangle$ & $\operatorname{Tr}[\hat{a}_{\ell} \hat{\rho}]$ & $\langle \alpha_{\ell} \rangle$\\
    \hline
    Photon number $\langle \hat{a}_{\ell}^\dagger \hat{a}_{\ell} \rangle$ & $\operatorname{Tr}[\hat{a}_{\ell}^\dagger \hat{a}_{\ell} \hat{\rho}]$ & $\langle|\alpha_{\ell}|^2\rangle -1/2$  \\
    \hline
    Spatial correlation $\langle \hat a_k^\dagger \hat a_\ell \rangle$ & $\operatorname{Tr}[\hat a_k^\dagger \hat a_\ell \hat \rho]$ & $\langle \alpha_k^*\alpha_\ell \rangle - \delta_{kl}/2 $ \\
    \hline
    Kerr nonlinearity  $\langle \hat{a}_{\ell}^{\dagger 2} \hat{a}_{\ell}^2 \rangle$ & $\operatorname{Tr}[\hat{a}_{\ell}^{\dagger2} \hat{a}_{\ell}^2 \hat{\rho}]$ & $\langle|\alpha_{\ell}|^4\rangle - 2 \langle|\alpha_{\ell}|^2\rangle +1/2$ \\
    \hline
\end{tabular}
\end{center}
\end{widetext}
where $\langle ... \rangle$ in the observable column is the standard quantum expectation value and, in the Wigner trajectories column, denotes the average with respect to Wigner trajectories.
The local Wigner function $W_\ell(t;\alpha, \alpha^*)$  is a phase-space representation of the reduced density matrix at site $\ell$, $\hat \rho_\ell(t) := \textrm{Tr}_{k\neq \ell} \, \hat \rho(t)$.
It can be simply reconstructed by generating the histogram of $\alpha_\ell(t)$ in the complex plane when sampling over Wigner trajectories.
When the NESS is reached, single-time observables converge to constant values and $W_\ell(t;\alpha, \alpha^*) \to W_\ell(\alpha, \alpha^*)$.
There, the statistics can be improved by also sampling the trajectories in time,  significantly reducing the computational overhead.

\subsection{Semiclassical OTOC}\label{sec:phase_OTOC}
\fil{
Here we describe in detail our probe of semiclassical chaos, the semiclassical OTOC of Eq.~\eqref{eqs:semiclassical_OTOC}.
We show how Eq.~\eqref{eqs:semiclassical_OTOC} can be derived from the quantum formulation of the out-of-time order correlators, }
\begin{align} \label{eq:SC-OTOC}
  C_{k,\ell}(t, \tau) \!:=\! \frac12  \textrm{Tr} \Big{(} [\hat B_\ell(t\!+\!\tau), \hat A_k(t)]^\dagger [\hat B_\ell(t\!+\!\tau), \hat A_k(t)]\hat \rho(t) \Big{)}.
\end{align}
The above OTOC involves a ``square commutator'' between site $k$ at time $t$ and site $\ell$ at a later time $t+\tau$. The operators $\hat A_\ell = \hat n_\ell := \sum_{n=0}^\infty n\, |n\rangle_\ell \langle n |_\ell $ and $\hat B_\ell = \rme^{\rmi \hat \varphi_\ell} :=\sum_{n=0}^\infty |n\rangle_\ell \langle n +1|_\ell $ are chosen to decompose the local field operator into $\hat a_\ell = \sqrt{\hat n_\ell} \, \rme^{\rmi \hat \varphi_\ell}$. 
They obey the quantum commutation relation $[\rme^{\rmi \hat \varphi_k}, \hat n_\ell] =  \delta_{k \ell} \, \rme^{\rmi \hat \varphi_\ell}$.
\cami{Quantum mechanically, while the absence of a well-defined phase operator is known, the operator $\rme^{\rmi \hat \varphi_\ell}$ is well defined.}
%\fab{Quantum mechanically, the operator $\hat{\varphi}_\ell$ presents defining issues due to the discrete, unbounded-from-above but bounded-from-below spectrum of $\hat{n}_\ell$.}

\fab{In a semiclassical approach, these operators are replaced with c-numbers, namely local number $n_\ell$ and phase $\varphi_\ell$, and \cami{the latter is} well defined.}
%In a semiclassical approach, these operators are replaced with c-numbers, namely local number $n_\ell$ and phase $\varphi_\ell$, 
The commutation relations are replaced with $\left\{ n_k, \varphi_\ell \right\} = \delta_{k \ell} $, where $\{ \cdot, \cdot\}$ denotes the Poisson brackets defined as
\begin{align}
    \{ f, g \} := \sum_{\ell=1}^L \left(\frac{\partial f}{\partial n_\ell} \frac{\partial g}{\partial \varphi_\ell}
    - 
    \frac{\partial g}{\partial n_\ell} \frac{\partial f}{\partial \varphi_\ell}\right)\,.
\end{align}
Carrying out this replacement in Eq.~(\ref{eq:SC-OTOC}), using the relation $ \{ \varphi_\ell(t') , n_k(t) \} = - \delta \varphi_\ell(t') / \delta \varphi_k(t) $, one obtains the semiclassical version of $C_{k,\ell}(t, \tau)$ which reads
\begin{align}\label{eqs:OTOC_definition2}
   D_{k,\ell}(t, \tau) & 
   = \frac{1}{2}\left\langle \left| \frac{\delta \rme^{\rmi \varphi_\ell(t+\tau)}}{\delta \varphi_{k}(t)} \right|^2\right\rangle \,,
\end{align}
where $\langle ... \rangle$ denote the average over the quantum noise and ${\delta }/{\delta \varphi_{k}(t)}$ implements an infinitesimal perturbation of the phase at site $k$ and time $t$.
In practice, this is implemented by cloning the system in two replicas $a$ and $b$, applying an infinitesimal perturbation at time $t$ on the phase at site $k$ of replica $b$, subsequently evolving both replicas subject to the same quantum noise, and finally averaging over realizations of the quantum noise. Thus, the semiclassical phase OTOC can be cast as
\begin{equation}
    D_{k, \ell}(t, \tau) = \frac{1}{2}\left\langle\left|\rme^{\rmi\varphi_\ell^{(a)}(t+\tau)} - \rme^{\rmi\varphi_\ell^{(b)}(t+\tau)}\right|^2\right\rangle\,,
\end{equation}
which boils down to the operational definition given in Eq.~(\ref{eqs:semiclassical_OTOC}).

Whenever the dynamics reaches a steady state, $\hat \rho := \lim\limits_{t\to\infty} \hat \rho(t) $, we can define the semiclassical steady-state phase OTOC as $D_{k,\ell}(\tau) := \lim\limits_{t\to\infty} D_{k,\ell}(t, \tau)$.
\fil{Throughout the paper, we focus on $D_{k,\ell}(\tau)$ only.}

\fil{We now show that our semiclassical OTOC is able to capture the basic features of quantum information spreading in extended many-body systems~\cite{xu_scrambling_2024}, as well as the initial exponential growth in the presence of chaos.
On general grounds, we have $D_{k,\ell}( \tau = 0) =0$, while $D_{k,\ell}( \tau > 0)$ is expected to increase with $\tau$ whenever trajectories in the two replicas start deviating, and, eventually, to saturate to a finite value $D_{k,\ell}(\tau \to \infty)$. Both the growth and the saturation regimes of $D_{k,\ell}(\tau)$ shed light on the chaotic versus regular nature of the dynamics in the NESS.
In Fig.~\ref{fig:OTOC} (a) we show the space-time dynamics of $D_{1, \ell}(\tau)$ in the chaotic regime of the boundary-driven dissipative Bose-Hubbard chain Eq.~\eqref{eqs:hamiltonian}, for a system's size equal to $L=80$.
We see how $D_{1, \ell}(\tau)$ exhibits a causal light-cone structure with a ballistic spreading of information characterized by a butterfly velocity $v=2J$~\cite{lieb_finite_1972}.
In Fig.~\ref{fig:OTOC} (b) and (c), we illustrate the growth of the steady-state semiclassical OTOC for two representative values of the drive strength $F$, one for the chaotic regime, one for the regular RNW regime, always for a chain with $L=80$.
In panel (b), $D_{1,\ell}(\tau)$ shows a rapid exponential growth of the form $D_{1,\ell}(\tau) \sim \exp[\lambda (t  - \ell/v) ]$ where $\lambda$ is a Lyapunov rate, followed by a saturation regime where $D_{1,\ell}(\tau\to\infty) \simeq 1$, \textit{i.e.}, maximal decorrelation.  
Altogether, the semiclassical OTOC $D_{1,\,\ell}(\tau)$ captures both the Lyapunov growth and the saturation regimes expected of quantum chaotic dynamics.
In panel (c), $D_{1,\ell}(\tau)$ instead displays early-time oscillations and the overall growth is slower than exponential. Eventually, the late-time saturation value $D_{1,\ell}(\tau)$ is significantly less than $1$. This is indicative of regular dynamics.
}

\bigskip

\begin{acknowledgments}
\fil{We are grateful to Andreas L\"auchli for invaluable suggestions on the analysis of local thermal equilibrium.} We also acknowledge enlightening discussions with Alberto Biella, \cami{Abhishek Dhar}, Lorenzo Fioroni, Luca Gravina,  \cami{Manas Kulkarni}, Alberto Mercurio,  and Pasquale Scarlino.
CA acknowledges the support from the French ANR ``MoMA'' project ANR-19-CE30-0020 and the support from the project 6004-1 of the Indo-French Centre for the Promotion of Advanced Research (IFCPAR). FF, FM, and VS acknowledge the support by the Swiss National Science Foundation through Projects No. 200020\_185015, 200020\_215172, and 20QU-1\_215928, and by the EPFL Science Seed Fund 2021.
\end{acknowledgments}

\bibliography{apssamp}

\begin{thebibliography}{103}%
\makeatletter
\providecommand \@ifxundefined [1]{%
 \@ifx{#1\undefined}
}%
\providecommand \@ifnum [1]{%
 \ifnum #1\expandafter \@firstoftwo
 \else \expandafter \@secondoftwo
 \fi
}%
\providecommand \@ifx [1]{%
 \ifx #1\expandafter \@firstoftwo
 \else \expandafter \@secondoftwo
 \fi
}%
\providecommand \natexlab [1]{#1}%
\providecommand \enquote  [1]{``#1''}%
\providecommand \bibnamefont  [1]{#1}%
\providecommand \bibfnamefont [1]{#1}%
\providecommand \citenamefont [1]{#1}%
\providecommand \href@noop [0]{\@secondoftwo}%
\providecommand \href [0]{\begingroup \@sanitize@url \@href}%
\providecommand \@href[1]{\@@startlink{#1}\@@href}%
\providecommand \@@href[1]{\endgroup#1\@@endlink}%
\providecommand \@sanitize@url [0]{\catcode `\\12\catcode `\$12\catcode
  `\&12\catcode `\#12\catcode `\^12\catcode `\_12\catcode `\%12\relax}%
\providecommand \@@startlink[1]{}%
\providecommand \@@endlink[0]{}%
\providecommand \url  [0]{\begingroup\@sanitize@url \@url }%
\providecommand \@url [1]{\endgroup\@href {#1}{\urlprefix }}%
\providecommand \urlprefix  [0]{URL }%
\providecommand \Eprint [0]{\href }%
\providecommand \doibase [0]{https://doi.org/}%
\providecommand \selectlanguage [0]{\@gobble}%
\providecommand \bibinfo  [0]{\@secondoftwo}%
\providecommand \bibfield  [0]{\@secondoftwo}%
\providecommand \translation [1]{[#1]}%
\providecommand \BibitemOpen [0]{}%
\providecommand \bibitemStop [0]{}%
\providecommand \bibitemNoStop [0]{.\EOS\space}%
\providecommand \EOS [0]{\spacefactor3000\relax}%
\providecommand \BibitemShut  [1]{\csname bibitem#1\endcsname}%
\let\auto@bib@innerbib\@empty
%</preamble>
\bibitem [{\citenamefont {Strogatz}(2018)}]{strogatz_nonlinear_2018}%
  \BibitemOpen
  \bibfield  {author} {\bibinfo {author} {\bibfnamefont {S.~H.}\ \bibnamefont
  {Strogatz}},\ }\href {https://doi.org/10.1201/9780429492563} {\emph {\bibinfo
  {title} {Nonlinear {Dynamics} and {Chaos}}}},\ \bibinfo {edition} {1st}\ ed.\
  (\bibinfo  {publisher} {CRC Press},\ \bibinfo {year} {2018})\BibitemShut
  {NoStop}%
\bibitem [{\citenamefont {D'Alessio}\ \emph {et~al.}(2016)\citenamefont
  {D'Alessio}, \citenamefont {Kafri}, \citenamefont {Polkovnikov},\ and\
  \citenamefont {Rigol}}]{dalessio_quantum_2016}%
  \BibitemOpen
  \bibfield  {author} {\bibinfo {author} {\bibfnamefont {L.}~\bibnamefont
  {D'Alessio}}, \bibinfo {author} {\bibfnamefont {Y.}~\bibnamefont {Kafri}},
  \bibinfo {author} {\bibfnamefont {A.}~\bibnamefont {Polkovnikov}},\ and\
  \bibinfo {author} {\bibfnamefont {M.}~\bibnamefont {Rigol}},\ }\bibfield
  {title} {\bibinfo {title} {From quantum chaos and eigenstate thermalization
  to statistical mechanics and thermodynamics},\ }\href
  {https://doi.org/10.1080/00018732.2016.1198134} {\bibfield  {journal}
  {\bibinfo  {journal} {Adv. Phys.}\ }\textbf {\bibinfo {volume} {65}},\
  \bibinfo {pages} {239} (\bibinfo {year} {2016})}\BibitemShut {NoStop}%
\bibitem [{\citenamefont {Hashimoto}\ \emph {et~al.}(2017)\citenamefont
  {Hashimoto}, \citenamefont {Murata},\ and\ \citenamefont
  {Yoshii}}]{hashimoto_out--time-order_2017}%
  \BibitemOpen
  \bibfield  {author} {\bibinfo {author} {\bibfnamefont {K.}~\bibnamefont
  {Hashimoto}}, \bibinfo {author} {\bibfnamefont {K.}~\bibnamefont {Murata}},\
  and\ \bibinfo {author} {\bibfnamefont {R.}~\bibnamefont {Yoshii}},\
  }\bibfield  {title} {\bibinfo {title} {Out-of-time-order correlators in
  quantum mechanics},\ }\href {https://doi.org/10.1007/JHEP10(2017)138}
  {\bibfield  {journal} {\bibinfo  {journal} {J. High Energy Phys.}\ }\textbf
  {\bibinfo {volume} {2017}}\bibinfo  {number} { (10)},\ \bibinfo {pages}
  {138}}\BibitemShut {NoStop}%
\bibitem [{\citenamefont {Berke}\ \emph {et~al.}(2022)\citenamefont {Berke},
  \citenamefont {Varvelis}, \citenamefont {Trebst}, \citenamefont {Altland},\
  and\ \citenamefont {DiVincenzo}}]{berke_transmon_2022}%
  \BibitemOpen
\bibfield  {number} {  }\bibfield  {author} {\bibinfo {author} {\bibfnamefont
  {C.}~\bibnamefont {Berke}}, \bibinfo {author} {\bibfnamefont
  {E.}~\bibnamefont {Varvelis}}, \bibinfo {author} {\bibfnamefont
  {S.}~\bibnamefont {Trebst}}, \bibinfo {author} {\bibfnamefont
  {A.}~\bibnamefont {Altland}},\ and\ \bibinfo {author} {\bibfnamefont {D.~P.}\
  \bibnamefont {DiVincenzo}},\ }\bibfield  {title} {\bibinfo {title} {Transmon
  platform for quantum computing challenged by chaotic fluctuations},\ }\href
  {https://doi.org/10.1038/s41467-022-29940-y} {\bibfield  {journal} {\bibinfo
  {journal} {Nat. Commun.}\ }\textbf {\bibinfo {volume} {13}},\ \bibinfo
  {pages} {2495} (\bibinfo {year} {2022})}\BibitemShut {NoStop}%
\bibitem [{\citenamefont {Shillito}\ \emph {et~al.}(2022)\citenamefont
  {Shillito}, \citenamefont {Petrescu}, \citenamefont {Cohen}, \citenamefont
  {Beall}, \citenamefont {Hauru}, \citenamefont {Ganahl}, \citenamefont
  {Lewis}, \citenamefont {Vidal},\ and\ \citenamefont
  {Blais}}]{shillito_dynamics_2022}%
  \BibitemOpen
  \bibfield  {author} {\bibinfo {author} {\bibfnamefont {R.}~\bibnamefont
  {Shillito}}, \bibinfo {author} {\bibfnamefont {A.}~\bibnamefont {Petrescu}},
  \bibinfo {author} {\bibfnamefont {J.}~\bibnamefont {Cohen}}, \bibinfo
  {author} {\bibfnamefont {J.}~\bibnamefont {Beall}}, \bibinfo {author}
  {\bibfnamefont {M.}~\bibnamefont {Hauru}}, \bibinfo {author} {\bibfnamefont
  {M.}~\bibnamefont {Ganahl}}, \bibinfo {author} {\bibfnamefont {A.~G.}\
  \bibnamefont {Lewis}}, \bibinfo {author} {\bibfnamefont {G.}~\bibnamefont
  {Vidal}},\ and\ \bibinfo {author} {\bibfnamefont {A.}~\bibnamefont {Blais}},\
  }\bibfield  {title} {\bibinfo {title} {Dynamics of {Transmon} {Ionization}},\
  }\href {https://doi.org/10.1103/PhysRevApplied.18.034031} {\bibfield
  {journal} {\bibinfo  {journal} {Phys. Rev. Appl.}\ }\textbf {\bibinfo
  {volume} {18}},\ \bibinfo {pages} {034031} (\bibinfo {year}
  {2022})}\BibitemShut {NoStop}%
\bibitem [{\citenamefont {Cohen}\ \emph {et~al.}(2023)\citenamefont {Cohen},
  \citenamefont {Petrescu}, \citenamefont {Shillito},\ and\ \citenamefont
  {Blais}}]{cohen_reminiscence_2023}%
  \BibitemOpen
  \bibfield  {author} {\bibinfo {author} {\bibfnamefont {J.}~\bibnamefont
  {Cohen}}, \bibinfo {author} {\bibfnamefont {A.}~\bibnamefont {Petrescu}},
  \bibinfo {author} {\bibfnamefont {R.}~\bibnamefont {Shillito}},\ and\
  \bibinfo {author} {\bibfnamefont {A.}~\bibnamefont {Blais}},\ }\bibfield
  {title} {\bibinfo {title} {Reminiscence of {Classical} {Chaos} in {Driven}
  {Transmons}},\ }\href {https://doi.org/10.1103/PRXQuantum.4.020312}
  {\bibfield  {journal} {\bibinfo  {journal} {PRX Quantum}\ }\textbf {\bibinfo
  {volume} {4}},\ \bibinfo {pages} {020312} (\bibinfo {year}
  {2023})}\BibitemShut {NoStop}%
\bibitem [{\citenamefont {Dumas}\ \emph {et~al.}(2024)\citenamefont {Dumas},
  \citenamefont {Groleau-Paré}, \citenamefont {McDonald}, \citenamefont
  {Muñoz-Arias}, \citenamefont {Lledó}, \citenamefont {D’Anjou},\ and\
  \citenamefont {Blais}}]{dumas_measurement-induced_2024}%
  \BibitemOpen
  \bibfield  {author} {\bibinfo {author} {\bibfnamefont {M.~F.}\ \bibnamefont
  {Dumas}}, \bibinfo {author} {\bibfnamefont {B.}~\bibnamefont
  {Groleau-Paré}}, \bibinfo {author} {\bibfnamefont {A.}~\bibnamefont
  {McDonald}}, \bibinfo {author} {\bibfnamefont {M.~H.}\ \bibnamefont
  {Muñoz-Arias}}, \bibinfo {author} {\bibfnamefont {C.}~\bibnamefont
  {Lledó}}, \bibinfo {author} {\bibfnamefont {B.}~\bibnamefont {D’Anjou}},\
  and\ \bibinfo {author} {\bibfnamefont {A.}~\bibnamefont {Blais}},\ }\bibfield
   {title} {\bibinfo {title} {Measurement-{Induced} {Transmon} {Ionization}},\
  }\href {https://doi.org/10.1103/PhysRevX.14.041023} {\bibfield  {journal}
  {\bibinfo  {journal} {Phys. Rev. X}\ }\textbf {\bibinfo {volume} {14}},\
  \bibinfo {pages} {041023} (\bibinfo {year} {2024})}\BibitemShut {NoStop}%
\bibitem [{\citenamefont {García-Mata}\ \emph {et~al.}(2025)\citenamefont
  {García-Mata}, \citenamefont {Reynoso}, \citenamefont {Cortiñas},
  \citenamefont {Chávez-Carlos}, \citenamefont {Batista}, \citenamefont
  {Santos},\ and\ \citenamefont {Wisniacki}}]{garcia-mata_impact_2025}%
  \BibitemOpen
  \bibfield  {author} {\bibinfo {author} {\bibfnamefont {I.}~\bibnamefont
  {García-Mata}}, \bibinfo {author} {\bibfnamefont {M.~A.~P.}\ \bibnamefont
  {Reynoso}}, \bibinfo {author} {\bibfnamefont {R.~G.}\ \bibnamefont
  {Cortiñas}}, \bibinfo {author} {\bibfnamefont {J.}~\bibnamefont
  {Chávez-Carlos}}, \bibinfo {author} {\bibfnamefont {V.~S.}\ \bibnamefont
  {Batista}}, \bibinfo {author} {\bibfnamefont {L.~F.}\ \bibnamefont
  {Santos}},\ and\ \bibinfo {author} {\bibfnamefont {D.~A.}\ \bibnamefont
  {Wisniacki}},\ }\bibfield  {title} {\bibinfo {title} {Impact of chaos on the
  excited-state quantum phase transition of the {Kerr} parametric oscillator},\
  }\href {https://doi.org/10.1103/PhysRevA.111.L031502} {\bibfield  {journal}
  {\bibinfo  {journal} {Phys. Rev. A}\ }\textbf {\bibinfo {volume} {111}},\
  \bibinfo {pages} {L031502} (\bibinfo {year} {2025})}\BibitemShut {NoStop}%
\bibitem [{\citenamefont {Chávez-Carlos}\ \emph {et~al.}(2025)\citenamefont
  {Chávez-Carlos}, \citenamefont {Prado~Reynoso}, \citenamefont {Cortiñas},
  \citenamefont {García-Mata}, \citenamefont {Batista}, \citenamefont
  {Pérez-Bernal}, \citenamefont {Wisniacki},\ and\ \citenamefont
  {Santos}}]{chavez-carlos_driving_2025}%
  \BibitemOpen
  \bibfield  {author} {\bibinfo {author} {\bibfnamefont {J.}~\bibnamefont
  {Chávez-Carlos}}, \bibinfo {author} {\bibfnamefont {M.~A.}\ \bibnamefont
  {Prado~Reynoso}}, \bibinfo {author} {\bibfnamefont {R.~G.}\ \bibnamefont
  {Cortiñas}}, \bibinfo {author} {\bibfnamefont {I.}~\bibnamefont
  {García-Mata}}, \bibinfo {author} {\bibfnamefont {V.~S.}\ \bibnamefont
  {Batista}}, \bibinfo {author} {\bibfnamefont {F.}~\bibnamefont
  {Pérez-Bernal}}, \bibinfo {author} {\bibfnamefont {D.~A.}\ \bibnamefont
  {Wisniacki}},\ and\ \bibinfo {author} {\bibfnamefont {L.~F.}\ \bibnamefont
  {Santos}},\ }\bibfield  {title} {\bibinfo {title} {Driving superconducting
  qubits into chaos},\ }\href {https://doi.org/10.1088/2058-9565/ad93fb}
  {\bibfield  {journal} {\bibinfo  {journal} {Quantum Sci. Technol.}\ }\textbf
  {\bibinfo {volume} {10}},\ \bibinfo {pages} {015039} (\bibinfo {year}
  {2025})}\BibitemShut {NoStop}%
\bibitem [{\citenamefont {Liu}\ and\ \citenamefont
  {Houck}(2017)}]{liu_quantum_2017}%
  \BibitemOpen
  \bibfield  {author} {\bibinfo {author} {\bibfnamefont {Y.}~\bibnamefont
  {Liu}}\ and\ \bibinfo {author} {\bibfnamefont {A.~A.}\ \bibnamefont
  {Houck}},\ }\bibfield  {title} {\bibinfo {title} {Quantum electrodynamics
  near a photonic bandgap},\ }\href {https://doi.org/10.1038/nphys3834}
  {\bibfield  {journal} {\bibinfo  {journal} {Nat. Phys.}\ }\textbf {\bibinfo
  {volume} {13}},\ \bibinfo {pages} {48} (\bibinfo {year} {2017})}\BibitemShut
  {NoStop}%
\bibitem [{\citenamefont {Sundaresan}\ \emph {et~al.}(2019)\citenamefont
  {Sundaresan}, \citenamefont {Lundgren}, \citenamefont {Zhu}, \citenamefont
  {Gorshkov},\ and\ \citenamefont {Houck}}]{sundaresan_interacting_2019}%
  \BibitemOpen
  \bibfield  {author} {\bibinfo {author} {\bibfnamefont {N.~M.}\ \bibnamefont
  {Sundaresan}}, \bibinfo {author} {\bibfnamefont {R.}~\bibnamefont
  {Lundgren}}, \bibinfo {author} {\bibfnamefont {G.}~\bibnamefont {Zhu}},
  \bibinfo {author} {\bibfnamefont {A.~V.}\ \bibnamefont {Gorshkov}},\ and\
  \bibinfo {author} {\bibfnamefont {A.~A.}\ \bibnamefont {Houck}},\ }\bibfield
  {title} {\bibinfo {title} {Interacting {Qubit}-{Photon} {Bound} {States} with
  {Superconducting} {Circuits}},\ }\href
  {https://doi.org/10.1103/PhysRevX.9.011021} {\bibfield  {journal} {\bibinfo
  {journal} {Phys. Rev. X}\ }\textbf {\bibinfo {volume} {9}},\ \bibinfo {pages}
  {011021} (\bibinfo {year} {2019})}\BibitemShut {NoStop}%
\bibitem [{\citenamefont {Scigliuzzo}\ \emph {et~al.}(2022)\citenamefont
  {Scigliuzzo}, \citenamefont {Calajò}, \citenamefont {Ciccarello},
  \citenamefont {Perez~Lozano}, \citenamefont {Bengtsson}, \citenamefont
  {Scarlino}, \citenamefont {Wallraff}, \citenamefont {Chang}, \citenamefont
  {Delsing},\ and\ \citenamefont {Gasparinetti}}]{scigliuzzo_controlling_2022}%
  \BibitemOpen
  \bibfield  {author} {\bibinfo {author} {\bibfnamefont {M.}~\bibnamefont
  {Scigliuzzo}}, \bibinfo {author} {\bibfnamefont {G.}~\bibnamefont {Calajò}},
  \bibinfo {author} {\bibfnamefont {F.}~\bibnamefont {Ciccarello}}, \bibinfo
  {author} {\bibfnamefont {D.}~\bibnamefont {Perez~Lozano}}, \bibinfo {author}
  {\bibfnamefont {A.}~\bibnamefont {Bengtsson}}, \bibinfo {author}
  {\bibfnamefont {P.}~\bibnamefont {Scarlino}}, \bibinfo {author}
  {\bibfnamefont {A.}~\bibnamefont {Wallraff}}, \bibinfo {author}
  {\bibfnamefont {D.}~\bibnamefont {Chang}}, \bibinfo {author} {\bibfnamefont
  {P.}~\bibnamefont {Delsing}},\ and\ \bibinfo {author} {\bibfnamefont
  {S.}~\bibnamefont {Gasparinetti}},\ }\bibfield  {title} {\bibinfo {title}
  {Controlling {Atom}-{Photon} {Bound} {States} in an {Array} of
  {Josephson}-{Junction} {Resonators}},\ }\href
  {https://doi.org/10.1103/PhysRevX.12.031036} {\bibfield  {journal} {\bibinfo
  {journal} {Phys. Rev. X}\ }\textbf {\bibinfo {volume} {12}},\ \bibinfo
  {pages} {031036} (\bibinfo {year} {2022})}\BibitemShut {NoStop}%
\bibitem [{\citenamefont {Karamlou}\ \emph {et~al.}(2024)\citenamefont
  {Karamlou}, \citenamefont {Rosen}, \citenamefont {Muschinske}, \citenamefont
  {Barrett}, \citenamefont {Di~Paolo}, \citenamefont {Ding}, \citenamefont
  {Harrington}, \citenamefont {Hays}, \citenamefont {Das}, \citenamefont {Kim},
  \citenamefont {Niedzielski}, \citenamefont {Schuldt}, \citenamefont
  {Serniak}, \citenamefont {Schwartz}, \citenamefont {Yoder}, \citenamefont
  {Gustavsson}, \citenamefont {Yanay}, \citenamefont {Grover},\ and\
  \citenamefont {Oliver}}]{karamlou_probing_2024}%
  \BibitemOpen
  \bibfield  {author} {\bibinfo {author} {\bibfnamefont {A.~H.}\ \bibnamefont
  {Karamlou}}, \bibinfo {author} {\bibfnamefont {I.~T.}\ \bibnamefont {Rosen}},
  \bibinfo {author} {\bibfnamefont {S.~E.}\ \bibnamefont {Muschinske}},
  \bibinfo {author} {\bibfnamefont {C.~N.}\ \bibnamefont {Barrett}}, \bibinfo
  {author} {\bibfnamefont {A.}~\bibnamefont {Di~Paolo}}, \bibinfo {author}
  {\bibfnamefont {L.}~\bibnamefont {Ding}}, \bibinfo {author} {\bibfnamefont
  {P.~M.}\ \bibnamefont {Harrington}}, \bibinfo {author} {\bibfnamefont
  {M.}~\bibnamefont {Hays}}, \bibinfo {author} {\bibfnamefont {R.}~\bibnamefont
  {Das}}, \bibinfo {author} {\bibfnamefont {D.~K.}\ \bibnamefont {Kim}},
  \bibinfo {author} {\bibfnamefont {B.~M.}\ \bibnamefont {Niedzielski}},
  \bibinfo {author} {\bibfnamefont {M.}~\bibnamefont {Schuldt}}, \bibinfo
  {author} {\bibfnamefont {K.}~\bibnamefont {Serniak}}, \bibinfo {author}
  {\bibfnamefont {M.~E.}\ \bibnamefont {Schwartz}}, \bibinfo {author}
  {\bibfnamefont {J.~L.}\ \bibnamefont {Yoder}}, \bibinfo {author}
  {\bibfnamefont {S.}~\bibnamefont {Gustavsson}}, \bibinfo {author}
  {\bibfnamefont {Y.}~\bibnamefont {Yanay}}, \bibinfo {author} {\bibfnamefont
  {J.~A.}\ \bibnamefont {Grover}},\ and\ \bibinfo {author} {\bibfnamefont
  {W.~D.}\ \bibnamefont {Oliver}},\ }\bibfield  {title} {\bibinfo {title}
  {Probing entanglement in a {2D} hard-core {Bose}–{Hubbard} lattice},\
  }\href {https://doi.org/10.1038/s41586-024-07325-z} {\bibfield  {journal}
  {\bibinfo  {journal} {Nature}\ }\textbf {\bibinfo {volume} {629}},\ \bibinfo
  {pages} {561} (\bibinfo {year} {2024})}\BibitemShut {NoStop}%
\bibitem [{\citenamefont {Braumüller}\ \emph {et~al.}(2022)\citenamefont
  {Braumüller}, \citenamefont {Karamlou}, \citenamefont {Yanay}, \citenamefont
  {Kannan}, \citenamefont {Kim}, \citenamefont {Kjaergaard}, \citenamefont
  {Melville}, \citenamefont {Niedzielski}, \citenamefont {Sung}, \citenamefont
  {Vepsäläinen}, \citenamefont {Winik}, \citenamefont {Yoder}, \citenamefont
  {Orlando}, \citenamefont {Gustavsson}, \citenamefont {Tahan},\ and\
  \citenamefont {Oliver}}]{braumuller_probing_2022}%
  \BibitemOpen
  \bibfield  {author} {\bibinfo {author} {\bibfnamefont {J.}~\bibnamefont
  {Braumüller}}, \bibinfo {author} {\bibfnamefont {A.~H.}\ \bibnamefont
  {Karamlou}}, \bibinfo {author} {\bibfnamefont {Y.}~\bibnamefont {Yanay}},
  \bibinfo {author} {\bibfnamefont {B.}~\bibnamefont {Kannan}}, \bibinfo
  {author} {\bibfnamefont {D.}~\bibnamefont {Kim}}, \bibinfo {author}
  {\bibfnamefont {M.}~\bibnamefont {Kjaergaard}}, \bibinfo {author}
  {\bibfnamefont {A.}~\bibnamefont {Melville}}, \bibinfo {author}
  {\bibfnamefont {B.~M.}\ \bibnamefont {Niedzielski}}, \bibinfo {author}
  {\bibfnamefont {Y.}~\bibnamefont {Sung}}, \bibinfo {author} {\bibfnamefont
  {A.}~\bibnamefont {Vepsäläinen}}, \bibinfo {author} {\bibfnamefont
  {R.}~\bibnamefont {Winik}}, \bibinfo {author} {\bibfnamefont {J.~L.}\
  \bibnamefont {Yoder}}, \bibinfo {author} {\bibfnamefont {T.~P.}\ \bibnamefont
  {Orlando}}, \bibinfo {author} {\bibfnamefont {S.}~\bibnamefont {Gustavsson}},
  \bibinfo {author} {\bibfnamefont {C.}~\bibnamefont {Tahan}},\ and\ \bibinfo
  {author} {\bibfnamefont {W.~D.}\ \bibnamefont {Oliver}},\ }\bibfield  {title}
  {\bibinfo {title} {Probing quantum information propagation with
  out-of-time-ordered correlators},\ }\href
  {https://doi.org/10.1038/s41567-021-01430-w} {\bibfield  {journal} {\bibinfo
  {journal} {Nat. Phys.}\ }\textbf {\bibinfo {volume} {18}},\ \bibinfo {pages}
  {172} (\bibinfo {year} {2022})}\BibitemShut {NoStop}%
\bibitem [{\citenamefont {Zhang}\ \emph {et~al.}(2023)\citenamefont {Zhang},
  \citenamefont {Kim}, \citenamefont {Mark}, \citenamefont {Choi},\ and\
  \citenamefont {Painter}}]{zhang_superconducting_2023}%
  \BibitemOpen
  \bibfield  {author} {\bibinfo {author} {\bibfnamefont {X.}~\bibnamefont
  {Zhang}}, \bibinfo {author} {\bibfnamefont {E.}~\bibnamefont {Kim}}, \bibinfo
  {author} {\bibfnamefont {D.~K.}\ \bibnamefont {Mark}}, \bibinfo {author}
  {\bibfnamefont {S.}~\bibnamefont {Choi}},\ and\ \bibinfo {author}
  {\bibfnamefont {O.}~\bibnamefont {Painter}},\ }\bibfield  {title} {\bibinfo
  {title} {A superconducting quantum simulator based on a photonic-bandgap
  metamaterial},\ }\href {https://doi.org/10.1126/science.ade7651} {\bibfield
  {journal} {\bibinfo  {journal} {Science}\ }\textbf {\bibinfo {volume}
  {379}},\ \bibinfo {pages} {278} (\bibinfo {year} {2023})}\BibitemShut
  {NoStop}%
\bibitem [{\citenamefont {Andersen}\ \emph {et~al.}(2025)\citenamefont
  {Andersen} \emph {et~al.}}]{andersen_thermalization_2025}%
  \BibitemOpen
  \bibfield  {author} {\bibinfo {author} {\bibfnamefont {T.~I.}\ \bibnamefont
  {Andersen}} \emph {et~al.},\ }\bibfield  {title} {\bibinfo {title}
  {Thermalization and criticality on an analogue–digital quantum simulator},\
  }\href {https://doi.org/10.1038/s41586-024-08460-3} {\bibfield  {journal}
  {\bibinfo  {journal} {Nature}\ }\textbf {\bibinfo {volume} {638}},\ \bibinfo
  {pages} {79} (\bibinfo {year} {2025})}\BibitemShut {NoStop}%
\bibitem [{\citenamefont {Putterman}\ \emph {et~al.}(2025)\citenamefont
  {Putterman} \emph {et~al.}}]{putterman_hardware-efficient_2025}%
  \BibitemOpen
  \bibfield  {author} {\bibinfo {author} {\bibfnamefont {H.}~\bibnamefont
  {Putterman}} \emph {et~al.},\ }\bibfield  {title} {\bibinfo {title}
  {Hardware-efficient quantum error correction via concatenated bosonic
  qubits},\ }\href {https://doi.org/10.1038/s41586-025-08642-7} {\bibfield
  {journal} {\bibinfo  {journal} {Nature}\ }\textbf {\bibinfo {volume} {638}},\
  \bibinfo {pages} {927} (\bibinfo {year} {2025})}\BibitemShut {NoStop}%
\bibitem [{\citenamefont {Boltzmann}(1872)}]{boltzmann_uber_1872}%
  \BibitemOpen
  \bibfield  {author} {\bibinfo {author} {\bibfnamefont {L.}~\bibnamefont
  {Boltzmann}},\ }\bibfield  {title} {\bibinfo {title} {Weitere {Studien} über
  das {Wärmegleichgewicht} unter {Gasmolekülen}},\ }\href@noop {} {\bibfield
  {journal} {\bibinfo  {journal} {Sitzungsberichte Akademie der
  Wissenschaften}\ }\textbf {\bibinfo {volume} {66}},\ \bibinfo {pages} {275}
  (\bibinfo {year} {1872})}\BibitemShut {NoStop}%
\bibitem [{\citenamefont {Brown}\ \emph {et~al.}(2009)\citenamefont {Brown},
  \citenamefont {Myrvold},\ and\ \citenamefont
  {Uffink}}]{brown_boltzmanns_2009}%
  \BibitemOpen
  \bibfield  {author} {\bibinfo {author} {\bibfnamefont {H.~R.}\ \bibnamefont
  {Brown}}, \bibinfo {author} {\bibfnamefont {W.}~\bibnamefont {Myrvold}},\
  and\ \bibinfo {author} {\bibfnamefont {J.}~\bibnamefont {Uffink}},\
  }\bibfield  {title} {\bibinfo {title} {Boltzmann's {H}-theorem, its
  discontents, and the birth of statistical mechanics},\ }\href
  {https://doi.org/10.1016/j.shpsb.2009.03.003} {\bibfield  {journal} {\bibinfo
   {journal} {Stud. Hist. Philos. Sci. Part B: Stud. Hist. Philos. Mod. Phys.}\
  }\textbf {\bibinfo {volume} {40}},\ \bibinfo {pages} {174} (\bibinfo {year}
  {2009})}\BibitemShut {NoStop}%
\bibitem [{\citenamefont {Deutsch}(1991)}]{deutsch_quantum_1991}%
  \BibitemOpen
  \bibfield  {author} {\bibinfo {author} {\bibfnamefont {J.~M.}\ \bibnamefont
  {Deutsch}},\ }\bibfield  {title} {\bibinfo {title} {Quantum statistical
  mechanics in a closed system},\ }\href
  {https://doi.org/10.1103/PhysRevA.43.2046} {\bibfield  {journal} {\bibinfo
  {journal} {Phys. Rev. A}\ }\textbf {\bibinfo {volume} {43}},\ \bibinfo
  {pages} {2046} (\bibinfo {year} {1991})}\BibitemShut {NoStop}%
\bibitem [{\citenamefont {Srednicki}(1994)}]{srednicki_chaos_1994}%
  \BibitemOpen
  \bibfield  {author} {\bibinfo {author} {\bibfnamefont {M.}~\bibnamefont
  {Srednicki}},\ }\bibfield  {title} {\bibinfo {title} {Chaos and quantum
  thermalization},\ }\href {https://doi.org/10.1103/PhysRevE.50.888} {\bibfield
   {journal} {\bibinfo  {journal} {Phys. Rev. E}\ }\textbf {\bibinfo {volume}
  {50}},\ \bibinfo {pages} {888} (\bibinfo {year} {1994})}\BibitemShut
  {NoStop}%
\bibitem [{\citenamefont {Srednicki}(1999)}]{srednicki_approach_1999}%
  \BibitemOpen
  \bibfield  {author} {\bibinfo {author} {\bibfnamefont {M.}~\bibnamefont
  {Srednicki}},\ }\bibfield  {title} {\bibinfo {title} {The approach to thermal
  equilibrium in quantized chaotic systems},\ }\href
  {https://doi.org/10.1088/0305-4470/32/7/007} {\bibfield  {journal} {\bibinfo
  {journal} {J. Phys. A: Math. Gen.}\ }\textbf {\bibinfo {volume} {32}},\
  \bibinfo {pages} {1163} (\bibinfo {year} {1999})}\BibitemShut {NoStop}%
\bibitem [{\citenamefont {Rigol}\ \emph {et~al.}(2008)\citenamefont {Rigol},
  \citenamefont {Dunjko},\ and\ \citenamefont
  {Olshanii}}]{rigol_thermalization_2008}%
  \BibitemOpen
  \bibfield  {author} {\bibinfo {author} {\bibfnamefont {M.}~\bibnamefont
  {Rigol}}, \bibinfo {author} {\bibfnamefont {V.}~\bibnamefont {Dunjko}},\ and\
  \bibinfo {author} {\bibfnamefont {M.}~\bibnamefont {Olshanii}},\ }\bibfield
  {title} {\bibinfo {title} {Thermalization and its mechanism for generic
  isolated quantum systems},\ }\href {https://doi.org/10.1038/nature06838}
  {\bibfield  {journal} {\bibinfo  {journal} {Nature}\ }\textbf {\bibinfo
  {volume} {452}},\ \bibinfo {pages} {854} (\bibinfo {year}
  {2008})}\BibitemShut {NoStop}%
\bibitem [{\citenamefont {Bertini}\ \emph {et~al.}(2015)\citenamefont
  {Bertini}, \citenamefont {Essler}, \citenamefont {Groha},\ and\ \citenamefont
  {Robinson}}]{bertini_prethermalization_2015}%
  \BibitemOpen
  \bibfield  {author} {\bibinfo {author} {\bibfnamefont {B.}~\bibnamefont
  {Bertini}}, \bibinfo {author} {\bibfnamefont {F.~H.}\ \bibnamefont {Essler}},
  \bibinfo {author} {\bibfnamefont {S.}~\bibnamefont {Groha}},\ and\ \bibinfo
  {author} {\bibfnamefont {N.~J.}\ \bibnamefont {Robinson}},\ }\bibfield
  {title} {\bibinfo {title} {Prethermalization and {Thermalization} in {Models}
  with {Weak} {Integrability} {Breaking}},\ }\href
  {https://doi.org/10.1103/PhysRevLett.115.180601} {\bibfield  {journal}
  {\bibinfo  {journal} {Phys. Rev. Lett.}\ }\textbf {\bibinfo {volume} {115}},\
  \bibinfo {pages} {180601} (\bibinfo {year} {2015})}\BibitemShut {NoStop}%
\bibitem [{\citenamefont {Babadi}\ \emph {et~al.}(2015)\citenamefont {Babadi},
  \citenamefont {Demler},\ and\ \citenamefont
  {Knap}}]{babadi_far--equilibrium_2015}%
  \BibitemOpen
  \bibfield  {author} {\bibinfo {author} {\bibfnamefont {M.}~\bibnamefont
  {Babadi}}, \bibinfo {author} {\bibfnamefont {E.}~\bibnamefont {Demler}},\
  and\ \bibinfo {author} {\bibfnamefont {M.}~\bibnamefont {Knap}},\ }\bibfield
  {title} {\bibinfo {title} {Far-from-{Equilibrium} {Field} {Theory} of
  {Many}-{Body} {Quantum} {Spin} {Systems}: {Prethermalization} and
  {Relaxation} of {Spin} {Spiral} {States} in {Three} {Dimensions}},\ }\href
  {https://doi.org/10.1103/PhysRevX.5.041005} {\bibfield  {journal} {\bibinfo
  {journal} {Phys. Rev. X}\ }\textbf {\bibinfo {volume} {5}},\ \bibinfo {pages}
  {041005} (\bibinfo {year} {2015})}\BibitemShut {NoStop}%
\bibitem [{\citenamefont {Birnkammer}\ \emph {et~al.}(2022)\citenamefont
  {Birnkammer}, \citenamefont {Bastianello},\ and\ \citenamefont
  {Knap}}]{birnkammer_prethermalization_2022}%
  \BibitemOpen
  \bibfield  {author} {\bibinfo {author} {\bibfnamefont {S.}~\bibnamefont
  {Birnkammer}}, \bibinfo {author} {\bibfnamefont {A.}~\bibnamefont
  {Bastianello}},\ and\ \bibinfo {author} {\bibfnamefont {M.}~\bibnamefont
  {Knap}},\ }\bibfield  {title} {\bibinfo {title} {Prethermalization in
  one-dimensional quantum many-body systems with confinement},\ }\href
  {https://doi.org/10.1038/s41467-022-35301-6} {\bibfield  {journal} {\bibinfo
  {journal} {Nat. Commun.}\ }\textbf {\bibinfo {volume} {13}},\ \bibinfo
  {pages} {7663} (\bibinfo {year} {2022})}\BibitemShut {NoStop}%
\bibitem [{\citenamefont {Abanin}\ \emph
  {et~al.}(2017{\natexlab{a}})\citenamefont {Abanin}, \citenamefont {De~Roeck},
  \citenamefont {Ho},\ and\ \citenamefont {Huveneers}}]{abanin_rigorous_2017}%
  \BibitemOpen
  \bibfield  {author} {\bibinfo {author} {\bibfnamefont {D.}~\bibnamefont
  {Abanin}}, \bibinfo {author} {\bibfnamefont {W.}~\bibnamefont {De~Roeck}},
  \bibinfo {author} {\bibfnamefont {W.~W.}\ \bibnamefont {Ho}},\ and\ \bibinfo
  {author} {\bibfnamefont {F.}~\bibnamefont {Huveneers}},\ }\bibfield  {title}
  {\bibinfo {title} {A {Rigorous} {Theory} of {Many}-{Body} {Prethermalization}
  for {Periodically} {Driven} and {Closed} {Quantum} {Systems}},\ }\href
  {https://doi.org/10.1007/s00220-017-2930-x} {\bibfield  {journal} {\bibinfo
  {journal} {Commun. Math. Phys.}\ }\textbf {\bibinfo {volume} {354}},\
  \bibinfo {pages} {809} (\bibinfo {year} {2017}{\natexlab{a}})}\BibitemShut
  {NoStop}%
\bibitem [{\citenamefont {Abanin}\ \emph
  {et~al.}(2017{\natexlab{b}})\citenamefont {Abanin}, \citenamefont {De~Roeck},
  \citenamefont {Ho},\ and\ \citenamefont {Huveneers}}]{abanin_effective_2017}%
  \BibitemOpen
  \bibfield  {author} {\bibinfo {author} {\bibfnamefont {D.~A.}\ \bibnamefont
  {Abanin}}, \bibinfo {author} {\bibfnamefont {W.}~\bibnamefont {De~Roeck}},
  \bibinfo {author} {\bibfnamefont {W.~W.}\ \bibnamefont {Ho}},\ and\ \bibinfo
  {author} {\bibfnamefont {F.}~\bibnamefont {Huveneers}},\ }\bibfield  {title}
  {\bibinfo {title} {Effective {Hamiltonians}, prethermalization, and slow
  energy absorption in periodically driven many-body systems},\ }\href
  {https://doi.org/10.1103/PhysRevB.95.014112} {\bibfield  {journal} {\bibinfo
  {journal} {Phys. Rev. B}\ }\textbf {\bibinfo {volume} {95}},\ \bibinfo
  {pages} {014112} (\bibinfo {year} {2017}{\natexlab{b}})}\BibitemShut
  {NoStop}%
\bibitem [{\citenamefont {Else}\ \emph {et~al.}(2017)\citenamefont {Else},
  \citenamefont {Bauer},\ and\ \citenamefont {Nayak}}]{else_prethermal_2017}%
  \BibitemOpen
  \bibfield  {author} {\bibinfo {author} {\bibfnamefont {D.~V.}\ \bibnamefont
  {Else}}, \bibinfo {author} {\bibfnamefont {B.}~\bibnamefont {Bauer}},\ and\
  \bibinfo {author} {\bibfnamefont {C.}~\bibnamefont {Nayak}},\ }\bibfield
  {title} {\bibinfo {title} {Prethermal {Phases} of {Matter} {Protected} by
  {Time}-{Translation} {Symmetry}},\ }\href
  {https://doi.org/10.1103/PhysRevX.7.011026} {\bibfield  {journal} {\bibinfo
  {journal} {Phys. Rev. X}\ }\textbf {\bibinfo {volume} {7}},\ \bibinfo {pages}
  {011026} (\bibinfo {year} {2017})}\BibitemShut {NoStop}%
\bibitem [{\citenamefont {Machado}\ \emph {et~al.}(2020)\citenamefont
  {Machado}, \citenamefont {Else}, \citenamefont {Kahanamoku-Meyer},
  \citenamefont {Nayak},\ and\ \citenamefont {Yao}}]{machado_long-range_2020}%
  \BibitemOpen
  \bibfield  {author} {\bibinfo {author} {\bibfnamefont {F.}~\bibnamefont
  {Machado}}, \bibinfo {author} {\bibfnamefont {D.~V.}\ \bibnamefont {Else}},
  \bibinfo {author} {\bibfnamefont {G.~D.}\ \bibnamefont {Kahanamoku-Meyer}},
  \bibinfo {author} {\bibfnamefont {C.}~\bibnamefont {Nayak}},\ and\ \bibinfo
  {author} {\bibfnamefont {N.~Y.}\ \bibnamefont {Yao}},\ }\bibfield  {title}
  {\bibinfo {title} {Long-{Range} {Prethermal} {Phases} of {Nonequilibrium}
  {Matter}},\ }\href {https://doi.org/10.1103/PhysRevX.10.011043} {\bibfield
  {journal} {\bibinfo  {journal} {Phys. Rev. X}\ }\textbf {\bibinfo {volume}
  {10}},\ \bibinfo {pages} {011043} (\bibinfo {year} {2020})}\BibitemShut
  {NoStop}%
\bibitem [{\citenamefont {Pizzi}\ \emph {et~al.}(2021)\citenamefont {Pizzi},
  \citenamefont {Nunnenkamp},\ and\ \citenamefont
  {Knolle}}]{pizzi_classical_2021}%
  \BibitemOpen
  \bibfield  {author} {\bibinfo {author} {\bibfnamefont {A.}~\bibnamefont
  {Pizzi}}, \bibinfo {author} {\bibfnamefont {A.}~\bibnamefont {Nunnenkamp}},\
  and\ \bibinfo {author} {\bibfnamefont {J.}~\bibnamefont {Knolle}},\
  }\bibfield  {title} {\bibinfo {title} {Classical {Prethermal} {Phases} of
  {Matter}},\ }\href {https://doi.org/10.1103/PhysRevLett.127.140602}
  {\bibfield  {journal} {\bibinfo  {journal} {Phys. Rev. Lett.}\ }\textbf
  {\bibinfo {volume} {127}},\ \bibinfo {pages} {140602} (\bibinfo {year}
  {2021})}\BibitemShut {NoStop}%
\bibitem [{\citenamefont {Vidmar}\ and\ \citenamefont
  {Rigol}(2016)}]{vidmar_generalized_2016}%
  \BibitemOpen
  \bibfield  {author} {\bibinfo {author} {\bibfnamefont {L.}~\bibnamefont
  {Vidmar}}\ and\ \bibinfo {author} {\bibfnamefont {M.}~\bibnamefont {Rigol}},\
  }\bibfield  {title} {\bibinfo {title} {Generalized {Gibbs} ensemble in
  integrable lattice models},\ }\href
  {https://doi.org/10.1088/1742-5468/2016/06/064007} {\bibfield  {journal}
  {\bibinfo  {journal} {J. Stat. Mech. Theory Exp.}\ }\textbf {\bibinfo
  {volume} {2016}},\ \bibinfo {pages} {064007} (\bibinfo {year}
  {2016})}\BibitemShut {NoStop}%
\bibitem [{\citenamefont {Mori}\ \emph {et~al.}(2018)\citenamefont {Mori},
  \citenamefont {Ikeda}, \citenamefont {Kaminishi},\ and\ \citenamefont
  {Ueda}}]{mori_thermalization_2018}%
  \BibitemOpen
  \bibfield  {author} {\bibinfo {author} {\bibfnamefont {T.}~\bibnamefont
  {Mori}}, \bibinfo {author} {\bibfnamefont {T.~N.}\ \bibnamefont {Ikeda}},
  \bibinfo {author} {\bibfnamefont {E.}~\bibnamefont {Kaminishi}},\ and\
  \bibinfo {author} {\bibfnamefont {M.}~\bibnamefont {Ueda}},\ }\bibfield
  {title} {\bibinfo {title} {Thermalization and prethermalization in isolated
  quantum systems: a theoretical overview},\ }\href
  {https://doi.org/10.1088/1361-6455/aabcdf} {\bibfield  {journal} {\bibinfo
  {journal} {J. Phys. B: At. Mol. Opt. Phys.}\ }\textbf {\bibinfo {volume}
  {51}},\ \bibinfo {pages} {112001} (\bibinfo {year} {2018})}\BibitemShut
  {NoStop}%
\bibitem [{\citenamefont {Berges}\ \emph {et~al.}(2004)\citenamefont {Berges},
  \citenamefont {Borsányi},\ and\ \citenamefont
  {Wetterich}}]{berges_prethermalization_2004}%
  \BibitemOpen
  \bibfield  {author} {\bibinfo {author} {\bibfnamefont {J.}~\bibnamefont
  {Berges}}, \bibinfo {author} {\bibfnamefont {S.}~\bibnamefont {Borsányi}},\
  and\ \bibinfo {author} {\bibfnamefont {C.}~\bibnamefont {Wetterich}},\
  }\bibfield  {title} {\bibinfo {title} {Prethermalization},\ }\href
  {https://doi.org/10.1103/PhysRevLett.93.142002} {\bibfield  {journal}
  {\bibinfo  {journal} {Phys. Rev. Lett.}\ }\textbf {\bibinfo {volume} {93}},\
  \bibinfo {pages} {142002} (\bibinfo {year} {2004})}\BibitemShut {NoStop}%
\bibitem [{\citenamefont {Akemann}\ \emph {et~al.}(2019)\citenamefont
  {Akemann}, \citenamefont {Kieburg}, \citenamefont {Mielke},\ and\
  \citenamefont {Prosen}}]{akemann_universal_2019}%
  \BibitemOpen
  \bibfield  {author} {\bibinfo {author} {\bibfnamefont {G.}~\bibnamefont
  {Akemann}}, \bibinfo {author} {\bibfnamefont {M.}~\bibnamefont {Kieburg}},
  \bibinfo {author} {\bibfnamefont {A.}~\bibnamefont {Mielke}},\ and\ \bibinfo
  {author} {\bibfnamefont {T.}~\bibnamefont {Prosen}},\ }\bibfield  {title}
  {\bibinfo {title} {Universal {Signature} from {Integrability} to {Chaos} in
  {Dissipative} {Open} {Quantum} {Systems}},\ }\href
  {https://doi.org/10.1103/PhysRevLett.123.254101} {\bibfield  {journal}
  {\bibinfo  {journal} {Phys. Rev. Lett.}\ }\textbf {\bibinfo {volume} {123}},\
  \bibinfo {pages} {254101} (\bibinfo {year} {2019})}\BibitemShut {NoStop}%
\bibitem [{\citenamefont {Hamazaki}\ \emph {et~al.}(2020)\citenamefont
  {Hamazaki}, \citenamefont {Kawabata}, \citenamefont {Kura},\ and\
  \citenamefont {Ueda}}]{hamazaki_universality_2020}%
  \BibitemOpen
  \bibfield  {author} {\bibinfo {author} {\bibfnamefont {R.}~\bibnamefont
  {Hamazaki}}, \bibinfo {author} {\bibfnamefont {K.}~\bibnamefont {Kawabata}},
  \bibinfo {author} {\bibfnamefont {N.}~\bibnamefont {Kura}},\ and\ \bibinfo
  {author} {\bibfnamefont {M.}~\bibnamefont {Ueda}},\ }\bibfield  {title}
  {\bibinfo {title} {Universality classes of non-{Hermitian} random matrices},\
  }\href {https://doi.org/10.1103/PhysRevResearch.2.023286} {\bibfield
  {journal} {\bibinfo  {journal} {Phys. Rev. Res.}\ }\textbf {\bibinfo {volume}
  {2}},\ \bibinfo {pages} {023286} (\bibinfo {year} {2020})}\BibitemShut
  {NoStop}%
\bibitem [{\citenamefont {Sá}\ \emph {et~al.}(2020)\citenamefont {Sá},
  \citenamefont {Ribeiro},\ and\ \citenamefont {Prosen}}]{sa_complex_2020}%
  \BibitemOpen
  \bibfield  {author} {\bibinfo {author} {\bibfnamefont {L.}~\bibnamefont
  {Sá}}, \bibinfo {author} {\bibfnamefont {P.}~\bibnamefont {Ribeiro}},\ and\
  \bibinfo {author} {\bibfnamefont {T.}~\bibnamefont {Prosen}},\ }\bibfield
  {title} {\bibinfo {title} {Complex {Spacing} {Ratios}: {A} {Signature} of
  {Dissipative} {Quantum} {Chaos}},\ }\href
  {https://doi.org/10.1103/PhysRevX.10.021019} {\bibfield  {journal} {\bibinfo
  {journal} {Phys. Rev. X}\ }\textbf {\bibinfo {volume} {10}},\ \bibinfo
  {pages} {021019} (\bibinfo {year} {2020})}\BibitemShut {NoStop}%
\bibitem [{\citenamefont {Dahan}\ \emph {et~al.}(2022)\citenamefont {Dahan},
  \citenamefont {Arwas},\ and\ \citenamefont
  {Grosfeld}}]{dahan_classical_2022}%
  \BibitemOpen
  \bibfield  {author} {\bibinfo {author} {\bibfnamefont {D.}~\bibnamefont
  {Dahan}}, \bibinfo {author} {\bibfnamefont {G.}~\bibnamefont {Arwas}},\ and\
  \bibinfo {author} {\bibfnamefont {E.}~\bibnamefont {Grosfeld}},\ }\bibfield
  {title} {\bibinfo {title} {Classical and quantum chaos in chirally-driven,
  dissipative {Bose}-{Hubbard} systems},\ }\href
  {https://doi.org/10.1038/s41534-022-00518-2} {\bibfield  {journal} {\bibinfo
  {journal} {npj Quantum Inf.}\ }\textbf {\bibinfo {volume} {8}},\ \bibinfo
  {pages} {14} (\bibinfo {year} {2022})}\BibitemShut {NoStop}%
\bibitem [{\citenamefont {Prasad}\ \emph {et~al.}(2022)\citenamefont {Prasad},
  \citenamefont {Yadalam}, \citenamefont {Aron},\ and\ \citenamefont
  {Kulkarni}}]{prasad_dissipative_2022}%
  \BibitemOpen
  \bibfield  {author} {\bibinfo {author} {\bibfnamefont {M.}~\bibnamefont
  {Prasad}}, \bibinfo {author} {\bibfnamefont {H.~K.}\ \bibnamefont {Yadalam}},
  \bibinfo {author} {\bibfnamefont {C.}~\bibnamefont {Aron}},\ and\ \bibinfo
  {author} {\bibfnamefont {M.}~\bibnamefont {Kulkarni}},\ }\bibfield  {title}
  {\bibinfo {title} {Dissipative quantum dynamics, phase transitions, and
  non-{Hermitian} random matrices},\ }\href
  {https://doi.org/10.1103/PhysRevA.105.L050201} {\bibfield  {journal}
  {\bibinfo  {journal} {Phys. Rev. A}\ }\textbf {\bibinfo {volume} {105}},\
  \bibinfo {pages} {L050201} (\bibinfo {year} {2022})}\BibitemShut {NoStop}%
\bibitem [{\citenamefont {Kawabata}\ \emph {et~al.}(2023)\citenamefont
  {Kawabata}, \citenamefont {Kulkarni}, \citenamefont {Li}, \citenamefont
  {Numasawa},\ and\ \citenamefont {Ryu}}]{kawabata_symmetry_2023}%
  \BibitemOpen
  \bibfield  {author} {\bibinfo {author} {\bibfnamefont {K.}~\bibnamefont
  {Kawabata}}, \bibinfo {author} {\bibfnamefont {A.}~\bibnamefont {Kulkarni}},
  \bibinfo {author} {\bibfnamefont {J.}~\bibnamefont {Li}}, \bibinfo {author}
  {\bibfnamefont {T.}~\bibnamefont {Numasawa}},\ and\ \bibinfo {author}
  {\bibfnamefont {S.}~\bibnamefont {Ryu}},\ }\bibfield  {title} {\bibinfo
  {title} {Symmetry of {Open} {Quantum} {Systems}: {Classification} of
  {Dissipative} {Quantum} {Chaos}},\ }\href
  {https://doi.org/10.1103/PRXQuantum.4.030328} {\bibfield  {journal} {\bibinfo
   {journal} {PRX Quantum}\ }\textbf {\bibinfo {volume} {4}},\ \bibinfo {pages}
  {030328} (\bibinfo {year} {2023})}\BibitemShut {NoStop}%
\bibitem [{\citenamefont {Villaseñor}\ \emph {et~al.}(2024)\citenamefont
  {Villaseñor}, \citenamefont {Santos},\ and\ \citenamefont
  {Barberis-Blostein}}]{villasenor_breakdown_2024}%
  \BibitemOpen
  \bibfield  {author} {\bibinfo {author} {\bibfnamefont {D.}~\bibnamefont
  {Villaseñor}}, \bibinfo {author} {\bibfnamefont {L.~F.}\ \bibnamefont
  {Santos}},\ and\ \bibinfo {author} {\bibfnamefont {P.}~\bibnamefont
  {Barberis-Blostein}},\ }\bibfield  {title} {\bibinfo {title} {Breakdown of
  the {Quantum} {Distinction} of {Regular} and {Chaotic} {Classical} {Dynamics}
  in {Dissipative} {Systems}},\ }\href
  {https://doi.org/10.1103/PhysRevLett.133.240404} {\bibfield  {journal}
  {\bibinfo  {journal} {Phys. Rev. Lett.}\ }\textbf {\bibinfo {volume} {133}},\
  \bibinfo {pages} {240404} (\bibinfo {year} {2024})}\BibitemShut {NoStop}%
\bibitem [{\citenamefont {Ferrari}\ \emph {et~al.}(2025)\citenamefont
  {Ferrari}, \citenamefont {Gravina}, \citenamefont {Eeltink}, \citenamefont
  {Scarlino}, \citenamefont {Savona},\ and\ \citenamefont
  {Minganti}}]{ferrari_dissipative_2025}%
  \BibitemOpen
  \bibfield  {author} {\bibinfo {author} {\bibfnamefont {F.}~\bibnamefont
  {Ferrari}}, \bibinfo {author} {\bibfnamefont {L.}~\bibnamefont {Gravina}},
  \bibinfo {author} {\bibfnamefont {D.}~\bibnamefont {Eeltink}}, \bibinfo
  {author} {\bibfnamefont {P.}~\bibnamefont {Scarlino}}, \bibinfo {author}
  {\bibfnamefont {V.}~\bibnamefont {Savona}},\ and\ \bibinfo {author}
  {\bibfnamefont {F.}~\bibnamefont {Minganti}},\ }\bibfield  {title} {\bibinfo
  {title} {Dissipative quantum chaos unveiled by stochastic quantum
  trajectories},\ }\href {https://doi.org/10.1103/PhysRevResearch.7.013276}
  {\bibfield  {journal} {\bibinfo  {journal} {Phys. Rev. Res.}\ }\textbf
  {\bibinfo {volume} {7}},\ \bibinfo {pages} {013276} (\bibinfo {year}
  {2025})}\BibitemShut {NoStop}%
\bibitem [{\citenamefont {Fitzpatrick}\ \emph {et~al.}(2017)\citenamefont
  {Fitzpatrick}, \citenamefont {Sundaresan}, \citenamefont {Li}, \citenamefont
  {Koch},\ and\ \citenamefont {Houck}}]{FitzpatrickPRX17}%
  \BibitemOpen
  \bibfield  {author} {\bibinfo {author} {\bibfnamefont {M.}~\bibnamefont
  {Fitzpatrick}}, \bibinfo {author} {\bibfnamefont {N.~M.}\ \bibnamefont
  {Sundaresan}}, \bibinfo {author} {\bibfnamefont {A.~C.~Y.}\ \bibnamefont
  {Li}}, \bibinfo {author} {\bibfnamefont {J.}~\bibnamefont {Koch}},\ and\
  \bibinfo {author} {\bibfnamefont {A.~A.}\ \bibnamefont {Houck}},\ }\bibfield
  {title} {\bibinfo {title} {Observation of a dissipative phase transition in a
  one-dimensional circuit qed lattice},\ }\href
  {https://doi.org/10.1103/PhysRevX.7.011016} {\bibfield  {journal} {\bibinfo
  {journal} {Phys. Rev. X}\ }\textbf {\bibinfo {volume} {7}},\ \bibinfo {pages}
  {011016} (\bibinfo {year} {2017})}\BibitemShut {NoStop}%
\bibitem [{\citenamefont {Fedorov}\ \emph {et~al.}(2021)\citenamefont
  {Fedorov}, \citenamefont {Remizov}, \citenamefont {Shapiro}, \citenamefont
  {Pogosov}, \citenamefont {Egorova}, \citenamefont {Tsitsilin}, \citenamefont
  {Andronik}, \citenamefont {Dobronosova}, \citenamefont {Rodionov},
  \citenamefont {Astafiev},\ and\ \citenamefont {Ustinov}}]{FedorovPRL21}%
  \BibitemOpen
  \bibfield  {author} {\bibinfo {author} {\bibfnamefont {G.~P.}\ \bibnamefont
  {Fedorov}}, \bibinfo {author} {\bibfnamefont {S.~V.}\ \bibnamefont
  {Remizov}}, \bibinfo {author} {\bibfnamefont {D.~S.}\ \bibnamefont
  {Shapiro}}, \bibinfo {author} {\bibfnamefont {W.~V.}\ \bibnamefont
  {Pogosov}}, \bibinfo {author} {\bibfnamefont {E.}~\bibnamefont {Egorova}},
  \bibinfo {author} {\bibfnamefont {I.}~\bibnamefont {Tsitsilin}}, \bibinfo
  {author} {\bibfnamefont {M.}~\bibnamefont {Andronik}}, \bibinfo {author}
  {\bibfnamefont {A.~A.}\ \bibnamefont {Dobronosova}}, \bibinfo {author}
  {\bibfnamefont {I.~A.}\ \bibnamefont {Rodionov}}, \bibinfo {author}
  {\bibfnamefont {O.~V.}\ \bibnamefont {Astafiev}},\ and\ \bibinfo {author}
  {\bibfnamefont {A.~V.}\ \bibnamefont {Ustinov}},\ }\bibfield  {title}
  {\bibinfo {title} {Photon transport in a {B}ose-{H}ubbard chain of
  superconducting artificial atoms},\ }\href
  {https://doi.org/10.1103/PhysRevLett.126.180503} {\bibfield  {journal}
  {\bibinfo  {journal} {Phys. Rev. Lett.}\ }\textbf {\bibinfo {volume} {126}},\
  \bibinfo {pages} {180503} (\bibinfo {year} {2021})}\BibitemShut {NoStop}%
\bibitem [{\citenamefont {Jouanny}\ \emph {et~al.}(2025)\citenamefont
  {Jouanny}, \citenamefont {Frasca}, \citenamefont {Weibel}, \citenamefont
  {Peyruchat}, \citenamefont {Scigliuzzo}, \citenamefont {Oppliger},
  \citenamefont {De~Palma}, \citenamefont {Sbroggiò}, \citenamefont
  {Beaulieu}, \citenamefont {Zilberberg},\ and\ \citenamefont
  {Scarlino}}]{jouanny_high_2025}%
  \BibitemOpen
  \bibfield  {author} {\bibinfo {author} {\bibfnamefont {V.}~\bibnamefont
  {Jouanny}}, \bibinfo {author} {\bibfnamefont {S.}~\bibnamefont {Frasca}},
  \bibinfo {author} {\bibfnamefont {V.~J.}\ \bibnamefont {Weibel}}, \bibinfo
  {author} {\bibfnamefont {L.}~\bibnamefont {Peyruchat}}, \bibinfo {author}
  {\bibfnamefont {M.}~\bibnamefont {Scigliuzzo}}, \bibinfo {author}
  {\bibfnamefont {F.}~\bibnamefont {Oppliger}}, \bibinfo {author}
  {\bibfnamefont {F.}~\bibnamefont {De~Palma}}, \bibinfo {author}
  {\bibfnamefont {D.}~\bibnamefont {Sbroggiò}}, \bibinfo {author}
  {\bibfnamefont {G.}~\bibnamefont {Beaulieu}}, \bibinfo {author}
  {\bibfnamefont {O.}~\bibnamefont {Zilberberg}},\ and\ \bibinfo {author}
  {\bibfnamefont {P.}~\bibnamefont {Scarlino}},\ }\bibfield  {title} {\bibinfo
  {title} {High kinetic inductance cavity arrays for compact band engineering
  and topology-based disorder meters},\ }\href
  {https://doi.org/10.1038/s41467-025-58595-8} {\bibfield  {journal} {\bibinfo
  {journal} {Nat. Commun.}\ }\textbf {\bibinfo {volume} {16}},\ \bibinfo
  {pages} {3396} (\bibinfo {year} {2025})}\BibitemShut {NoStop}%
\bibitem [{\citenamefont {Mei}\ \emph {et~al.}(2022)\citenamefont {Mei},
  \citenamefont {Li}, \citenamefont {Wu}, \citenamefont {Cai}, \citenamefont
  {Wang}, \citenamefont {Yao}, \citenamefont {Zhou},\ and\ \citenamefont
  {Duan}}]{MeiPRL22}%
  \BibitemOpen
  \bibfield  {author} {\bibinfo {author} {\bibfnamefont {Q.-X.}\ \bibnamefont
  {Mei}}, \bibinfo {author} {\bibfnamefont {B.-W.}\ \bibnamefont {Li}},
  \bibinfo {author} {\bibfnamefont {Y.-K.}\ \bibnamefont {Wu}}, \bibinfo
  {author} {\bibfnamefont {M.-L.}\ \bibnamefont {Cai}}, \bibinfo {author}
  {\bibfnamefont {Y.}~\bibnamefont {Wang}}, \bibinfo {author} {\bibfnamefont
  {L.}~\bibnamefont {Yao}}, \bibinfo {author} {\bibfnamefont {Z.-C.}\
  \bibnamefont {Zhou}},\ and\ \bibinfo {author} {\bibfnamefont {L.-M.}\
  \bibnamefont {Duan}},\ }\bibfield  {title} {\bibinfo {title} {Experimental
  realization of the {R}abi-{H}ubbard model with trapped ions},\ }\href
  {https://doi.org/10.1103/PhysRevLett.128.160504} {\bibfield  {journal}
  {\bibinfo  {journal} {Phys. Rev. Lett.}\ }\textbf {\bibinfo {volume} {128}},\
  \bibinfo {pages} {160504} (\bibinfo {year} {2022})}\BibitemShut {NoStop}%
\bibitem [{\citenamefont {Rodriguez}\ \emph {et~al.}(2016)\citenamefont
  {Rodriguez}, \citenamefont {Amo}, \citenamefont {Sagnes}, \citenamefont
  {Le~Gratiet}, \citenamefont {Galopin}, \citenamefont {Lemaître},\ and\
  \citenamefont {Bloch}}]{rodriguez_interaction-induced_2016}%
  \BibitemOpen
  \bibfield  {author} {\bibinfo {author} {\bibfnamefont {S.~R.~K.}\
  \bibnamefont {Rodriguez}}, \bibinfo {author} {\bibfnamefont {A.}~\bibnamefont
  {Amo}}, \bibinfo {author} {\bibfnamefont {I.}~\bibnamefont {Sagnes}},
  \bibinfo {author} {\bibfnamefont {L.}~\bibnamefont {Le~Gratiet}}, \bibinfo
  {author} {\bibfnamefont {E.}~\bibnamefont {Galopin}}, \bibinfo {author}
  {\bibfnamefont {A.}~\bibnamefont {Lemaître}},\ and\ \bibinfo {author}
  {\bibfnamefont {J.}~\bibnamefont {Bloch}},\ }\bibfield  {title} {\bibinfo
  {title} {Interaction-induced hopping phase in driven-dissipative coupled
  photonic microcavities},\ }\href {https://doi.org/10.1038/ncomms11887}
  {\bibfield  {journal} {\bibinfo  {journal} {Nat. Commun.}\ }\textbf {\bibinfo
  {volume} {7}},\ \bibinfo {pages} {11887} (\bibinfo {year}
  {2016})}\BibitemShut {NoStop}%
\bibitem [{\citenamefont {Benary}\ \emph {et~al.}(2022)\citenamefont {Benary},
  \citenamefont {Baals}, \citenamefont {Bernhart}, \citenamefont {Jiang},
  \citenamefont {R\"{o}hrle},\ and\ \citenamefont {Ott}}]{Benary2022}%
  \BibitemOpen
  \bibfield  {author} {\bibinfo {author} {\bibfnamefont {J.}~\bibnamefont
  {Benary}}, \bibinfo {author} {\bibfnamefont {C.}~\bibnamefont {Baals}},
  \bibinfo {author} {\bibfnamefont {E.}~\bibnamefont {Bernhart}}, \bibinfo
  {author} {\bibfnamefont {J.}~\bibnamefont {Jiang}}, \bibinfo {author}
  {\bibfnamefont {M.}~\bibnamefont {R\"{o}hrle}},\ and\ \bibinfo {author}
  {\bibfnamefont {H.}~\bibnamefont {Ott}},\ }\bibfield  {title} {\bibinfo
  {title} {Experimental observation of a dissipative phase transition in a
  multi-mode many-body quantum system},\ }\href
  {https://doi.org/10.1088/1367-2630/ac97b6} {\bibfield  {journal} {\bibinfo
  {journal} {New J. Phys.}\ }\textbf {\bibinfo {volume} {24}},\ \bibinfo
  {pages} {103034} (\bibinfo {year} {2022})}\BibitemShut {NoStop}%
\bibitem [{\citenamefont {Debnath}\ \emph {et~al.}(2017)\citenamefont
  {Debnath}, \citenamefont {Mascarenhas},\ and\ \citenamefont
  {Savona}}]{debnath_nonequilibrium_2017}%
  \BibitemOpen
  \bibfield  {author} {\bibinfo {author} {\bibfnamefont {K.}~\bibnamefont
  {Debnath}}, \bibinfo {author} {\bibfnamefont {E.}~\bibnamefont
  {Mascarenhas}},\ and\ \bibinfo {author} {\bibfnamefont {V.}~\bibnamefont
  {Savona}},\ }\bibfield  {title} {\bibinfo {title} {Nonequilibrium photonic
  transport and phase transition in an array of optical cavities},\ }\href
  {https://doi.org/10.1088/1367-2630/aa969e} {\bibfield  {journal} {\bibinfo
  {journal} {New J. Phys.}\ }\textbf {\bibinfo {volume} {19}},\ \bibinfo
  {pages} {115006} (\bibinfo {year} {2017})}\BibitemShut {NoStop}%
\bibitem [{\citenamefont {Prem}\ \emph {et~al.}(2023)\citenamefont {Prem},
  \citenamefont {Bulchandani},\ and\ \citenamefont
  {Sondhi}}]{prem_dynamics_2023}%
  \BibitemOpen
  \bibfield  {author} {\bibinfo {author} {\bibfnamefont {A.}~\bibnamefont
  {Prem}}, \bibinfo {author} {\bibfnamefont {V.~B.}\ \bibnamefont
  {Bulchandani}},\ and\ \bibinfo {author} {\bibfnamefont {S.~L.}\ \bibnamefont
  {Sondhi}},\ }\bibfield  {title} {\bibinfo {title} {Dynamics and transport in
  the boundary-driven dissipative {Klein}-{Gordon} chain},\ }\href
  {https://doi.org/10.1103/PhysRevB.107.104304} {\bibfield  {journal} {\bibinfo
   {journal} {Phys. Rev. B}\ }\textbf {\bibinfo {volume} {107}},\ \bibinfo
  {pages} {104304} (\bibinfo {year} {2023})}\BibitemShut {NoStop}%
\bibitem [{\citenamefont {Kumar}\ \emph {et~al.}(2024)\citenamefont {Kumar},
  \citenamefont {Mishra}, \citenamefont {Kundu},\ and\ \citenamefont
  {Dhar}}]{Abhishek2024}%
  \BibitemOpen
  \bibfield  {author} {\bibinfo {author} {\bibfnamefont {U.}~\bibnamefont
  {Kumar}}, \bibinfo {author} {\bibfnamefont {S.}~\bibnamefont {Mishra}},
  \bibinfo {author} {\bibfnamefont {A.}~\bibnamefont {Kundu}},\ and\ \bibinfo
  {author} {\bibfnamefont {A.}~\bibnamefont {Dhar}},\ }\bibfield  {title}
  {\bibinfo {title} {Observation of multiple attractors and diffusive transport
  in a periodically driven klein-gordon chain},\ }\href
  {https://doi.org/10.1103/PhysRevE.109.064124} {\bibfield  {journal} {\bibinfo
   {journal} {Phys. Rev. E}\ }\textbf {\bibinfo {volume} {109}},\ \bibinfo
  {pages} {064124} (\bibinfo {year} {2024})}\BibitemShut {NoStop}%
\bibitem [{\citenamefont {Parisi}(1997)}]{parisi_approach_1997}%
  \BibitemOpen
  \bibfield  {author} {\bibinfo {author} {\bibfnamefont {G.}~\bibnamefont
  {Parisi}},\ }\bibfield  {title} {\bibinfo {title} {On the approach to
  equilibrium of a {Hamiltonian} chain of anharmonic oscillators},\ }\href
  {https://doi.org/10.1209/epl/i1997-00471-9} {\bibfield  {journal} {\bibinfo
  {journal} {Eur. Phys. Lett.}\ }\textbf {\bibinfo {volume} {40}},\ \bibinfo
  {pages} {357} (\bibinfo {year} {1997})}\BibitemShut {NoStop}%
\bibitem [{\citenamefont {Kolovsky}\ and\ \citenamefont
  {Buchleitner}(2004)}]{kolovsky_quantum_2004}%
  \BibitemOpen
  \bibfield  {author} {\bibinfo {author} {\bibfnamefont {A.~R.}\ \bibnamefont
  {Kolovsky}}\ and\ \bibinfo {author} {\bibfnamefont {A.}~\bibnamefont
  {Buchleitner}},\ }\bibfield  {title} {\bibinfo {title} {Quantum chaos in the
  {Bose}-{Hubbard} model},\ }\href {https://doi.org/10.1209/epl/i2004-10265-7}
  {\bibfield  {journal} {\bibinfo  {journal} {Eur. Phys. Lett.}\ }\textbf
  {\bibinfo {volume} {68}},\ \bibinfo {pages} {632} (\bibinfo {year}
  {2004})}\BibitemShut {NoStop}%
\bibitem [{\citenamefont {Kollath}\ \emph {et~al.}(2007)\citenamefont
  {Kollath}, \citenamefont {L\"auchli},\ and\ \citenamefont
  {Altman}}]{KollathPRL07}%
  \BibitemOpen
  \bibfield  {author} {\bibinfo {author} {\bibfnamefont {C.}~\bibnamefont
  {Kollath}}, \bibinfo {author} {\bibfnamefont {A.~M.}\ \bibnamefont
  {L\"auchli}},\ and\ \bibinfo {author} {\bibfnamefont {E.}~\bibnamefont
  {Altman}},\ }\bibfield  {title} {\bibinfo {title} {Quench {Dynamics} and
  {Nonequilibrium} {Phase} {Diagram} of the {Bose}-{Hubbard} {Model}},\ }\href
  {https://doi.org/10.1103/PhysRevLett.98.180601} {\bibfield  {journal}
  {\bibinfo  {journal} {Phys. Rev. Lett.}\ }\textbf {\bibinfo {volume} {98}},\
  \bibinfo {pages} {180601} (\bibinfo {year} {2007})}\BibitemShut {NoStop}%
\bibitem [{\citenamefont {Schlagheck}\ and\ \citenamefont
  {Shepelyansky}(2016)}]{SchlagheckPRE16}%
  \BibitemOpen
  \bibfield  {author} {\bibinfo {author} {\bibfnamefont {P.}~\bibnamefont
  {Schlagheck}}\ and\ \bibinfo {author} {\bibfnamefont {D.~L.}\ \bibnamefont
  {Shepelyansky}},\ }\bibfield  {title} {\bibinfo {title} {Dynamical
  thermalization in {Bose}-{Hubbard} systems},\ }\href
  {https://doi.org/10.1103/PhysRevE.93.012126} {\bibfield  {journal} {\bibinfo
  {journal} {Phys. Rev. E}\ }\textbf {\bibinfo {volume} {93}},\ \bibinfo
  {pages} {012126} (\bibinfo {year} {2016})}\BibitemShut {NoStop}%
\bibitem [{\citenamefont {Carmichael}(1999)}]{carmichael_statistical_1999}%
  \BibitemOpen
  \bibfield  {author} {\bibinfo {author} {\bibfnamefont {H.~J.}\ \bibnamefont
  {Carmichael}},\ }\href {https://doi.org/10.1007/978-3-662-03875-8} {\emph
  {\bibinfo {title} {Statistical {Methods} in {Quantum} {Optics} 1}}}\
  (\bibinfo  {publisher} {Springer Berlin Heidelberg},\ \bibinfo {address}
  {Berlin, Heidelberg},\ \bibinfo {year} {1999})\BibitemShut {NoStop}%
\bibitem [{\citenamefont {Polkovnikov}(2010)}]{polkovnikov_phase_2010}%
  \BibitemOpen
  \bibfield  {author} {\bibinfo {author} {\bibfnamefont {A.}~\bibnamefont
  {Polkovnikov}},\ }\bibfield  {title} {\bibinfo {title} {Phase space
  representation of quantum dynamics},\ }\href
  {https://doi.org/10.1016/j.aop.2010.02.006} {\bibfield  {journal} {\bibinfo
  {journal} {Ann. Phys.}\ }\textbf {\bibinfo {volume} {325}},\ \bibinfo {pages}
  {1790} (\bibinfo {year} {2010})}\BibitemShut {NoStop}%
\bibitem [{\citenamefont {Iubini}\ \emph {et~al.}(2012)\citenamefont {Iubini},
  \citenamefont {Lepri},\ and\ \citenamefont {Politi}}]{Politi}%
  \BibitemOpen
  \bibfield  {author} {\bibinfo {author} {\bibfnamefont {S.}~\bibnamefont
  {Iubini}}, \bibinfo {author} {\bibfnamefont {S.}~\bibnamefont {Lepri}},\ and\
  \bibinfo {author} {\bibfnamefont {A.}~\bibnamefont {Politi}},\ }\bibfield
  {title} {\bibinfo {title} {Nonequilibrium discrete nonlinear {Schrödinger}
  equation},\ }\href {https://doi.org/10.1103/PhysRevE.86.011108} {\bibfield
  {journal} {\bibinfo  {journal} {Phys. Rev. E}\ }\textbf {\bibinfo {volume}
  {86}},\ \bibinfo {pages} {011108} (\bibinfo {year} {2012})}\BibitemShut
  {NoStop}%
\bibitem [{\citenamefont {Iubini}\ \emph {et~al.}(2016)\citenamefont {Iubini},
  \citenamefont {Lepri}, \citenamefont {Livi},\ and\ \citenamefont
  {Politi}}]{iubini_coupled_2016}%
  \BibitemOpen
  \bibfield  {author} {\bibinfo {author} {\bibfnamefont {S.}~\bibnamefont
  {Iubini}}, \bibinfo {author} {\bibfnamefont {S.}~\bibnamefont {Lepri}},
  \bibinfo {author} {\bibfnamefont {R.}~\bibnamefont {Livi}},\ and\ \bibinfo
  {author} {\bibfnamefont {A.}~\bibnamefont {Politi}},\ }\bibfield  {title}
  {\bibinfo {title} {Coupled transport in rotor models},\ }\href
  {https://doi.org/10.1088/1367-2630/18/8/083023} {\bibfield  {journal}
  {\bibinfo  {journal} {New J. Phys.}\ }\textbf {\bibinfo {volume} {18}},\
  \bibinfo {pages} {083023} (\bibinfo {year} {2016})}\BibitemShut {NoStop}%
\bibitem [{\citenamefont {Iubini}(2019)}]{iubini_coupled_2019}%
  \BibitemOpen
  \bibfield  {author} {\bibinfo {author} {\bibfnamefont {S.}~\bibnamefont
  {Iubini}},\ }\bibfield  {title} {\bibinfo {title} {Coupled transport in a
  linear-stochastic {Schrödinger} equation},\ }\href
  {https://doi.org/10.1088/1742-5468/ab3aec} {\bibfield  {journal} {\bibinfo
  {journal} {J. Stat. Mech. Theory Exp.}\ }\textbf {\bibinfo {volume} {2019}},\
  \bibinfo {pages} {094016} (\bibinfo {year} {2019})}\BibitemShut {NoStop}%
\bibitem [{\citenamefont {Chatterjee}\ \emph {et~al.}(2020)\citenamefont
  {Chatterjee}, \citenamefont {Kundu},\ and\ \citenamefont
  {Kulkarni}}]{manas_spatiotemporal_2020}%
  \BibitemOpen
  \bibfield  {author} {\bibinfo {author} {\bibfnamefont {A.~K.}\ \bibnamefont
  {Chatterjee}}, \bibinfo {author} {\bibfnamefont {A.}~\bibnamefont {Kundu}},\
  and\ \bibinfo {author} {\bibfnamefont {M.}~\bibnamefont {Kulkarni}},\
  }\bibfield  {title} {\bibinfo {title} {Spatiotemporal spread of perturbations
  in a driven dissipative {Duffing} chain: {An} out-of-time-ordered correlator
  approach},\ }\href {https://doi.org/10.1103/PhysRevE.102.052103} {\bibfield
  {journal} {\bibinfo  {journal} {Phys. Rev. E}\ }\textbf {\bibinfo {volume}
  {102}},\ \bibinfo {pages} {052103} (\bibinfo {year} {2020})}\BibitemShut
  {NoStop}%
\bibitem [{\citenamefont {Komorowski}\ \emph {et~al.}(2023)\citenamefont
  {Komorowski}, \citenamefont {Lebowitz},\ and\ \citenamefont
  {Olla}}]{komorowski_heat_2023}%
  \BibitemOpen
  \bibfield  {author} {\bibinfo {author} {\bibfnamefont {T.}~\bibnamefont
  {Komorowski}}, \bibinfo {author} {\bibfnamefont {J.~L.}\ \bibnamefont
  {Lebowitz}},\ and\ \bibinfo {author} {\bibfnamefont {S.}~\bibnamefont
  {Olla}},\ }\bibfield  {title} {\bibinfo {title} {Heat {Flow} in a
  {Periodically} {Forced}, {Thermostatted} {Chain}},\ }\href
  {https://doi.org/10.1007/s00220-023-04654-4} {\bibfield  {journal} {\bibinfo
  {journal} {Commun. Math. Phys.}\ }\textbf {\bibinfo {volume} {400}},\
  \bibinfo {pages} {2181} (\bibinfo {year} {2023})}\BibitemShut {NoStop}%
\bibitem [{\citenamefont {Dagvadorj}\ \emph {et~al.}(2015)\citenamefont
  {Dagvadorj}, \citenamefont {Fellows}, \citenamefont {Matyjaśkiewicz},
  \citenamefont {Marchetti}, \citenamefont {Carusotto},\ and\ \citenamefont
  {Szymańska}}]{dagvadorj_nonequilibrium_2015}%
  \BibitemOpen
  \bibfield  {author} {\bibinfo {author} {\bibfnamefont {G.}~\bibnamefont
  {Dagvadorj}}, \bibinfo {author} {\bibfnamefont {J.}~\bibnamefont {Fellows}},
  \bibinfo {author} {\bibfnamefont {S.}~\bibnamefont {Matyjaśkiewicz}},
  \bibinfo {author} {\bibfnamefont {F.}~\bibnamefont {Marchetti}}, \bibinfo
  {author} {\bibfnamefont {I.}~\bibnamefont {Carusotto}},\ and\ \bibinfo
  {author} {\bibfnamefont {M.}~\bibnamefont {Szymańska}},\ }\bibfield  {title}
  {\bibinfo {title} {Nonequilibrium {Phase} {Transition} in a
  {Two}-{Dimensional} {Driven} {Open} {Quantum} {System}},\ }\href
  {https://doi.org/10.1103/PhysRevX.5.041028} {\bibfield  {journal} {\bibinfo
  {journal} {Phys. Rev. X}\ }\textbf {\bibinfo {volume} {5}},\ \bibinfo {pages}
  {041028} (\bibinfo {year} {2015})}\BibitemShut {NoStop}%
\bibitem [{\citenamefont {Dujardin}\ \emph {et~al.}(2015)\citenamefont
  {Dujardin}, \citenamefont {Argüelles},\ and\ \citenamefont
  {Schlagheck}}]{dujardin_elastic_2015}%
  \BibitemOpen
  \bibfield  {author} {\bibinfo {author} {\bibfnamefont {J.}~\bibnamefont
  {Dujardin}}, \bibinfo {author} {\bibfnamefont {A.}~\bibnamefont
  {Argüelles}},\ and\ \bibinfo {author} {\bibfnamefont {P.}~\bibnamefont
  {Schlagheck}},\ }\bibfield  {title} {\bibinfo {title} {Elastic and inelastic
  transmission in guided atom lasers: {A} truncated {Wigner} approach},\ }\href
  {https://doi.org/10.1103/PhysRevA.91.033614} {\bibfield  {journal} {\bibinfo
  {journal} {Phys. Rev. A}\ }\textbf {\bibinfo {volume} {91}},\ \bibinfo
  {pages} {033614} (\bibinfo {year} {2015})}\BibitemShut {NoStop}%
\bibitem [{\citenamefont {Dujardin}\ \emph {et~al.}(2016)\citenamefont
  {Dujardin}, \citenamefont {Engl},\ and\ \citenamefont
  {Schlagheck}}]{dujardin_breakdown_2016}%
  \BibitemOpen
  \bibfield  {author} {\bibinfo {author} {\bibfnamefont {J.}~\bibnamefont
  {Dujardin}}, \bibinfo {author} {\bibfnamefont {T.}~\bibnamefont {Engl}},\
  and\ \bibinfo {author} {\bibfnamefont {P.}~\bibnamefont {Schlagheck}},\
  }\bibfield  {title} {\bibinfo {title} {Breakdown of {Anderson} localization
  in the transport of {Bose}-{Einstein} condensates through one-dimensional
  disordered potentials},\ }\href {https://doi.org/10.1103/PhysRevA.93.013612}
  {\bibfield  {journal} {\bibinfo  {journal} {Phys. Rev. A}\ }\textbf {\bibinfo
  {volume} {93}},\ \bibinfo {pages} {013612} (\bibinfo {year}
  {2016})}\BibitemShut {NoStop}%
\bibitem [{\citenamefont {Vicentini}\ \emph {et~al.}(2018)\citenamefont
  {Vicentini}, \citenamefont {Minganti}, \citenamefont {Rota}, \citenamefont
  {Orso},\ and\ \citenamefont {Ciuti}}]{vicentini_critical_2018}%
  \BibitemOpen
  \bibfield  {author} {\bibinfo {author} {\bibfnamefont {F.}~\bibnamefont
  {Vicentini}}, \bibinfo {author} {\bibfnamefont {F.}~\bibnamefont {Minganti}},
  \bibinfo {author} {\bibfnamefont {R.}~\bibnamefont {Rota}}, \bibinfo {author}
  {\bibfnamefont {G.}~\bibnamefont {Orso}},\ and\ \bibinfo {author}
  {\bibfnamefont {C.}~\bibnamefont {Ciuti}},\ }\bibfield  {title} {\bibinfo
  {title} {Critical slowing down in driven-dissipative {Bose}-{Hubbard}
  lattices},\ }\href {https://doi.org/10.1103/PhysRevA.97.013853} {\bibfield
  {journal} {\bibinfo  {journal} {Phys. Rev. A}\ }\textbf {\bibinfo {volume}
  {97}},\ \bibinfo {pages} {013853} (\bibinfo {year} {2018})}\BibitemShut
  {NoStop}%
\bibitem [{\citenamefont {Vicentini}\ \emph {et~al.}(2019)\citenamefont
  {Vicentini}, \citenamefont {Minganti}, \citenamefont {Biella}, \citenamefont
  {Orso},\ and\ \citenamefont {Ciuti}}]{vicentini_optimal_2019}%
  \BibitemOpen
  \bibfield  {author} {\bibinfo {author} {\bibfnamefont {F.}~\bibnamefont
  {Vicentini}}, \bibinfo {author} {\bibfnamefont {F.}~\bibnamefont {Minganti}},
  \bibinfo {author} {\bibfnamefont {A.}~\bibnamefont {Biella}}, \bibinfo
  {author} {\bibfnamefont {G.}~\bibnamefont {Orso}},\ and\ \bibinfo {author}
  {\bibfnamefont {C.}~\bibnamefont {Ciuti}},\ }\bibfield  {title} {\bibinfo
  {title} {Optimal stochastic unraveling of disordered open quantum systems:
  {Application} to driven-dissipative photonic lattices},\ }\href
  {https://doi.org/10.1103/PhysRevA.99.032115} {\bibfield  {journal} {\bibinfo
  {journal} {Phys. Rev. A}\ }\textbf {\bibinfo {volume} {99}},\ \bibinfo
  {pages} {032115} (\bibinfo {year} {2019})}\BibitemShut {NoStop}%
\bibitem [{\citenamefont {Schlagheck}\ \emph {et~al.}(2019)\citenamefont
  {Schlagheck}, \citenamefont {Ullmo}, \citenamefont {Urbina}, \citenamefont
  {Richter},\ and\ \citenamefont {Tomsovic}}]{schlagheck_enhancement_2019}%
  \BibitemOpen
  \bibfield  {author} {\bibinfo {author} {\bibfnamefont {P.}~\bibnamefont
  {Schlagheck}}, \bibinfo {author} {\bibfnamefont {D.}~\bibnamefont {Ullmo}},
  \bibinfo {author} {\bibfnamefont {J.~D.}\ \bibnamefont {Urbina}}, \bibinfo
  {author} {\bibfnamefont {K.}~\bibnamefont {Richter}},\ and\ \bibinfo {author}
  {\bibfnamefont {S.}~\bibnamefont {Tomsovic}},\ }\bibfield  {title} {\bibinfo
  {title} {Enhancement of {Many}-{Body} {Quantum} {Interference} in {Chaotic}
  {Bosonic} {Systems}: {The} {Role} of {Symmetry} and {Dynamics}},\ }\href
  {https://doi.org/10.1103/PhysRevLett.123.215302} {\bibfield  {journal}
  {\bibinfo  {journal} {Phys. Rev. Lett.}\ }\textbf {\bibinfo {volume} {123}},\
  \bibinfo {pages} {215302} (\bibinfo {year} {2019})}\BibitemShut {NoStop}%
\bibitem [{\citenamefont {Seibold}\ \emph {et~al.}(2020)\citenamefont
  {Seibold}, \citenamefont {Rota},\ and\ \citenamefont
  {Savona}}]{seibold_dissipative_2020}%
  \BibitemOpen
  \bibfield  {author} {\bibinfo {author} {\bibfnamefont {K.}~\bibnamefont
  {Seibold}}, \bibinfo {author} {\bibfnamefont {R.}~\bibnamefont {Rota}},\ and\
  \bibinfo {author} {\bibfnamefont {V.}~\bibnamefont {Savona}},\ }\bibfield
  {title} {\bibinfo {title} {Dissipative time crystal in an asymmetric
  nonlinear photonic dimer},\ }\href
  {https://doi.org/10.1103/PhysRevA.101.033839} {\bibfield  {journal} {\bibinfo
   {journal} {Phys. Rev. A}\ }\textbf {\bibinfo {volume} {101}},\ \bibinfo
  {pages} {033839} (\bibinfo {year} {2020})}\BibitemShut {NoStop}%
\bibitem [{\citenamefont {Deuar}\ \emph {et~al.}(2021)\citenamefont {Deuar},
  \citenamefont {Ferrier}, \citenamefont {Matuszewski}, \citenamefont {Orso},\
  and\ \citenamefont {Szymańska}}]{deuar_fully_2021}%
  \BibitemOpen
  \bibfield  {author} {\bibinfo {author} {\bibfnamefont {P.}~\bibnamefont
  {Deuar}}, \bibinfo {author} {\bibfnamefont {A.}~\bibnamefont {Ferrier}},
  \bibinfo {author} {\bibfnamefont {M.}~\bibnamefont {Matuszewski}}, \bibinfo
  {author} {\bibfnamefont {G.}~\bibnamefont {Orso}},\ and\ \bibinfo {author}
  {\bibfnamefont {M.~H.}\ \bibnamefont {Szymańska}},\ }\bibfield  {title}
  {\bibinfo {title} {Fully {Quantum} {Scalable} {Description} of
  {Driven}-{Dissipative} {Lattice} {Models}},\ }\href
  {https://doi.org/10.1103/PRXQuantum.2.010319} {\bibfield  {journal} {\bibinfo
   {journal} {PRX Quantum}\ }\textbf {\bibinfo {volume} {2}},\ \bibinfo {pages}
  {010319} (\bibinfo {year} {2021})}\BibitemShut {NoStop}%
\bibitem [{\citenamefont {Blais}\ \emph {et~al.}(2021)\citenamefont {Blais},
  \citenamefont {Grimsmo}, \citenamefont {Girvin},\ and\ \citenamefont
  {Wallraff}}]{blais_circuit_2021}%
  \BibitemOpen
  \bibfield  {author} {\bibinfo {author} {\bibfnamefont {A.}~\bibnamefont
  {Blais}}, \bibinfo {author} {\bibfnamefont {A.~L.}\ \bibnamefont {Grimsmo}},
  \bibinfo {author} {\bibfnamefont {S.}~\bibnamefont {Girvin}},\ and\ \bibinfo
  {author} {\bibfnamefont {A.}~\bibnamefont {Wallraff}},\ }\bibfield  {title}
  {\bibinfo {title} {Circuit quantum electrodynamics},\ }\href
  {https://doi.org/10.1103/RevModPhys.93.025005} {\bibfield  {journal}
  {\bibinfo  {journal} {Rev. Mod. Phys.}\ }\textbf {\bibinfo {volume} {93}},\
  \bibinfo {pages} {025005} (\bibinfo {year} {2021})}\BibitemShut {NoStop}%
\bibitem [{Note1()}]{Note1}%
  \BibitemOpen
  \bibinfo {note} {Compared to the conventional indicator $g^{(2)}$, $\delta
  n_\ell $ defined in Eq.~(\ref {eqs:delta_n}) is less prone to numerical
  errors caused by a finite sampling of Wigner trajectories.}\BibitemShut
  {Stop}%
\bibitem [{\citenamefont {Scully}\ and\ \citenamefont
  {Zubairy}(1997)}]{scully_quantum_1997}%
  \BibitemOpen
  \bibfield  {author} {\bibinfo {author} {\bibfnamefont {M.~O.}\ \bibnamefont
  {Scully}}\ and\ \bibinfo {author} {\bibfnamefont {M.~S.}\ \bibnamefont
  {Zubairy}},\ }\href {https://doi.org/10.1017/CBO9780511813993} {\emph
  {\bibinfo {title} {Quantum {Optics}}}},\ \bibinfo {edition} {1st}\ ed.\
  (\bibinfo  {publisher} {Cambridge University Press},\ \bibinfo {year}
  {1997})\BibitemShut {NoStop}%
\bibitem [{\citenamefont {Das}\ \emph {et~al.}(2018)\citenamefont {Das},
  \citenamefont {Chakrabarty}, \citenamefont {Dhar}, \citenamefont {Kundu},
  \citenamefont {Huse}, \citenamefont {Moessner}, \citenamefont {Ray},\ and\
  \citenamefont {Bhattacharjee}}]{das_light-cone_2018}%
  \BibitemOpen
  \bibfield  {author} {\bibinfo {author} {\bibfnamefont {A.}~\bibnamefont
  {Das}}, \bibinfo {author} {\bibfnamefont {S.}~\bibnamefont {Chakrabarty}},
  \bibinfo {author} {\bibfnamefont {A.}~\bibnamefont {Dhar}}, \bibinfo {author}
  {\bibfnamefont {A.}~\bibnamefont {Kundu}}, \bibinfo {author} {\bibfnamefont
  {D.~A.}\ \bibnamefont {Huse}}, \bibinfo {author} {\bibfnamefont
  {R.}~\bibnamefont {Moessner}}, \bibinfo {author} {\bibfnamefont {S.~S.}\
  \bibnamefont {Ray}},\ and\ \bibinfo {author} {\bibfnamefont {S.}~\bibnamefont
  {Bhattacharjee}},\ }\bibfield  {title} {\bibinfo {title} {Light-{Cone}
  {Spreading} of {Perturbations} and the {Butterfly} {Effect} in a {Classical}
  {Spin} {Chain}},\ }\href {https://doi.org/10.1103/PhysRevLett.121.024101}
  {\bibfield  {journal} {\bibinfo  {journal} {Phys. Rev. Lett.}\ }\textbf
  {\bibinfo {volume} {121}},\ \bibinfo {pages} {024101} (\bibinfo {year}
  {2018})}\BibitemShut {NoStop}%
\bibitem [{\citenamefont {Bilitewski}\ \emph {et~al.}(2018)\citenamefont
  {Bilitewski}, \citenamefont {Bhattacharjee},\ and\ \citenamefont
  {Moessner}}]{bilitewski_temperature2018}%
  \BibitemOpen
  \bibfield  {author} {\bibinfo {author} {\bibfnamefont {T.}~\bibnamefont
  {Bilitewski}}, \bibinfo {author} {\bibfnamefont {S.}~\bibnamefont
  {Bhattacharjee}},\ and\ \bibinfo {author} {\bibfnamefont {R.}~\bibnamefont
  {Moessner}},\ }\bibfield  {title} {\bibinfo {title} {Temperature dependence
  of the butterfly effect in a classical many-body system},\ }\href
  {https://doi.org/10.1103/PhysRevLett.121.250602} {\bibfield  {journal}
  {\bibinfo  {journal} {Phys. Rev. Lett.}\ }\textbf {\bibinfo {volume} {121}},\
  \bibinfo {pages} {250602} (\bibinfo {year} {2018})}\BibitemShut {NoStop}%
\bibitem [{\citenamefont {Schuckert}\ and\ \citenamefont
  {Knap}(2019)}]{schuckert_thermal2019}%
  \BibitemOpen
  \bibfield  {author} {\bibinfo {author} {\bibfnamefont {A.}~\bibnamefont
  {Schuckert}}\ and\ \bibinfo {author} {\bibfnamefont {M.}~\bibnamefont
  {Knap}},\ }\bibfield  {title} {\bibinfo {title} {{Many-body chaos near a
  thermal phase transition}},\ }\href
  {https://doi.org/10.21468/SciPostPhys.7.2.022} {\bibfield  {journal}
  {\bibinfo  {journal} {SciPost Phys.}\ }\textbf {\bibinfo {volume} {7}},\
  \bibinfo {pages} {022} (\bibinfo {year} {2019})}\BibitemShut {NoStop}%
\bibitem [{\citenamefont {Bilitewski}\ \emph {et~al.}(2021)\citenamefont
  {Bilitewski}, \citenamefont {Bhattacharjee},\ and\ \citenamefont
  {Moessner}}]{bilitewski_classical2021}%
  \BibitemOpen
  \bibfield  {author} {\bibinfo {author} {\bibfnamefont {T.}~\bibnamefont
  {Bilitewski}}, \bibinfo {author} {\bibfnamefont {S.}~\bibnamefont
  {Bhattacharjee}},\ and\ \bibinfo {author} {\bibfnamefont {R.}~\bibnamefont
  {Moessner}},\ }\bibfield  {title} {\bibinfo {title} {Classical many-body
  chaos with and without quasiparticles},\ }\href
  {https://doi.org/10.1103/PhysRevB.103.174302} {\bibfield  {journal} {\bibinfo
   {journal} {Phys. Rev. B}\ }\textbf {\bibinfo {volume} {103}},\ \bibinfo
  {pages} {174302} (\bibinfo {year} {2021})}\BibitemShut {NoStop}%
\bibitem [{\citenamefont {Ruidas}\ and\ \citenamefont
  {Banerjee}(2021)}]{sibaram_thermal2021}%
  \BibitemOpen
  \bibfield  {author} {\bibinfo {author} {\bibfnamefont {S.}~\bibnamefont
  {Ruidas}}\ and\ \bibinfo {author} {\bibfnamefont {S.}~\bibnamefont
  {Banerjee}},\ }\bibfield  {title} {\bibinfo {title} {{Many-body chaos and
  anomalous diffusion across thermal phase transitions in two dimensions}},\
  }\href {https://doi.org/10.21468/SciPostPhys.11.5.087} {\bibfield  {journal}
  {\bibinfo  {journal} {SciPost Phys.}\ }\textbf {\bibinfo {volume} {11}},\
  \bibinfo {pages} {087} (\bibinfo {year} {2021})}\BibitemShut {NoStop}%
\bibitem [{\citenamefont {Deger}\ \emph {et~al.}(2022)\citenamefont {Deger},
  \citenamefont {Roy},\ and\ \citenamefont {Lazarides}}]{deger_arresting2022}%
  \BibitemOpen
  \bibfield  {author} {\bibinfo {author} {\bibfnamefont {A.}~\bibnamefont
  {Deger}}, \bibinfo {author} {\bibfnamefont {S.}~\bibnamefont {Roy}},\ and\
  \bibinfo {author} {\bibfnamefont {A.}~\bibnamefont {Lazarides}},\ }\bibfield
  {title} {\bibinfo {title} {Arresting classical many-body chaos by kinetic
  constraints},\ }\href {https://doi.org/10.1103/PhysRevLett.129.160601}
  {\bibfield  {journal} {\bibinfo  {journal} {Phys. Rev. Lett.}\ }\textbf
  {\bibinfo {volume} {129}},\ \bibinfo {pages} {160601} (\bibinfo {year}
  {2022})}\BibitemShut {NoStop}%
\bibitem [{\citenamefont {Biondi}\ \emph {et~al.}(2017)\citenamefont {Biondi},
  \citenamefont {Blatter}, \citenamefont {Türeci},\ and\ \citenamefont
  {Schmidt}}]{biondi_nonequilibrium2017}%
  \BibitemOpen
  \bibfield  {author} {\bibinfo {author} {\bibfnamefont {M.}~\bibnamefont
  {Biondi}}, \bibinfo {author} {\bibfnamefont {G.}~\bibnamefont {Blatter}},
  \bibinfo {author} {\bibfnamefont {H.~E.}\ \bibnamefont {Türeci}},\ and\
  \bibinfo {author} {\bibfnamefont {S.}~\bibnamefont {Schmidt}},\ }\bibfield
  {title} {\bibinfo {title} {Nonequilibrium gas-liquid transition in the
  driven-dissipative photonic lattice},\ }\href
  {https://doi.org/10.1103/PhysRevA.96.043809} {\bibfield  {journal} {\bibinfo
  {journal} {Phys. Rev. A}\ }\textbf {\bibinfo {volume} {96}},\ \bibinfo
  {pages} {043809} (\bibinfo {year} {2017})}\BibitemShut {NoStop}%
\bibitem [{\citenamefont {Minganti}\ \emph {et~al.}(2020)\citenamefont
  {Minganti}, \citenamefont {Arkhipov}, \citenamefont {Miranowicz},\ and\
  \citenamefont {Nori}}]{minganti_correspondance_2020}%
  \BibitemOpen
  \bibfield  {author} {\bibinfo {author} {\bibfnamefont {F.}~\bibnamefont
  {Minganti}}, \bibinfo {author} {\bibfnamefont {I.~I.}\ \bibnamefont
  {Arkhipov}}, \bibinfo {author} {\bibfnamefont {A.}~\bibnamefont
  {Miranowicz}},\ and\ \bibinfo {author} {\bibfnamefont {F.}~\bibnamefont
  {Nori}},\ }\href {https://arxiv.org/abs/2008.08075} {\bibinfo {title}
  {Correspondence between dissipative phase transitions of light and time
  crystals}} (\bibinfo {year} {2020}),\ \Eprint
  {https://arxiv.org/abs/2008.08075} {arXiv:2008.08075 [quant-ph]} \BibitemShut
  {NoStop}%
\bibitem [{\citenamefont {Minganti}\ \emph
  {et~al.}(2021{\natexlab{a}})\citenamefont {Minganti}, \citenamefont
  {Arkhipov}, \citenamefont {Miranowicz},\ and\ \citenamefont
  {Nori}}]{minganti_continuous_2021}%
  \BibitemOpen
  \bibfield  {author} {\bibinfo {author} {\bibfnamefont {F.}~\bibnamefont
  {Minganti}}, \bibinfo {author} {\bibfnamefont {I.~I.}\ \bibnamefont
  {Arkhipov}}, \bibinfo {author} {\bibfnamefont {A.}~\bibnamefont
  {Miranowicz}},\ and\ \bibinfo {author} {\bibfnamefont {F.}~\bibnamefont
  {Nori}},\ }\bibfield  {title} {\bibinfo {title} {Continuous dissipative phase
  transitions with or without symmetry breaking},\ }\href
  {https://doi.org/10.1088/1367-2630/ac3db8} {\bibfield  {journal} {\bibinfo
  {journal} {New J. Phys.}\ }\textbf {\bibinfo {volume} {23}},\ \bibinfo
  {pages} {122001} (\bibinfo {year} {2021}{\natexlab{a}})}\BibitemShut
  {NoStop}%
\bibitem [{\citenamefont {Minganti}\ \emph
  {et~al.}(2021{\natexlab{b}})\citenamefont {Minganti}, \citenamefont
  {Arkhipov}, \citenamefont {Miranowicz},\ and\ \citenamefont
  {Nori}}]{minganti_liouvillian_2021}%
  \BibitemOpen
  \bibfield  {author} {\bibinfo {author} {\bibfnamefont {F.}~\bibnamefont
  {Minganti}}, \bibinfo {author} {\bibfnamefont {I.~I.}\ \bibnamefont
  {Arkhipov}}, \bibinfo {author} {\bibfnamefont {A.}~\bibnamefont
  {Miranowicz}},\ and\ \bibinfo {author} {\bibfnamefont {F.}~\bibnamefont
  {Nori}},\ }\bibfield  {title} {\bibinfo {title} {Liouvillian spectral
  collapse in the {Scully}-{Lamb} laser model},\ }\href
  {https://doi.org/10.1103/PhysRevResearch.3.043197} {\bibfield  {journal}
  {\bibinfo  {journal} {Phys. Rev. Res.}\ }\textbf {\bibinfo {volume} {3}},\
  \bibinfo {pages} {043197} (\bibinfo {year} {2021}{\natexlab{b}})}\BibitemShut
  {NoStop}%
\bibitem [{\citenamefont {Scully}\ and\ \citenamefont
  {Lamb}(1967)}]{scully_quantum_1967}%
  \BibitemOpen
  \bibfield  {author} {\bibinfo {author} {\bibfnamefont {M.~O.}\ \bibnamefont
  {Scully}}\ and\ \bibinfo {author} {\bibfnamefont {W.~E.}\ \bibnamefont
  {Lamb}},\ }\bibfield  {title} {\bibinfo {title} {Quantum {Theory} of an
  {Optical} {Maser}. {I}. {General} {Theory}},\ }\href
  {https://doi.org/10.1103/PhysRev.159.208} {\bibfield  {journal} {\bibinfo
  {journal} {Phys. Rev.}\ }\textbf {\bibinfo {volume} {159}},\ \bibinfo {pages}
  {208} (\bibinfo {year} {1967})}\BibitemShut {NoStop}%
\bibitem [{\citenamefont {Yamamoto}\ and\ \citenamefont {$\dot{\rm
  I}$mamo$\breve{\rm g}$lu}(1999)}]{yamamoto_mesoscopic_1999}%
  \BibitemOpen
  \bibfield  {author} {\bibinfo {author} {\bibfnamefont {Y.}~\bibnamefont
  {Yamamoto}}\ and\ \bibinfo {author} {\bibfnamefont {A.}~\bibnamefont
  {$\dot{\rm I}$mamo$\breve{\rm g}$lu}},\ }\href@noop {} {\emph {\bibinfo
  {title} {Mesoscopic quantum optics}}}\ (\bibinfo  {publisher} {John Wiley},\
  \bibinfo {address} {New York},\ \bibinfo {year} {1999})\BibitemShut {NoStop}%
\bibitem [{\citenamefont {Minganti}\ \emph {et~al.}(2018)\citenamefont
  {Minganti}, \citenamefont {Biella}, \citenamefont {Bartolo},\ and\
  \citenamefont {Ciuti}}]{minganti_spectral_2018}%
  \BibitemOpen
  \bibfield  {author} {\bibinfo {author} {\bibfnamefont {F.}~\bibnamefont
  {Minganti}}, \bibinfo {author} {\bibfnamefont {A.}~\bibnamefont {Biella}},
  \bibinfo {author} {\bibfnamefont {N.}~\bibnamefont {Bartolo}},\ and\ \bibinfo
  {author} {\bibfnamefont {C.}~\bibnamefont {Ciuti}},\ }\bibfield  {title}
  {\bibinfo {title} {Spectral theory of {Liouvillians} for dissipative phase
  transitions},\ }\href {https://doi.org/10.1103/PhysRevA.98.042118} {\bibfield
   {journal} {\bibinfo  {journal} {Phys. Rev. A}\ }\textbf {\bibinfo {volume}
  {98}},\ \bibinfo {pages} {042118} (\bibinfo {year} {2018})}\BibitemShut
  {NoStop}%
\bibitem [{\citenamefont {Garbe}\ \emph {et~al.}(2024)\citenamefont {Garbe},
  \citenamefont {Minoguchi}, \citenamefont {Huber},\ and\ \citenamefont
  {Rabl}}]{garbe_bosonic_2024}%
  \BibitemOpen
  \bibfield  {author} {\bibinfo {author} {\bibfnamefont {L.}~\bibnamefont
  {Garbe}}, \bibinfo {author} {\bibfnamefont {Y.}~\bibnamefont {Minoguchi}},
  \bibinfo {author} {\bibfnamefont {J.}~\bibnamefont {Huber}},\ and\ \bibinfo
  {author} {\bibfnamefont {P.}~\bibnamefont {Rabl}},\ }\bibfield  {title}
  {\bibinfo {title} {The bosonic skin effect: {Boundary} condensation in
  asymmetric transport},\ }\href
  {https://doi.org/10.21468/SciPostPhys.16.1.029} {\bibfield  {journal}
  {\bibinfo  {journal} {SciPost Phys.}\ }\textbf {\bibinfo {volume} {16}},\
  \bibinfo {pages} {029} (\bibinfo {year} {2024})}\BibitemShut {NoStop}%
\bibitem [{\citenamefont {Muraev}\ \emph {et~al.}(2024)\citenamefont {Muraev},
  \citenamefont {Maksimov},\ and\ \citenamefont
  {Kolovsky}}]{muraev_signatures_2024}%
  \BibitemOpen
  \bibfield  {author} {\bibinfo {author} {\bibfnamefont {P.~S.}\ \bibnamefont
  {Muraev}}, \bibinfo {author} {\bibfnamefont {D.~N.}\ \bibnamefont
  {Maksimov}},\ and\ \bibinfo {author} {\bibfnamefont {A.~R.}\ \bibnamefont
  {Kolovsky}},\ }\bibfield  {title} {\bibinfo {title} {Signatures of quantum
  chaos and fermionization in the incoherent transport of bosonic carriers in
  the {Bose}-{Hubbard} chain},\ }\href
  {https://doi.org/10.1103/PhysRevE.109.L032107} {\bibfield  {journal}
  {\bibinfo  {journal} {Phys. Rev. E}\ }\textbf {\bibinfo {volume} {109}},\
  \bibinfo {pages} {L032107} (\bibinfo {year} {2024})}\BibitemShut {NoStop}%
\bibitem [{\citenamefont {Marković}\ and\ \citenamefont
  {Čubrović}(2024)}]{markovic_chaos_2024}%
  \BibitemOpen
  \bibfield  {author} {\bibinfo {author} {\bibfnamefont {D.}~\bibnamefont
  {Marković}}\ and\ \bibinfo {author} {\bibfnamefont {M.}~\bibnamefont
  {Čubrović}},\ }\bibfield  {title} {\bibinfo {title} {Chaos and anomalous
  transport in a semiclassical {Bose}-{Hubbard} chain},\ }\href
  {https://doi.org/10.1103/PhysRevE.109.034213} {\bibfield  {journal} {\bibinfo
   {journal} {Phys. Rev. E}\ }\textbf {\bibinfo {volume} {109}},\ \bibinfo
  {pages} {034213} (\bibinfo {year} {2024})}\BibitemShut {NoStop}%
\bibitem [{\citenamefont {Xu}\ and\ \citenamefont
  {Swingle}(2024)}]{xu_scrambling_2024}%
  \BibitemOpen
  \bibfield  {author} {\bibinfo {author} {\bibfnamefont {S.}~\bibnamefont
  {Xu}}\ and\ \bibinfo {author} {\bibfnamefont {B.}~\bibnamefont {Swingle}},\
  }\bibfield  {title} {\bibinfo {title} {Scrambling {Dynamics} and
  {Out}-of-{Time}-{Ordered} {Correlators} in {Quantum} {Many}-{Body}
  {Systems}},\ }\href {https://doi.org/10.1103/PRXQuantum.5.010201} {\bibfield
  {journal} {\bibinfo  {journal} {PRX Quantum}\ }\textbf {\bibinfo {volume}
  {5}},\ \bibinfo {pages} {010201} (\bibinfo {year} {2024})}\BibitemShut
  {NoStop}%
\bibitem [{\citenamefont {Lieb}\ and\ \citenamefont
  {Robinson}(1972)}]{lieb_finite_1972}%
  \BibitemOpen
  \bibfield  {author} {\bibinfo {author} {\bibfnamefont {E.~H.}\ \bibnamefont
  {Lieb}}\ and\ \bibinfo {author} {\bibfnamefont {D.~W.}\ \bibnamefont
  {Robinson}},\ }\bibfield  {title} {\bibinfo {title} {The finite group
  velocity of quantum spin systems},\ }\href
  {https://doi.org/10.1007/BF01645779} {\bibfield  {journal} {\bibinfo
  {journal} {Commun. Math. Phys.}\ }\textbf {\bibinfo {volume} {28}},\ \bibinfo
  {pages} {251} (\bibinfo {year} {1972})}\BibitemShut {NoStop}%
\bibitem [{\citenamefont {Wiseman}\ and\ \citenamefont
  {Milburn}(2009)}]{wiseman_quantum_2009}%
  \BibitemOpen
  \bibfield  {author} {\bibinfo {author} {\bibfnamefont {H.~M.}\ \bibnamefont
  {Wiseman}}\ and\ \bibinfo {author} {\bibfnamefont {G.~J.}\ \bibnamefont
  {Milburn}},\ }\href {https://doi.org/10.1017/CBO9780511813948} {\emph
  {\bibinfo {title} {Quantum {Measurement} and {Control}}}},\ \bibinfo
  {edition} {1st}\ ed.\ (\bibinfo  {publisher} {Cambridge University Press},\
  \bibinfo {year} {2009})\BibitemShut {NoStop}%
\bibitem [{\citenamefont {Jacobs}(2014)}]{jacobs_2014}%
  \BibitemOpen
  \bibfield  {author} {\bibinfo {author} {\bibfnamefont {K.}~\bibnamefont
  {Jacobs}},\ }\href {https://doi.org/10.1017/CBO9781139179027} {\emph
  {\bibinfo {title} {Quantum Measurement Theory and its Applications}}}\
  (\bibinfo  {publisher} {Cambridge University Press},\ \bibinfo {year}
  {2014})\BibitemShut {NoStop}%
\bibitem [{\citenamefont {Daley}(2014)}]{daley_quantum_2014}%
  \BibitemOpen
  \bibfield  {author} {\bibinfo {author} {\bibfnamefont {A.~J.}\ \bibnamefont
  {Daley}},\ }\bibfield  {title} {\bibinfo {title} {Quantum trajectories and
  open many-body quantum systems},\ }\href
  {https://doi.org/10.1080/00018732.2014.933502} {\bibfield  {journal}
  {\bibinfo  {journal} {Adv. Phys.}\ }\textbf {\bibinfo {volume} {63}},\
  \bibinfo {pages} {77} (\bibinfo {year} {2014})}\BibitemShut {NoStop}%
\bibitem [{\citenamefont {Carmichael}(1993)}]{carmichael_open_1993}%
  \BibitemOpen
  \bibfield  {author} {\bibinfo {author} {\bibfnamefont {H.}~\bibnamefont
  {Carmichael}},\ }\href {https://doi.org/10.1007/978-3-540-47620-7} {\emph
  {\bibinfo {title} {An {Open} {Systems} {Approach} to {Quantum} {Optics}:
  {Lectures} {Presented} at the {Université} {Libre} de {Bruxelles} {October}
  28 to {November} 4, 1991}}},\ edited by\ \bibinfo {editor} {\bibfnamefont
  {H.}~\bibnamefont {Araki}}, \bibinfo {editor} {\bibfnamefont
  {E.}~\bibnamefont {Brézin}}, \bibinfo {editor} {\bibfnamefont
  {J.}~\bibnamefont {Ehlers}}, \bibinfo {editor} {\bibfnamefont
  {U.}~\bibnamefont {Frisch}}, \bibinfo {editor} {\bibfnamefont
  {K.}~\bibnamefont {Hepp}}, \bibinfo {editor} {\bibfnamefont {R.~L.}\
  \bibnamefont {Jaffe}}, \bibinfo {editor} {\bibfnamefont {R.}~\bibnamefont
  {Kippenhahn}}, \bibinfo {editor} {\bibfnamefont {H.~A.}\ \bibnamefont
  {Weidenm\"uller}}, \bibinfo {editor} {\bibfnamefont {J.}~\bibnamefont
  {Wess}}, \bibinfo {editor} {\bibfnamefont {J.}~\bibnamefont {Zittartz}},\
  and\ \bibinfo {editor} {\bibfnamefont {W.}~\bibnamefont {Beiglb\"ock}},\
  \bibinfo {series} {Lecture {Notes} in {Physics} {Monographs}}, Vol.~\bibinfo
  {volume} {18}\ (\bibinfo  {publisher} {Springer Berlin Heidelberg},\ \bibinfo
  {address} {Berlin, Heidelberg},\ \bibinfo {year} {1993})\BibitemShut
  {NoStop}%
\bibitem [{\citenamefont {Mølmer}\ \emph {et~al.}(1993)\citenamefont
  {Mølmer}, \citenamefont {Castin},\ and\ \citenamefont
  {Dalibard}}]{molmer_monte_1993}%
  \BibitemOpen
  \bibfield  {author} {\bibinfo {author} {\bibfnamefont {K.}~\bibnamefont
  {Mølmer}}, \bibinfo {author} {\bibfnamefont {Y.}~\bibnamefont {Castin}},\
  and\ \bibinfo {author} {\bibfnamefont {J.}~\bibnamefont {Dalibard}},\
  }\bibfield  {title} {\bibinfo {title} {Monte {Carlo} wave-function method in
  quantum optics},\ }\href {https://doi.org/10.1364/JOSAB.10.000524} {\bibfield
   {journal} {\bibinfo  {journal} {J. Opt. Soc. Am. B}\ }\textbf {\bibinfo
  {volume} {10}},\ \bibinfo {pages} {524} (\bibinfo {year} {1993})}\BibitemShut
  {NoStop}%
\bibitem [{\citenamefont {Mercurio}\ \emph {et~al.}(2025)\citenamefont
  {Mercurio}, \citenamefont {Huang}, \citenamefont {Cai}, \citenamefont {Chen},
  \citenamefont {Savona},\ and\ \citenamefont {Nori}}]{mercurio_ARXIV_2025}%
  \BibitemOpen
  \bibfield  {author} {\bibinfo {author} {\bibfnamefont {A.}~\bibnamefont
  {Mercurio}}, \bibinfo {author} {\bibfnamefont {Y.-T.}\ \bibnamefont {Huang}},
  \bibinfo {author} {\bibfnamefont {L.-X.}\ \bibnamefont {Cai}}, \bibinfo
  {author} {\bibfnamefont {Y.-N.}\ \bibnamefont {Chen}}, \bibinfo {author}
  {\bibfnamefont {V.}~\bibnamefont {Savona}},\ and\ \bibinfo {author}
  {\bibfnamefont {F.}~\bibnamefont {Nori}},\ }\href
  {https://arxiv.org/abs/2504.21440} {\bibinfo {title} {Quantumtoolbox.jl: An
  efficient julia framework for simulating open quantum systems}} (\bibinfo
  {year} {2025}),\ \Eprint {https://arxiv.org/abs/2504.21440} {arXiv:2504.21440
  [quant-ph]} \BibitemShut {NoStop}%
\bibitem [{\citenamefont {Kubo}(1962)}]{kubo_generalized_1962}%
  \BibitemOpen
  \bibfield  {author} {\bibinfo {author} {\bibfnamefont {R.}~\bibnamefont
  {Kubo}},\ }\bibfield  {title} {\bibinfo {title} {Generalized {Cumulant}
  {Expansion} {Method}},\ }\href {https://doi.org/10.1143/JPSJ.17.1100}
  {\bibfield  {journal} {\bibinfo  {journal} {J. Phys. Soc. Jpn.}\ }\textbf
  {\bibinfo {volume} {17}},\ \bibinfo {pages} {1100} (\bibinfo {year}
  {1962})}\BibitemShut {NoStop}%
\bibitem [{\citenamefont {Plankensteiner}\ \emph {et~al.}(2022)\citenamefont
  {Plankensteiner}, \citenamefont {Hotter},\ and\ \citenamefont
  {Ritsch}}]{plankensteiner_quantumcumulantsjl_2022}%
  \BibitemOpen
  \bibfield  {author} {\bibinfo {author} {\bibfnamefont {D.}~\bibnamefont
  {Plankensteiner}}, \bibinfo {author} {\bibfnamefont {C.}~\bibnamefont
  {Hotter}},\ and\ \bibinfo {author} {\bibfnamefont {H.}~\bibnamefont
  {Ritsch}},\ }\bibfield  {title} {\bibinfo {title} {{QuantumCumulants}.jl: {A}
  {Julia} framework for generalized mean-field equations in open quantum
  systems},\ }\href {https://doi.org/10.22331/q-2022-01-04-617} {\bibfield
  {journal} {\bibinfo  {journal} {Quantum}\ }\textbf {\bibinfo {volume} {6}},\
  \bibinfo {pages} {617} (\bibinfo {year} {2022})}\BibitemShut {NoStop}%
\bibitem [{\citenamefont {Yariv}(1989)}]{yariv_quantum_1989}%
  \BibitemOpen
  \bibfield  {author} {\bibinfo {author} {\bibfnamefont {A.}~\bibnamefont
  {Yariv}},\ }\href@noop {} {\emph {\bibinfo {title} {Quantum electronics}}},\
  \bibinfo {edition} {3rd}\ ed.\ (\bibinfo  {publisher} {Wiley},\ \bibinfo
  {address} {New York},\ \bibinfo {year} {1989})\BibitemShut {NoStop}%
\bibitem [{\citenamefont {Milonni}\ and\ \citenamefont
  {Eberly}(2010)}]{milonni_laser_2010}%
  \BibitemOpen
  \bibfield  {author} {\bibinfo {author} {\bibfnamefont {P.~W.}\ \bibnamefont
  {Milonni}}\ and\ \bibinfo {author} {\bibfnamefont {J.~H.}\ \bibnamefont
  {Eberly}},\ }\href {https://doi.org/10.1002/9780470409718} {\emph {\bibinfo
  {title} {Laser {Physics}}}},\ \bibinfo {edition} {1st}\ ed.\ (\bibinfo
  {publisher} {Wiley},\ \bibinfo {year} {2010})\BibitemShut {NoStop}%
\bibitem [{\citenamefont {Maiman}(1960)}]{maiman_stimulated_1960}%
  \BibitemOpen
  \bibfield  {author} {\bibinfo {author} {\bibfnamefont {T.~H.}\ \bibnamefont
  {Maiman}},\ }\bibfield  {title} {\bibinfo {title} {Stimulated {Optical}
  {Radiation} in {Ruby}},\ }\href {https://doi.org/10.1038/187493a0} {\bibfield
   {journal} {\bibinfo  {journal} {Nature}\ }\textbf {\bibinfo {volume}
  {187}},\ \bibinfo {pages} {493} (\bibinfo {year} {1960})}\BibitemShut
  {NoStop}%
\bibitem [{\citenamefont {Beaulieu}\ \emph {et~al.}(2025)\citenamefont
  {Beaulieu}, \citenamefont {Minganti}, \citenamefont {Frasca}, \citenamefont
  {Savona}, \citenamefont {Felicetti}, \citenamefont {Di~Candia},\ and\
  \citenamefont {Scarlino}}]{beaulieu_observation_2025}%
  \BibitemOpen
  \bibfield  {author} {\bibinfo {author} {\bibfnamefont {G.}~\bibnamefont
  {Beaulieu}}, \bibinfo {author} {\bibfnamefont {F.}~\bibnamefont {Minganti}},
  \bibinfo {author} {\bibfnamefont {S.}~\bibnamefont {Frasca}}, \bibinfo
  {author} {\bibfnamefont {V.}~\bibnamefont {Savona}}, \bibinfo {author}
  {\bibfnamefont {S.}~\bibnamefont {Felicetti}}, \bibinfo {author}
  {\bibfnamefont {R.}~\bibnamefont {Di~Candia}},\ and\ \bibinfo {author}
  {\bibfnamefont {P.}~\bibnamefont {Scarlino}},\ }\bibfield  {title} {\bibinfo
  {title} {Observation of first- and second-order dissipative phase transitions
  in a two-photon driven {Kerr} resonator},\ }\href
  {https://doi.org/10.1038/s41467-025-56830-w} {\bibfield  {journal} {\bibinfo
  {journal} {Nat. Commun.}\ }\textbf {\bibinfo {volume} {16}},\ \bibinfo
  {pages} {1954} (\bibinfo {year} {2025})}\BibitemShut {NoStop}%
\end{thebibliography}%

\clearpage
\onecolumngrid

\section*{Supplementary information}

\subsection*{1. Numerical approaches}\label{sec:numerics}

\subsubsection*{Quantum trajectories}\label{sec:ED}

Our starting point is the Lindblad master equation written in its most general form
\begin{equation}\label{eqs:lindblad_gen}
    \frac{\partial\hat{\rho}}{\partial t} = -\rmi[\hat{H}, \hat{\rho}] + \sum_j  \left(\hat{L}_j\hat{\rho}\hat{L}_j^{\dagger} - \frac{1}{2}\left\{ \hat{L}_j^{\dagger}\hat{L}_j, \hat{\rho}\right\}\right),
\end{equation}
where $\{\hat{L}_j\}$ is a collection of Lindblad jump operators.
Equation~(\ref{eqs:lindblad_gen}) admits a stochastic unraveling in terms of quantum trajectories $\ket{\psi(t)}$, combining the Hamiltonian dynamics with a continuous monitoring of the environment~\cite{wiseman_quantum_2009, jacobs_2014, daley_quantum_2014}.
Different measurement's protocols (\textit{i.e.}, different unraveling protocols) are possible, and they lead to different quantum trajectories. 
The most popular protocols are the homodyne measurement, which results in a stochastic Wiener process for the system's state $\ket{\psi(t)}$~\cite{carmichael_open_1993}, and the photon-counting measurement, giving rise to the Monte Carlo quantum trajectories~\cite{molmer_monte_1993}.
In the latter case, a quantum jump occurs in a time step $\rmd t$ with probability $\rmd p =  \sum_j \bra{\psi(t)}\hat{L}_j^{\dagger}\hat{L}_j\ket{\psi(t)}\rmd t$, and $\ket{\psi(t)}$ evolves into
\begin{equation}\label{eqs:jump_operators}
    \ket{\psi(t +\rmd t)}  \propto \hat{L}_j\ket{\psi(t)},
\end{equation}
where the jump operator $\hat{L}_j$ is sampled from the probability distribution 
\begin{equation}
    p_j = \frac{ \bra{\psi(t)}\hat{L}_j^{\dagger}\hat{L}_j\ket{\psi(t)}}{\sum_k\bra{\psi(t)}\hat{L}_k^{\dagger}\hat{L}_k\ket{\psi(t)}} .
\end{equation}
No quantum jump occurs in the time $\rmd t$ with probability $1-\rmd p$, and $\ket{\psi(t)}$ evolves into
\begin{equation}\label{eqs:non_hermitian_hamiltonian}
    \ket{\psi(t + \rmd t)} \propto (\hat 1 - \rmi \, \rmd t\hat{H}_{\rm nh})\ket{\psi(t)},
\end{equation}
where $\hat{H}_{\rm nh} := \hat{H} -\rmi \sum_j \hat{L}_j^{\dagger}\hat{L}_j/2$ is the associated non-Hermitian Hamiltonian. 
After each time step, the state is renormalized according to $\ket{\psi(t+\rmd t)} \mapsto \ket{\psi(t+\rmd t)}/\langle\psi(t+\rmd t)|\psi(t+\rmd t)\rangle$. 

The overall process leads to a stochastic Schr\"odinger equation for the wave function $\ket{\psi(t)}$. 
Expectation values of operators can be obtained by averaging over many independent quantum trajectories (see the discussion below).
The numerical results for exact dynamics have been obtained with the \textit{QuantumToolbox.jl} package \cite{mercurio_ARXIV_2025} available in Julia.

\subsubsection*{Truncated Wigner approximation}\label{sec:TWA}

In what follows, we mainly adapt the discussion on phase-space representation presented in Ref.~\cite{polkovnikov_phase_2010}. To simplify the discussion, we initially consider a single-mode resonator  described by a Hamiltonian $\hat{H}$ and a collection of Lindblad jump operators $\hat{L}_j$ that can be written as polynomials of single-mode creation and annihilation operators $\hat{a}^{\dagger}$ and $\hat{a}$. 
The phase-space representation maps operators (defined in an $M$-dimensional Hilbert space) into functions (defined in the phase space).
Here, we work in the coherent-state basis $\{\ket{\alpha}\}$, where $\ket{\alpha}$ is the eigenstate of the annihilation operator $\hat{a}$ with eigenvalue $\alpha$. 
The one-to-one mapping between an operator $\hat{O}$ in the Hilbert space and a function $O_W$ in phase space can be achieved by introducing the Weyl symbol
\begin{equation}\label{eqs:weyl_symbol}
     O_W(\alpha, \alpha^*) := \frac{1}{2^M}\int \rmd\zeta\rmd\zeta^*\bra{\alpha-\frac{1}{2}\zeta}\hat{O}(\hat{a}, \hat{a}^{\dagger})\ket{\alpha+\frac{1}{2}\zeta} \rme^{-\left(\alpha^*-\frac{\zeta^*}{2}\right)\left(\alpha+\frac{\zeta}{2}\right)}.
\end{equation}
If the operator $\hat{O}(\hat{a}, \hat{a}^{\dagger})$ is symmetric in $\hat{a}$ and $\hat{a}^{\dagger}$, then the Weyl symbol can be obtained with the simple substitution $\hat{a}\rightarrow \alpha$, $\hat{a}^{\dagger}\rightarrow \alpha^*$. 
The Weyl symbol of the density matrix $\hat{\rho}$ is called the Wigner function,
\begin{equation}\label{eqs:wigner_function}
    W(\alpha, \alpha^*) := \int \frac{\rmd\zeta \rmd\zeta^*}{2\pi}\bra{\alpha-\frac{1}{2}\zeta}\hat{\rho}\ket{\alpha+\frac{1}{2}\zeta} \rme^{-\left(\alpha^*-\frac{\zeta^*}{2}\right)\left(\alpha+\frac{\zeta}{2}\right)}.
\end{equation}
Within this formalism, expectation values of operators can be expressed as
\begin{equation}\label{eqs:expectation_value}
    \langle \hat{O}(\hat{a}, \hat{a}^{\dagger}) \rangle = \int \rmd\alpha \rmd\alpha^*\,O_W(\alpha, \alpha^*) W(\alpha, \alpha^*),
\end{equation}
\textit{i.e.}, classical statistical expectation values weighted over the Wigner function.
The Weyl symbol corresponding to a product of operators can be expressed as
\begin{equation}\label{eqs:moyal_product}
    (\hat{O}_1\hat{O}_2)_W(\alpha, \alpha^*) = O_{1W}(\alpha, \alpha^*)\, \rme^{\vecdouble{\Lambda}/2}\, O_{2W}(\alpha, \alpha^*),
\end{equation}
where $\rme^{\vecdouble{\Lambda}/2}$ is a Moyal product based on the derivative operator associated with the Poisson bracket
\begin{equation}\label{eqs:simplectic_poisson}
    \vecdouble{\Lambda} := \frac{\cev{\partial}}{\partial \alpha}\frac{\vec{\partial}}{\partial \alpha^*} - \frac{\cev{\partial}}{\partial \alpha^*}\frac{\vec{\partial}}{\partial \alpha}.
\end{equation}
The derivative $\vec{\partial}$ acts on the right, while $\cev{\partial}$ acts on the left. 
The non-commutativity of the Moyal product corresponds to the non-commutativity of the operator product in the Hilbert space.
The commutator between two operators is expressed in terms of phase-space variables as
\begin{align}\label{eqs:commutator}
    &([\hat{O}_1, \hat{O}_2])_W(\alpha, \alpha^*) = 2 \, O_{1W}(\alpha, \alpha^*)\sinh\left(\frac{1}{2}\vecdouble{\Lambda}\right)O_{2W}(\alpha, \alpha^*).
\end{align}
Weyl symbols of any operator $\hat{O}(\hat{a}, \hat{a}^{\dagger})$ can be computed using the Moyal product in Eq.~\eqref{eqs:moyal_product}. For example,
\begin{equation}\label{eqs:Moyal_n}
    (\hat{a}^{\dagger}\hat{a})_W = \alpha^*\left(1 - \frac{1}{2}\frac{\cev{\partial}}{\partial \alpha^*}\frac{\vec{\partial}}{\partial \alpha}\right)\alpha = |\alpha|^2 - \frac{1}{2},
\end{equation}

\begin{equation}\label{eqs:Moyal_delta_n}
    (\hat{a}^{\dagger}\hat{a}^{\dagger}\hat{a}\hat{a})_W = \alpha^{*2}\left(1 - \frac{1}{2}\frac{\cev{\partial}}{\partial \alpha^*}\frac{\vec{\partial}}{\partial \alpha} + \frac{1}{8}\frac{\cev{\partial}^2}{\partial \alpha^{*2}}\frac{\vec{\partial}^2}{\partial \alpha^2}\right)\alpha^2 = |\alpha|^4 - 2|\alpha|^2 + \frac{1}{2}.
\end{equation}
Let us notice that the zeroth-order expansion of the Moyal product would have led to $(\hat{a}^\dagger\hat{a})_W = |\alpha|^2$ and $(\hat{a}^\dagger\hat{a}^\dagger\hat{a}\hat{a})_W = |\alpha|^4$, \textit{i.e.}, treating quantum operators classical commuting quantities. 
The remaining terms in Eqs.~(\ref{eqs:Moyal_n}) and (\ref{eqs:Moyal_delta_n}) originate from the quantum fluctuations that are  captured by the systematic expansion of the Moyal product.

The phase-space representation can be easily generalized to spatially extended systems, where the full Hilbert space is now the tensor product of, say, $L$ local Hilbert spaces.
The Wigner function associated with the density matrix $\hat \rho(t)$ is now defined as 
\begin{align}\label{eq:wigner_function}
    W(t;\alpha_1,\alpha_1^*,..., \alpha_L, \alpha_L^*) := &\int \frac{\rmd\zeta_1 \rmd\zeta_1^* ...\rmd\zeta_L \rmd\zeta_L^*}{(2\pi)^L}\big{\langle} \alpha_1-\frac{\zeta_1}{2},...,\alpha_L-\frac{\zeta_L}{2} \big{|} \hat{\rho}(t) \big{|} \alpha_1+\frac{\zeta_1}{2},..., \alpha_L+\frac{\zeta_L}{2} \big{\rangle} \nonumber \\
    & \quad \times \rme^{-\left(\alpha_1^*-\frac{1}{2}\zeta_1^*\right)\left(\alpha_1+\frac{1}{2}\zeta_1\right)}...\,\rme^{-\left(\alpha_L^*-\frac{1}{2}\zeta_L^*\right)\left(\alpha_L+\frac{1}{2}\zeta_L\right)}.
\end{align}
Similarly, the expressions for the Weyl symbols and expectation values are easily generalized from the single-mode case in Eqs.~\eqref{eqs:weyl_symbol} and~\eqref{eqs:expectation_value}.
Finally, the Poisson bracket in Eq.~\eqref{eqs:simplectic_poisson} generalizes to 
\begin{equation}
    \vecdouble{\Lambda} = \sum_{\ell=1}^L\left(\frac{\cev{\partial}}{\partial \alpha_\ell}\frac{\vec{\partial}}{\partial \alpha_\ell^*} - \frac{\cev{\partial}}{\partial \alpha_\ell^*}\frac{\vec{\partial}}{\partial \alpha_\ell}\right).
\end{equation}

Within this theoretical framework, the Lindblad equation~\eqref{eqs:lindblad} can be mapped into a partial differential equation (PDE) for the Wigner function $W(t;\alpha_1, \alpha^*_1, ..., \alpha_L, \alpha^*_L)$ that reads
\begin{align}\label{eqs:PDE}
   \rmi \frac{\partial W}{\partial t} = & - \Delta\sum_{\ell=1}^L\left(\alpha_\ell^*\frac{\partial}{\partial\alpha_\ell^*} - \alpha_\ell\frac{\partial}{\partial\alpha_\ell}\right)W 
    - J\sum_{\ell=1}^{L-1}\left(\alpha_{\ell+1}^*\frac{\partial}{\partial\alpha_\ell^*} - \alpha_{\ell+1}\frac{\partial}{\partial\alpha_\ell}\right)W
    - F\left(\frac{\partial}{\partial\alpha_1^*} - \frac{\partial}{\partial\alpha_1}\right)W \nonumber
   \\
   & 
   + U\sum_{\ell=1}^L(|\alpha_\ell|^2-1)\left(\alpha_\ell^*\frac{\partial}{\partial\alpha_\ell^*} - \alpha_\ell\frac{\partial}{\partial\alpha_\ell}\right)W - \frac{U}{4}\sum_{\ell=1}^L\left(\alpha_\ell^*\frac{\partial}{\partial\alpha_\ell^*} - \alpha_\ell\frac{\partial}{\partial\alpha_\ell}\right)\frac{\partial^2}{\partial\alpha_\ell\partial\alpha_\ell^*}W \nonumber \\
&   + \frac{\rmi\gamma}{2}  \left[ 
\frac{\partial}{\partial\alpha_1}(\alpha_1 W)+ \frac{\partial}{\partial\alpha_1^*}(\alpha_1^*W) + \frac{\partial^2}{\partial\alpha_1\partial\alpha_1^*}W
+ \frac{\partial}{\partial\alpha_L}(\alpha_L W)+ \frac{\partial}{\partial\alpha_L^*}(\alpha_L^*W) + \frac{\partial^2}{\partial\alpha_L\partial\alpha_L^*}W
\right].
\end{align}
The mapping of Eq.~\eqref{eqs:lindblad} to Eq.~\eqref{eqs:PDE} is exact.
The TWA consists in expanding the Moyal product $\rme^{\vecdouble{\Lambda}/2}$ up to second order and neglecting higher-order quantum fluctuations.
In Eq.~\eqref{eqs:PDE}, this amounts to discarding the third-order derivative terms that stem from the Kerr nonlinearity while keeping lower-order contributions.
This semiclassical approximation is valid in the limit of weak Kerr nonlinearity~\cite{vicentini_critical_2018}.
Within the TWA, the PDE becomes a Fokker-Planck equation~\cite{carmichael_statistical_1999} and $W$ can be interpreted as a well-defined probability distribution of the phase space variables.
Finally, the Fokker-Planck equation can be mapped to a Langevin equation~\cite{carmichael_statistical_1999}, yielding the set of stochastic differential equations 
\begin{align}\label{eqs:stochastic_differential_equations_app}
    \rmi \frac{\partial  \alpha_1}{\partial t} &=   -(\Delta + \rmi\gamma/2) \, \alpha_1 + U\,(|\alpha_1|^2-1)\alpha_1 - J  \alpha_{2} + F + \sqrt{\frac{\gamma}{2}} \xi_1(t)
    \,, \nonumber \\
    \rmi \frac{\partial  \alpha_\ell}{\partial t} &= -\Delta \, \alpha_\ell + U\,(|\alpha_\ell|^2-1)\alpha_\ell   - J (\alpha_{\ell-1} + \alpha_{\ell+1})
    \,,\quad \ell=2,...,L-1\\
    \rmi \frac{\partial  \alpha_L}{\partial t} &=  -(\Delta + \rmi\gamma/2) \, \alpha_L + U\,(|\alpha_L|^2-1)\alpha_L   - J  \alpha_{L-1}  + \sqrt{\frac{\gamma}{2}} \xi_L(t)
    \,,\nonumber
\end{align}
coinciding with Eqs.~\eqref{eqs:stochastic_differential_equations}. 
Expectation values of operators can be mapped from integrals over the Wigner function~\eqref{eqs:expectation_value} into statistical expectation values of the Weyl symbols averaged over many Wigner trajectories solutions of Eqs.~\eqref{eqs:stochastic_differential_equations},
\begin{equation}
    \int\! \rmd\alpha_1\rmd\alpha^*_1 ... \rmd\alpha_L\rmd\alpha^*_L O_W(\alpha_1, \alpha^*_1, ..., \alpha_L,\alpha^*_L)W(t;\alpha_1, \alpha^*_1, ..., \alpha_L,\alpha^*_L) =\frac{1}{N_{\textrm{traj}}}  \! \sum_{j=1}^{N_{\textrm{traj}}} O_W\big{(}\alpha_1^{(j)}(t), \alpha_1^{* (j)}(t), ..., \alpha_L^{(j)}(t), \alpha_L^{* (j)}(t)\big{)}.
\end{equation}
Finally, the Wigner trajectories also give access to the local Wigner function $W_\ell(t; \alpha, \alpha^*)$ that is defined as
\begin{equation}
    W_\ell(t; \alpha_\ell, \alpha^*_\ell) := \int\! \rmd\alpha_1\alpha_1^* ... \rmd\alpha_{\ell-1}\rmd\alpha_{\ell-1}^*\rmd\alpha_{\ell+1}\rmd\alpha_{\ell+1}^* ...\rmd\alpha_L\rmd\alpha^*_L \, W(t;\alpha_1, \alpha^*_1, ..., \alpha_L,\alpha^*_L).
\end{equation}
To construct the steady-state quantity $W_\ell(\alpha, \alpha^*):=\lim\limits_{t\to\infty}W_\ell(t;\alpha, \alpha^*)$, we realize a histogram from the data of $5 \times 10^7$ fields $\alpha_\ell(t)$.
The fields have been obtained from the time evolution of $10^2$ Wigner trajectories after average quantities have reached steady-state values, sampling over a time window $\Delta\tau = 10^4$.
Those numbers are fixed for all the Wigner functions presented in the manuscript.

The numerical results have been obtained by numerically solving Eqs.~\eqref{eqs:stochastic_differential_equations} using the solver SOSRI of the package \textit{Stochastic Differential Equations} that is available in Julia.
The initial conditions $\alpha_\ell(0)$ for each trajectory are complex random numbers sampled from a zero-mean Gaussian distribution with variance $1/2$ for $\ell = 1, ..., L$.
The variance $1/2$ guarantees that the initial conditions correspond to the normalized vacuum state $\ketbra{0}$.

\subsubsection*{Quantum cumulant expansion}\label{sec:QCE}

The many-body dynamics generated by Eq.~\eqref{eqs:lindblad} can in principle be explored by transforming the master equation for the density operator $\hat{\rho}$ into a set of coupled differential equations for the bosonic fields $\langle\hat{a}_\ell\rangle$.
By exploiting the cyclic property of the trace, the Lindblad master equation \eqref{eqs:lindblad} can be exactly mapped to the following set of equations,
\begin{align}\label{eqs:lindblad_cumulants}
    &\rmi \frac{\partial}{\partial t} \langle\hat{a}_1\rangle = -\left(\Delta + \rmi \gamma/2\right)\langle\hat{a}_1\rangle + U\langle\hat{a}_1^{\dagger}\hat{a}_1\hat{a}_1\rangle\ - J\langle\hat{a}_2\rangle +F\,\nonumber,
    \\
    &\rmi \frac{\partial}{\partial t} \langle\hat{a}_\ell\rangle = - \Delta\langle\hat{a}_\ell\rangle + U\langle\hat{a}_\ell^{\dagger}\hat{a}_\ell\hat{a}_\ell\rangle - J \left(\langle\hat{a}_{\ell+1}\rangle + \langle\hat{a}_{\ell-1}\rangle \right)\,,\quad \ell=2,...,L-1\\
    &\rmi \frac{\partial}{\partial t} \langle\hat{a}_L\rangle = -\left(\Delta + \rmi\gamma/2\right)\langle\hat{a}_L\rangle + U\langle\hat{a}_L^{\dagger}\hat{a}_L\hat{a}_L\rangle - J\langle\hat{a}_{L-1}\rangle\,.\nonumber
\end{align}

Here, the issue is that Eqs.~\eqref{eqs:lindblad_cumulants} are not closed, since one needs differential equations for the time evolution of $\langle\hat{a}_\ell^{\dagger}\hat{a}_\ell\hat{a}_\ell\rangle$.
This generates an infinite BBGKY hierarchy of equations involving higher and higher-order correlations.
A possible approach to obtain a closed (and solvable) set of differential equations is to truncate this infinite set by keeping only expectation values with products involving a maximum of $d$ operators.
A systematic way to perform this truncation is the quantum cumulant expansion (QCE), that decomposes the expectation values of products of $D>d$ operators into a sum of expectation values of products of 1, 2, ..., $d$ operators.
The integer $d$ is the quantum cumulant order.
A complete discussion on the truncation rules, with several examples, can be found in Ref.~\cite{kubo_generalized_1962}.

The QCE approach allows exploring quantum correlations beyond the classical limit, but the set of equations for large system sizes becomes numerically intractable as we increase $d$. 
In particular, in the case of Eqs.~\eqref{eqs:lindblad_cumulants}, the number of cumulant equations scales as $L^d$.
In this work, we focus on $d=2$, that captures Gaussian correlations. 
Calculations have been performed with the help of symbolic calculators for QCE equations available in Julia~\cite{plankensteiner_quantumcumulantsjl_2022}.

\subsubsection*{Gross-Pitaevksii classical equations of motion}
\cami{The classical limit of the Lindblad equation Eq.~\eqref{eqs:lindblad} is described in terms of the zero-noise limit of the Langevin equations \eqref{eqs:stochastic_differential_equations_app}. From the TWA, the classical limit is obtained by reducing the Moyal product $\rme^{\vecdouble{\Lambda}/2} $ to the standard (commutative) product.
The equations reduce to the classical Gross-Pitaevskii equations of motion reading~\cite{debnath_nonequilibrium_2017}
}
%\fil{The classical limit of the Lindblad equation Eq.~\eqref{eqs:lindblad}, or, equivalently, of Eqs.~\eqref{eqs:lindblad_cumulants}, can be taken by replacing quantum operators with c-numbers. Many-body correlators factorize in products of complex amplitudes $\alpha_\ell$ and Eqs.~\eqref{eqs:lindblad_cumulants} reduce to the classical Gross-Pitaevskii equations of motion~\cite{debnath_nonequilibrium_2017}
\fil{
\begin{align}\label{eqs:GP_equations}
    &\rmi \frac{\partial\alpha_1}{\partial t} = -\left(\Delta + \rmi \gamma/2\right)\alpha_1 + U|\alpha_1|^2\alpha_1\ - J\alpha_2 +F\,\nonumber,
    \\
    &\rmi \frac{\partial\alpha_\ell}{\partial t} = - \Delta\alpha_\ell + U|\alpha_\ell|^2\alpha_\ell - J \left(\alpha_{\ell+1} + \alpha_{\ell-1} \right)\,,\quad \ell=2,...,L-1 \\
    &\rmi \frac{\partial\alpha_L}{\partial t} = -\left(\Delta + \rmi\gamma/2\right)\alpha_L + U|\alpha_L|^2\alpha_L - J\alpha_{L-1}\,.\nonumber
\end{align}
They describe the time evolution of the field's coherence $\alpha_\ell(t)$, starting initial conditions $\alpha_\ell(0)$ in the complex plane.
%Eqs.~\eqref{eqs:GP_equations} can be seen also as the classical limit of the Langevin equations \eqref{eqs:stochastic_differential_equations_app} once the quantum noise is neglected and bosonic operators on the same site commute.
The numerical solution of Eqs.~\eqref{eqs:GP_equations} is obtained using the package \textit{Differential Equations} that is available in Julia.}

\subsection*{Benchmarking the TWA}

\begin{figure*}[t!]
\includegraphics[width=0.95 \textwidth]{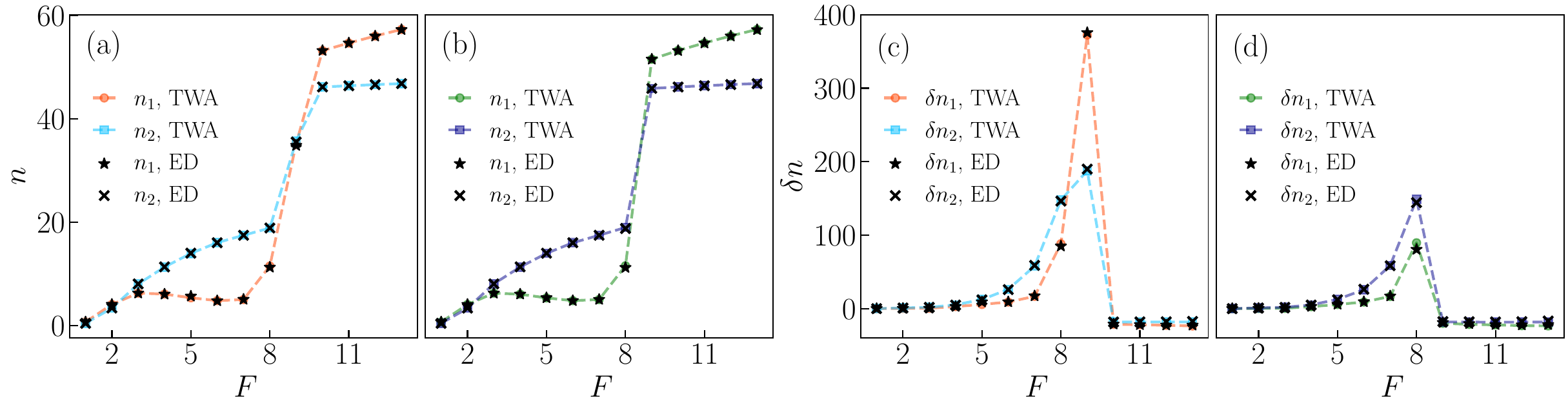}\vspace{0.8em}
\caption{Benchmark of TWA against exact dynamics (ED) for a driven-dissipative Bose-Hubbard chain described in Eq.~\eqref{eqs:hamiltonian} with length $L=2$. 
(a) Comparison between steady-state photon number $n$ obtained with TWA (colored lines) and Monte Carlo quantum trajectories (black markers) obtained by time evolving Eqs.~\eqref{eqs:stochastic_differential_equations} and the stochastic Schr\"odinger equation up to $t = 100$. 
Averages have been performed over the time interval $[50, 100]$ and over $500$ trajectories both for TWA and ED. 
The initial condition for TWA was $\alpha_1(0) = \alpha_2(0) = 0$, while the initial state for ED was $\ket{\psi(0)} = \ket{0}\otimes\ket{0}$. 
(b) Same analysis as in~(a) but considering as initial states the coherent states $\alpha_1(0) = \alpha_1$, $\alpha_2(0) = \alpha_2$ and $\ket{\psi(0)} = \ket{\alpha_1}\otimes\ket{\alpha_1}$, where $\alpha_1 = \langle\hat{a}_1\rangle$ and $\alpha_2 =\langle\hat{a}_2\rangle$ are computed with the TWA in the steady state.
(c) Same analysis as in~(a) but for the photon number statistics $\delta n_\ell$.
(d) Same analysis as in~(b) but for the photon number statistics $\delta n_\ell$.
ED and TWA show an excellent agreement in the vacuum, chaotic and high-photon number phases. The other parameters are set as in Fig.~\ref{fig:phase_diagram}.}
\label{fig:benchmark}
\end{figure*}

We first benchmark the TWA equations~\eqref{eqs:stochastic_differential_equations} against the Lindblad master equation~\eqref{eqs:lindblad}. 
For the parameters considered in the main manuscript, the photon number is around $60$ in the regular coherent regime. 
This implies that a direct solution of the Lindblad equation is difficult even in the smallest chain of $L=2$ resonators, because of the large required cutoff in the Hilbert space. 
We therefore exploit the Monte-Carlo wave function method described in the previous section for the solution of Eq.~\eqref{eqs:lindblad}.

Expectation values of operators are obtained upon averaging over many independent quantum trajectories. In particular, photon number and its fluctuations can be respectively obtained as
\begin{align}\label{eqs:ED_equation}
    n = \frac{1}{N_{\textrm{traj}}}\sum_{j=1}^{N_{\textrm{traj}}} {\bra{\psi_j}\hat{a}^{\dagger}\hat{a}\ket{\psi_j}},\qquad
    \delta n = \frac{1}{N_{\textrm{traj}}} \sum_{j=1}^{N_{\textrm{traj}}}{\bra{\psi_j}\hat{a}^{\dagger 2}\hat{a}^2\ket{\psi_j}} - \frac{1}{N_{\textrm{traj}}^2} \left( \sum_{j=1}^{N_{\textrm{traj}}}{\bra{\psi_j}\hat{a}^{\dagger}\hat{a}\ket{\psi_j}}\right)^2.
\end{align}

In Fig.~\ref{fig:benchmark}, we present the comparison between TWA dynamics and exact dynamics (ED) for a driven-dissipative Bose-Hubbard dimer, \textit{i.e.} $L=2$.
Due to the large size of the Hilbert space, we limit the exact time evolution to $t=100$ and the number of Monte Carlo trajectories to $N_{\rm traj} = 500$.
Furthermore, we average over the time window $[50, 100]$.
The same protocol is applied to the Wigner trajectories.
In Fig.~\ref{fig:benchmark} (a), we plot the photon number $n$ for the modes $\hat{a}_1$ and $\hat{a}_2$ as a function of the drive amplitude $F$. The initial conditions are chosen as $\ket{\psi(0)} = \ket{0}\otimes\ket{0}$ for the ED, and $\alpha_1(0) = \alpha_2(0) = 0$ for the TWA dynamics.
In Fig.~\ref{fig:benchmark} (b), we again plot $n$ for both the modes, but the initial conditions are now $\ket{\psi(0)} = \ket{\alpha_1}\otimes\ket{\alpha_2}$ for the ED and $\alpha_1(0) = \alpha_1$, $\alpha_2(0) = \alpha_2$ for the TWA dynamics, where $\alpha_1$ and $\alpha_2$ are the steady-state values of the complex fields in each resonator computed with the TWA.
In both the panels, TWA and ED show an excellent agreement.
In particular, panel (b) shows that TWA and ED coincides in the NESS.
In Figs.~\ref{fig:benchmark} (c) and (d), we repeat the same analysis of Figs.~\ref{fig:benchmark} (a) and (b) but for $\delta n$ instead of $n$.
Again, TWA and ED show an excellent agreement.

\begin{figure*}[t!]
\centering
\includegraphics[width= 0.65 \textwidth]{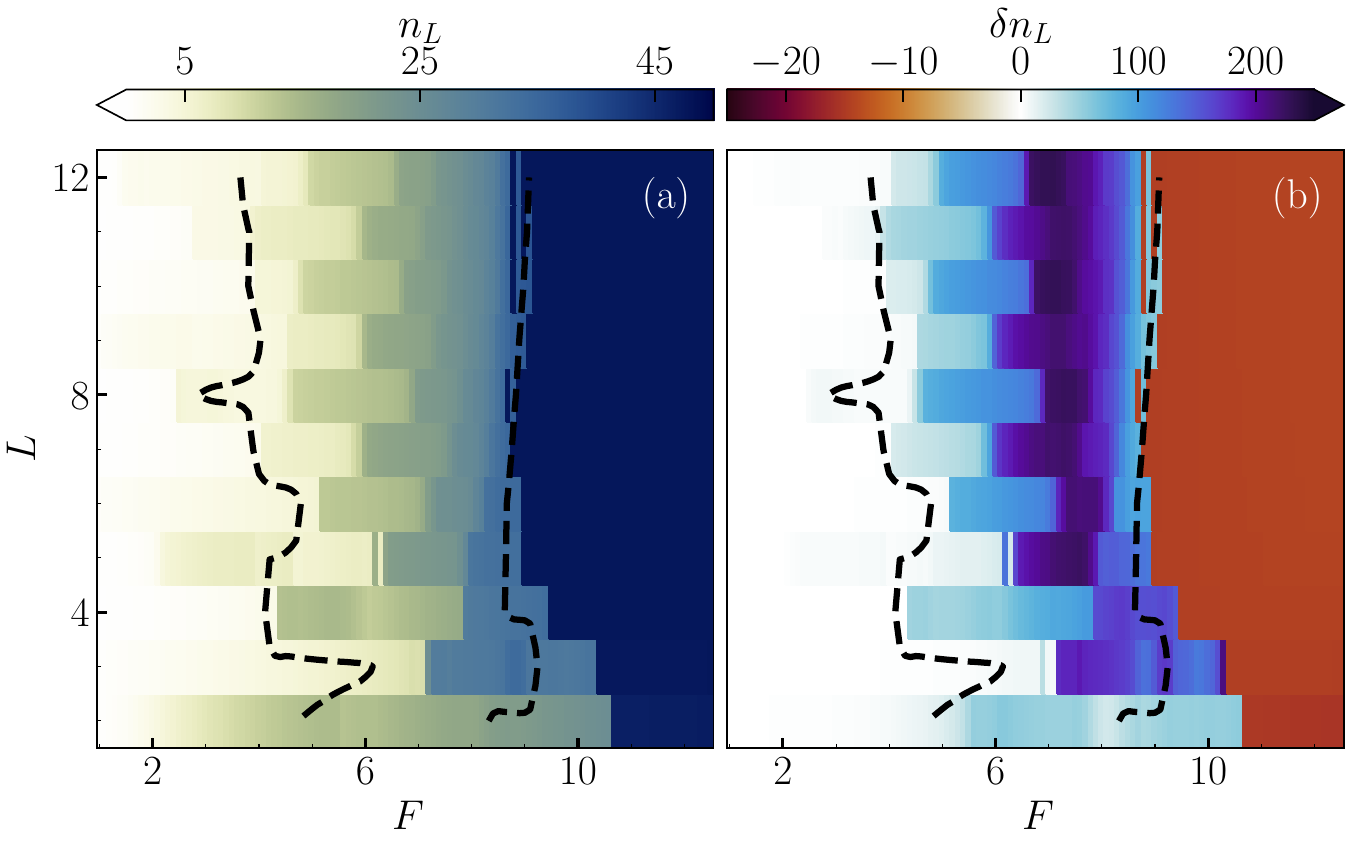}\vspace{0.8em}
\caption{Comparison between TWA and the quantum cumulant expansion. 
(a) Photon number and (b) photon number statistics in the last resonator computed with the quantum cumulant expansion at second order as a function of the drive strength $F$ and the chain's length $L$.
Each point has been obtained upon averaging over a time window equal to $\Delta\tau=10^3$ in the steady state.
The black dashed line indicates the boundaries of the chaotic region identified by the TWA.
The other parameters are set as in Fig.~\ref{fig:phase_diagram}.}
\label{fig:QCE}
\end{figure*}

We now compare TWA and the quantum cumulant expansion. In particular, we compare $n$ and $\delta n$ for the $L$-th resonator of the driven-dissipative Bose-Hubbard chain. We work with a second-order QCE approximation for which the expression of the photon number simply reads $n_{\ell}=\langle\hat{a}_\ell^\dagger\hat{a}_\ell\rangle$, while $\delta n$ is now expressed as
\begin{align}\label{eqs:methods}
    \delta n_{\ell} = \langle\hat{a}_\ell^{\dagger 2}\rangle\langle\hat{a}_\ell^2\rangle + \langle\hat{a}_\ell^{\dagger}\hat{a}_\ell\rangle^2 - 2|\langle\hat{a}_\ell\rangle|^4\,.
\end{align}

In Fig.~\ref{fig:QCE} we present the comparison between the TWA and the QCE. Figs.~\ref{fig:QCE} (a) and (b) show $n_L$ and $\delta n_L$ computed solving cumulant equations truncated at order two. 
To compute the steady-state photon number and its fluctuations at each point in the phase diagram, we let the QCE dynamics evolve to the steady state, and we subsequently averaged the dynamics over a time window equal to $\Delta\tau=10^3$.
The black line indicates the boundaries of the chaotic region computed from TWA equations.
Data show a good qualitative agreement between TWA and QCE.
However, as it is already evident from a rapid comparison between Figs.~\ref{fig:phase_diagram} (b) and (c) and Fig.~\ref{fig:QCE}, the two methods do not show a quantitative agreement.

\subsection*{Additional phase diagrams}\label{sec:add_pd}

\begin{figure*}[t!]
\centering
\includegraphics[width=0.85 \textwidth]{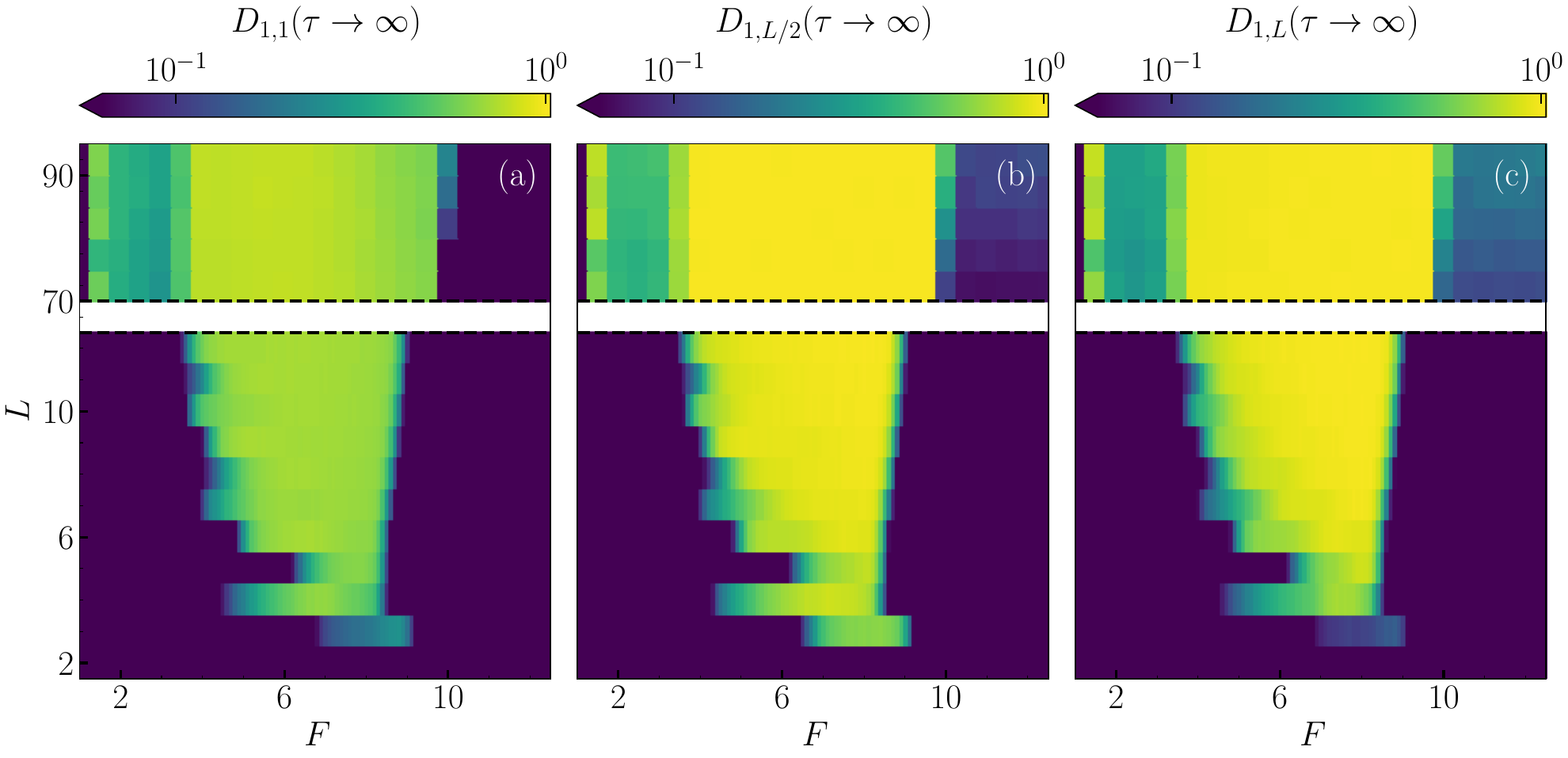}
\caption{
Saturation value of the steady-state phase OTOC, $D_{1, \ell}(\tau\to\infty)$ computed at different representative sites $\ell$ across the chain, as a function of the drive amplitude $F$ and the chain's length $L$: (a) $\ell=1$, (b) $\ell=L/2$ , and (c) $\ell=L$ [same data as in Fig.~\ref{fig:phase_diagram} (c)].
The other parameters are set as in Fig.~\ref{fig:phase_diagram}.
}
\label{fig:additional_phase_diagram}
\end{figure*}

In this section, we provide additional plots to support the claim that the presence of semiclassical chaos is spread over the whole chain, and not only at the right boundary $\ell = L$ which is used in the main manuscript as a representative site.
In Fig.~\ref{fig:additional_phase_diagram}, we plot the saturation value of the phase OTOC $D_{1, \ell}(\tau\to\infty)$ as a function of the drive amplitude $F$ and the chain's length $L$, when evaluated at three representative sites across the chain:
 first $\ell=1$, middle $\ell=L/2$, and last site $\ell=L$ (that was already presented in the main text).

When $\ell=1$, the saturation value never reaches 1, signaling that the first site does not reach maximal decorrelation. 
This is in agreement with the OTOC's dynamics studied in Fig.~\ref{fig:OTOC} (a) and (b) for a specific value of $F$ and $L$.
The phase diagram for $\ell=L/2$ is instead essentially identical to the one realized for $\ell=L$. 
This analysis shows that the prethermal and thermal domains are equally chaotic.

\subsection*{Comparison with the uniformly driven-dissipative 1D chain}\label{sec:comparison}

\begin{figure*}[t!]
\centering
\includegraphics[width=0.95 \textwidth]{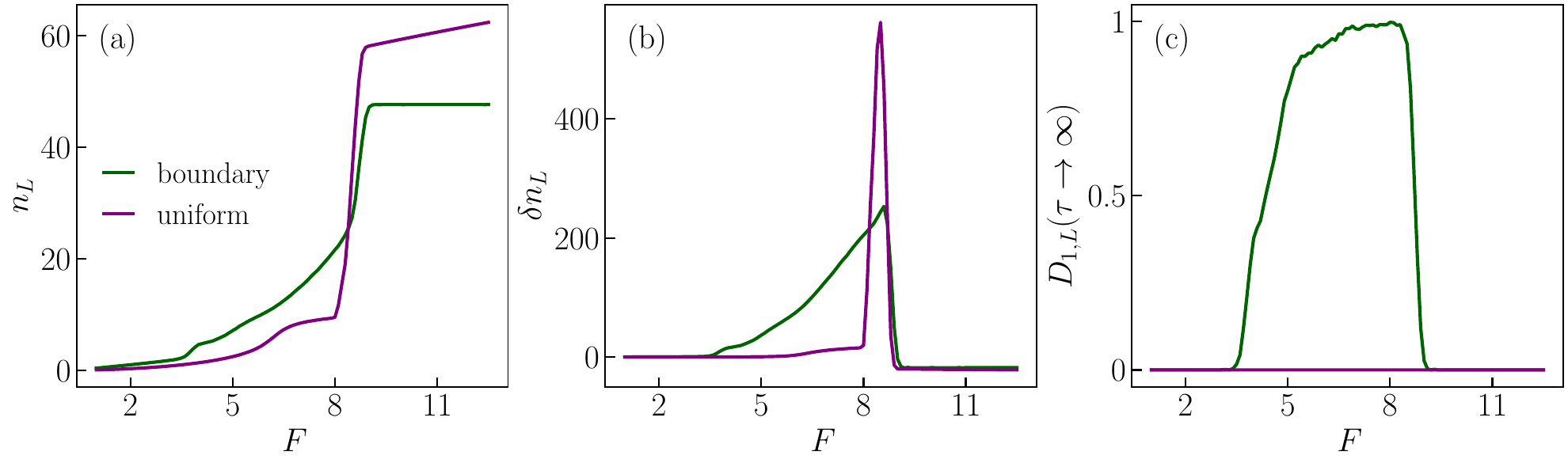}
\caption{
Comparison between the lattices with boundary and uniform drive and dissipation for $L=10$.
(a) Photon number in the last site as a function of $F$,
(b) photon number statistics in the last site as a function of $F$,
(c) saturation value of the steady-state OTOC in the last site as a function of $F$.
The results are obtained upon averaging over $10^3$ independent Wigner trajectories. The other parameters are set as in Fig.~\ref{fig:phase_diagram}.
}
\label{fig:boundary_vs_uniform}
\end{figure*}

Here, we compare the boundary-driven, boundary-dissipative Bose-Hubbard lattice~\eqref{eqs:lindblad}, for which we have found a wide chaotic region in parameter space, with the uniformly-driven uniformly-dissipative Bose-Hubbard chain. The Hamiltonian of the latter model is given by
\begin{equation}\label{eqs:hamiltonian_add}
    \hat{H} = \sum_{\ell=1}^L\left[-\Delta \hat{a}_\ell^{\dagger}\hat{a}_\ell + \frac{1}{2}U\hat{a}_\ell^{\dagger}\hat{a}_\ell^{\dagger}\hat{a}_\ell\hat{a}_\ell + F(\hat{a}_\ell^{\dagger} + \hat{a}_\ell)\right] - J\sum_{\ell=1}^{L-1}\left(\hat{a}_{\ell+1}^{\dagger}\hat{a}_\ell + \hat{a}_\ell^{\dagger}\hat{a}_{\ell+1}\right),
\end{equation}
while the Lindblad master equation reads
\begin{equation}\label{eqs:lindblad_add}
\frac{\partial\hat{\rho}}{\partial t} = -\rmi [\hat{H}, \hat{\rho}] + \sum_{\ell=1}^L\gamma\left(\hat{a}_\ell\hat{\rho}\hat{a}_\ell^{\dagger} - \frac{1}{2}\acomm{\hat{a}_\ell^{\dagger}\hat{a}_\ell}{\hat{\rho}}\right).
\end{equation}

In Fig.~\ref{fig:boundary_vs_uniform}, we compare the boundary-driven boundary-dissipative chain studied in this paper and the uniformly-driven uniformly-dissipative chain, studied, \textit{e.g.}, in Ref.~\cite{vicentini_critical_2018}, showing that semiclassical chaos is a feature peculiar of boundary drive and dissipation mechanisms. 
We focus on the last cavities of the two chains in the steady-state regime. 
In Fig.~\ref{fig:boundary_vs_uniform} (a), we show that the photon number $n_L$ exhibits a similar behavior with an abrupt jump around the same critical drive amplitude $F$. 
In Fig.~\ref{fig:boundary_vs_uniform} (b), we present the photon number statistics $\delta n_L$. 
In the case of the uniformly-driven uniformly-dissipative chain, large fluctuations are concentrated at the value of $F$ where the jump occurs, as expected.
In contrast, in the case of boundary drive and dissipation, large fluctuations are spread over a much larger region. 
In Fig.~\ref{fig:boundary_vs_uniform} (c), we plot the saturation value of the steady-state semiclassical phase  OTOC defined in Eq.~\eqref{eqs:semiclassical_OTOC}, showing final evidence that only the boundary-driven, boundary-dissipative lattice hosts semiclassical chaos, while the dynamics in the uniformly-driven dissipative chain is regular.

\subsection*{Classical thermodynamics within the chaotic chain: application of the equipartition theorem}\label{sec:impurity_model}

\begin{figure*}[t!]
\includegraphics[width=1 \textwidth]{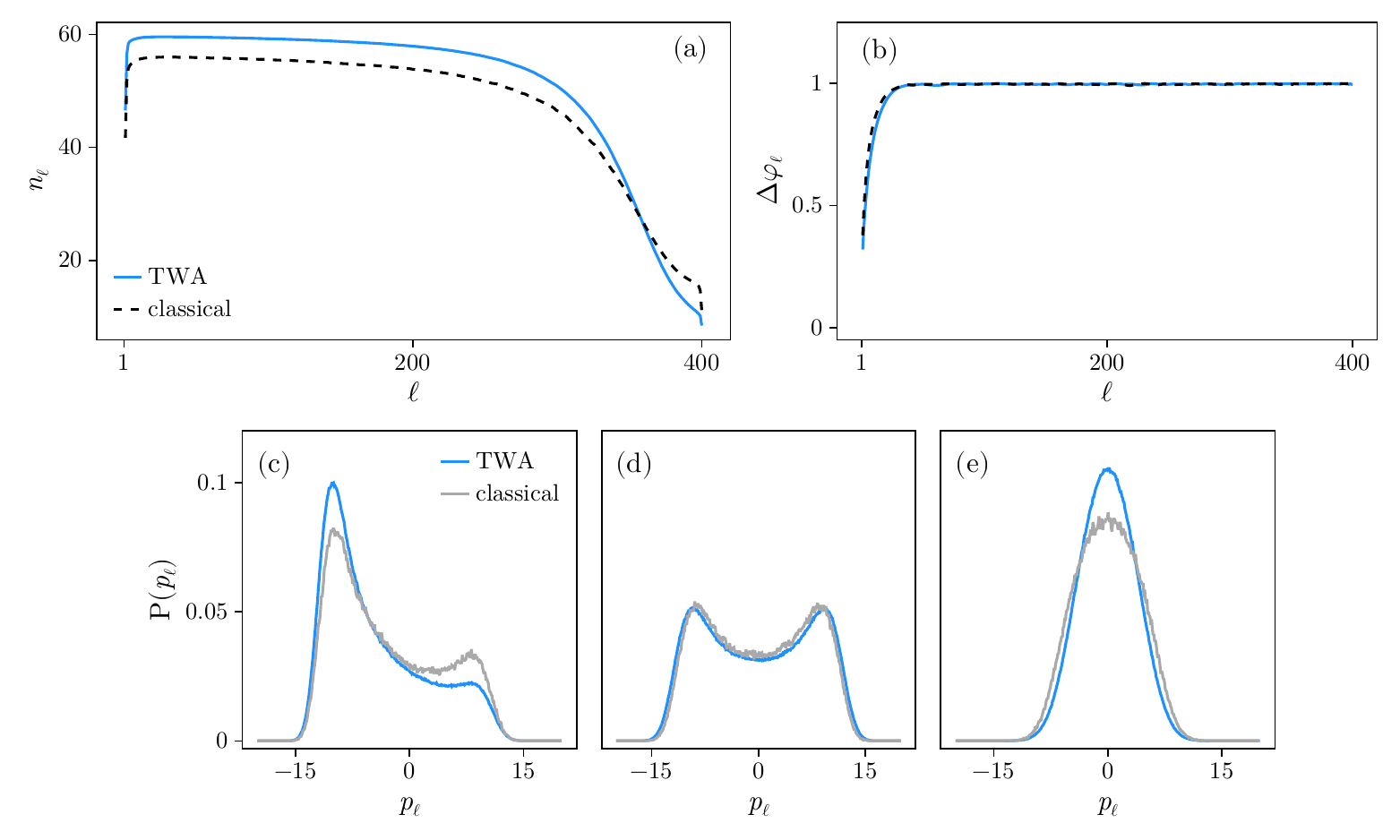}\vspace{0.8em}
\caption{\fil{Comparison between the TWA equations \eqref{eqs:stochastic_differential_equations_app} and the classical equations of motion \eqref{eqs:GP_equations} in the chaotic regime. (a) TWA (blue line) and classical (black-dashed line) steady-state photon number $n_\ell$ as a function of $\ell$ for a $L = 400$ chain. (b) Same as in panel (a) but for the circular variance $\Delta\varphi_\ell$. (c-e) Local momentum distribution $\textrm{P}(p_\ell)$ computed from the TWA (blue lines) and the mean-field (gray line) equations. The site indices are (c) $\ell = 5$, (d) $\ell = 200$, (e) $\ell = 390$. The dashed lines curves in panel (e) are the Maxwell-Boltzmann distributions associated with the TWA temperature (black-dashed line) and the classical temperature (red-dashed line). 
Results are computed by averaging over $N_{\rm traj} = 10^2$ independent Wigner trajectories and over a time window $\Delta\tau=10^4$ once the steady state is reached.
The drive amplitude is fixed to $F = 7.5$. 
Results are computed by averaging over $N_{\rm traj} = 10^2$ independent Wigner trajectories and over a time window $\Delta\tau=10^4$ once the steady state is reached.
The other parameters are set as in Fig.~\ref{fig:phase_diagram}.}
}
\label{fig:TWA_vs_GP}
\end{figure*}

\begin{figure*}[t!]
\includegraphics[width=1 \textwidth]{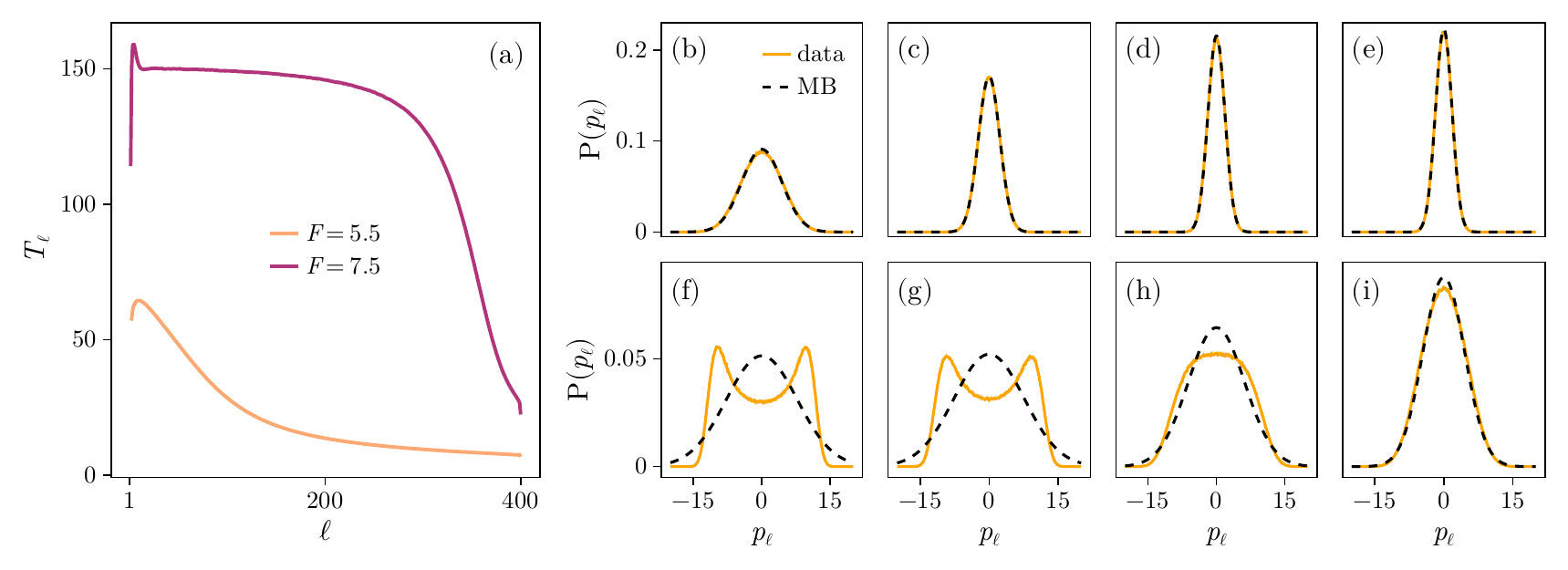}\vspace{0.8em}
\caption{\fil{Application of the equipartition theorem in the chaotic regime. (a) Temperature profile $T_\ell$ as a function of $\ell$ in a $L=400$ chain computed according to the equipartition theorem $T_\ell = |\Delta| \langle p_\ell^2\rangle$ in at low drive amplitude $F=5.5$ (pink line) and intermediate drive amplitude $F=7.5$ (purple line). (b-e) Momentum distributions $\textrm{P}(p_\ell)$ as a function of $p_\ell$ (orange lines) compared to the thermal Maxwell-Boltzmann distribution in Eq.~\eqref{eqs:equipartition} with $T_\ell = |\Delta| \langle p_\ell^2\rangle$ at low drive amplitude $F=5.5$ at the sites (b) $\ell=50$, (c) $\ell=200$, (d) $\ell=340$, and (e) $\ell=370$. (f-i) Same as in panels (b-e) but for intermediate drive amplitude $F=7.5$. The other parameters are set as in Fig.~\ref{fig:phase_diagram}.}
}
\label{fig:equipartition}
\end{figure*}

\fil{The goal of this section is to discuss the similarities and differences between the TWA and classical Gross-Pitaevskii approaches in the chaotic regime, and the applicability of the equipartition theorem, a result derived from classical thermodynamics and largely applied in Ref.~\cite{prem_dynamics_2023}.
The equipartition theorem states that at thermal equilibrium, each quadratic degree of freedom contributes $T/2$ ($k_B=1$) to the average energy of the system.
The equipartition theorem holds if: (i) quantum effects are negligible and the system is well described by classical mechanics, (ii) the energy is quadratic in the generalized coordinates like momentum and position, (iii) the system reached thermal equilibrium.
For the Bose-Hubbard model, equipartition theorem does not apply in principle, even in semiclassical regimes, because of the presence of the Kerr term.
In very diluite limits, however, nonlinearities play a negligible role and only quadratic terms contribute to the average energy.
In Eq.~\eqref{eqs:hamiltonian}, $\Delta$ fixes the energy scale of the quadratic Hamiltonian, and assuming that the Kerr term is negligible, one obtains that the distribution of momenta $\hat p :=i(\hat{a}^\dagger - \hat{a})/\sqrt{2}$ follows the thermal Maxwell-Boltzmann distribution
\begin{equation}\label{eqs:equipartition}
    \textrm{P}_{\rm MB}(p) = \sqrt{ \frac{|\Delta|}{2\pi T}}e^{-|\Delta| p^2/2T}.
\end{equation}
The temperature can be estimated from the second moment of the distribution as $T = |\Delta| \langle p^2\rangle$.
Within the TWA framework, $\langle p\rangle$ and $\langle p^2\rangle$ are given by $\sqrt{2}\langle\textrm{Im}(\alpha)\rangle$ and $2\langle\textrm{Im}^2(\alpha)\rangle$, respectively.}

\fil{We start by studying the comparison between the TWA and Gross-Pitaevskii solutions for the chaotic regime.
Results are collected in Fig.~\ref{fig:TWA_vs_GP} for a $L=400$ chain at intermediate drive amplitude, $F=7.5$.
In Figs.~\ref{fig:TWA_vs_GP} (a) and (b) we present the photon number $n_\ell$ and the circular variance $\Delta\varphi_\ell$ as a function of $\ell$.
Both quantities show the qualitative agreement between the two approaches: the circular variance rapidly saturates to one and the photon number remains flat in the bulk of the chain.
The three distinct domains, nonthermal, prethermal and thermal are captured by both the approaches, as indicated by the momentum distributions in Figs.~\ref{fig:TWA_vs_GP} (c-e).
This indicates that in the chaotic regime, quantum fluctuations have a marginal role and the relevant physical phenomena can be attributed to classical nonlinear fluctuations.}

\fil{We then apply the equipartition theorem.
In Fig.~\ref{fig:equipartition} (a) we show the temperature estimated with the equipartition theorem in the chaotic regime for $F=5.5$ (pink line) and $F=7.5$ (purple line) in a $L=400$ chain. 
At $F=5.5$, $T_\ell$ decreases monotonically throughout the chain, in agreement with the behavior of the entropy density in Fig.~\ref{fig:pre_thermalization} (c). 
Moreover, the momentum distributions $\textrm{P}(p_\ell)$ reported in Figs.~\ref{fig:equipartition} (b-e) for $\ell=50$, $\ell=200$, $\ell=340$ and $\ell=370$ show a Gaussian profile, in agreement with the Maxwell-Boltzmann distribution in Eq.~\eqref{eqs:equipartition}.
At low drive amplitude, the effect of the nonlinear term reduces to breaking the model's integrability, without shaping the local states.
At $F=7.5$, $T_\ell$ saturates in the bulk of the chain, and decreases at the end of the chain, in correspondence of the thermal domain.
This behavior is in sharp contrast with the temperature profile estimated with the Gibbs state, reported in Fig.~\ref{fig:pre_thermalization} (a), and with the non-monotonic behavior of the steady-state entropy in Fig.~\ref{fig:pre_thermalization} (c).
As expected, the momentum distributions $\textrm{P}(p_\ell)$ reported in Figs.~\ref{fig:equipartition} (f-i) for $\ell=50$, $\ell=200$, $\ell=340$ and $\ell=370$ show a strong non-Gaussian behavior in the bulk of the chain.
At intermediate drive amplitudes, nonlinear fluctuations not only induce chaotic behavior, but also shape significantly the local states throughout the chain.
The equipartition theorem breaks down because of the non negligible Kerr nonlinearity, activated by the stronger driving field.
These nonlinear contributions are instead captured by the Gibbs ansatz, which unveils the anomalous heating in the prethermal domain.
In the thermal domain ($\ell=370$), the shape of $\textrm{P}(p_\ell)$ has a bell shape and its profile is reproduced by a thermal Maxwell-Boltzmann distribution, with a small deviation due again to the Kerr nonlinearity.}

\subsection*{Modeling the local state of the chaotic Bose-Hubbard chain with $\mathbb{U}(1)$-symmetric impurity models}\label{sec:impurity_model}

\subsubsection*{Models and effective temperature}

We consider two single-site driven-dissipative impurity models that describe $\mathbb{U}(1)$-symmetric quantum states exhibiting, across their parameter space, thermal features and population inversion.
Population inversion is achieved when higher energy states are more populated than lower energy states, in contrast to a thermal population, and requires an external energy source~\cite{yariv_quantum_1989, yamamoto_mesoscopic_1999, milonni_laser_2010}.
Population inversion is an essential ingredient to obtain lasing states, where stimulated emission dominates over absorption~\cite{maiman_stimulated_1960}.
The Hamiltonian of both the models is taken as the local Hamiltonian of the Bose-Hubbard chain in Eq.~\eqref{eqs:hamiltonian}, $\hat{H}_{\rm imp} = \omega_0\hat{a}^{\dagger}\hat{a} + U\hat{a}^{\dagger 2}\hat{a}^2/2$.
The first impurity model, which we refer as the 2-photon decay model, is defined by the set of Lindblad dissipators described in the main text,
\begin{equation}\label{eqs:impurity_1}
    \hat{L}_\uparrow = \sqrt{\gamma^\uparrow} \hat a^\dagger,\qquad 
    \hat{L}_\downarrow = \sqrt{\gamma^\downarrow}\hat a,\qquad 
    \hat{L}_\phi = \sqrt{\gamma^\phi}\hat{a}\hat a^{\dagger} ,\qquad 
    \hat{L}_{\rm s} = \sqrt{\gamma^{\rm s}} \hat a^2\,.
\end{equation}
The non-negatives parameters  $\gamma^\uparrow$, $\gamma^\downarrow$, $\gamma^\phi$, and $\gamma^{\rm s}$ are effective rates of incoherent pumping, decay, dephasing, and 2-photon decay.
The second impurity model is the generalized Scully-Lamb model, introduced in Ref.~\cite{minganti_liouvillian_2021} and characterized by the Lindblad dissipators 
\begin{equation}\label{eqs:impurity_2}
    \hat{L}_\uparrow = \hat{a}^{\dagger}(\gamma^{\uparrow} - \mathcal{S}\hat{a}\hat{a}^{\dagger})/\sqrt{\gamma^{\uparrow}},\qquad \hat{L}_\phi = \sqrt{\gamma^{\phi}}\hat{a}\hat{a}^{\dagger} = \sqrt{\eta + \frac{3\mathcal{S}}{2}}\hat{a}\hat a^{\dagger},\qquad \hat{L}_\downarrow = \sqrt{\gamma^{\downarrow}}\hat{a}.
\end{equation}
The non-negatives parameters $\gamma^\uparrow$, $\gamma^\downarrow$, $\gamma^\phi$, and $\mathcal{S}$ are effective rates of incoherent pumping, decay, dephasing, and saturation.
If $\gamma^{\phi}=3\mathcal{S}/2$ the above model reduces to the standard Scully-Lamb model of Ref.~\cite{scully_quantum_1967}.
The Scully-Lamb model is valid if~\cite{minganti_continuous_2021}
\begin{equation}\label{eqs:SC_conditions}
    \gamma^{\uparrow} \sim \mathcal{O}(\gamma^{\downarrow}),\qquad \mathcal{S}\langle\hat{a}\hat{a}^{\dagger}\rangle\ll\gamma^{\uparrow},
\end{equation}
while for the first model one requires $\gamma^{\rm s}\langle\hat{a}\hat{a}^{\dagger}\rangle\ll\gamma^{\uparrow}$.
The choice of $\hat{H}_{\rm imp}$ and $\gamma^{\phi}$ does not affect the steady-state properties, including the steady-state Wigner function $W(\alpha, \alpha^*)$.
In the steady state, they have an effect only on time-correlation functions.
In particular, by imposing $\gamma^{\phi}=3\mathcal{S}/2$ one finds oscillating time-correlation functions with a usually very small decay rate (that is $3\mathcal{S}/2$ which is by assumption much smaller than $2\gamma^{\uparrow}/\langle\hat{a}\hat{a}^{\dagger}\rangle$).
If instead $\gamma^\phi \gg \mathcal{S}$ one obtains the generalized version of the Scully-Lamb model exhibiting rapidly decaying time-correlation functions in the lasing phase. A similar phenomenology occurs in the 2-photon decay impurity model, where the phase coherence is suppressed if $\gamma^\phi \gg \gamma^{\rm s}$.

While the two impurity models above exhibits very similar steady-state properties, a finite effective temperature $T$ can, in principle, only be straightfowardly associated with the 2-photon decay  model. 
Indeed, the set of Lindblad operators in Eq.~(\ref{eqs:impurity_1}) can be easily interpreted as coupling to a set of thermal reservoirs. $\hat L_{\rm s}$ corresponds to a zero-temperature 2-photon reservoir. $\hat L_\uparrow$ and $\hat L_\downarrow$ correspond to  finite-temperature 1-photon reservoir at a temperature set by the detailed balance
\fil{
\begin{equation}\label{eqs:detailed_balance}
    \frac{\gamma^\uparrow}{\gamma^\downarrow} =: \rme^{- (\omega_0-\mu)/T} =: \rme^{\mu/T}, \quad \textit{i.e.} \quad \mu/T := \log \left({\gamma_\uparrow}\big{/}{\gamma_\downarrow} \right),
\end{equation}
}
where we neglected $U \ll \omega_0$ since, in typical circuit QED setups, the resonator frequency is of the order of GHz while the Kerr nonlinearity ranges from kHz to MHz~\cite{blais_circuit_2021}. The dephasing operator $\hat L_\phi$ is self-adjoint and corresponds to infinite temperature.

In contrast, inspecting of the set of Lindblad operators of the generalized Scully-Lamb model given in Eq.~(\ref{eqs:impurity_2}), one sees that $\hat{L}_1$ is not the adjoint  of $\hat L_\downarrow$ because of the presence of the nonlinear saturation $\mathcal{S}$. Thus, the balance between $\hat{L}_1$ and $\hat L_\downarrow$ cannot be used to define an effective temperature.

\subsubsection*{Fitting procedure for the Wigner function}

\begin{figure*}[t!]
\includegraphics[width=1 \textwidth]{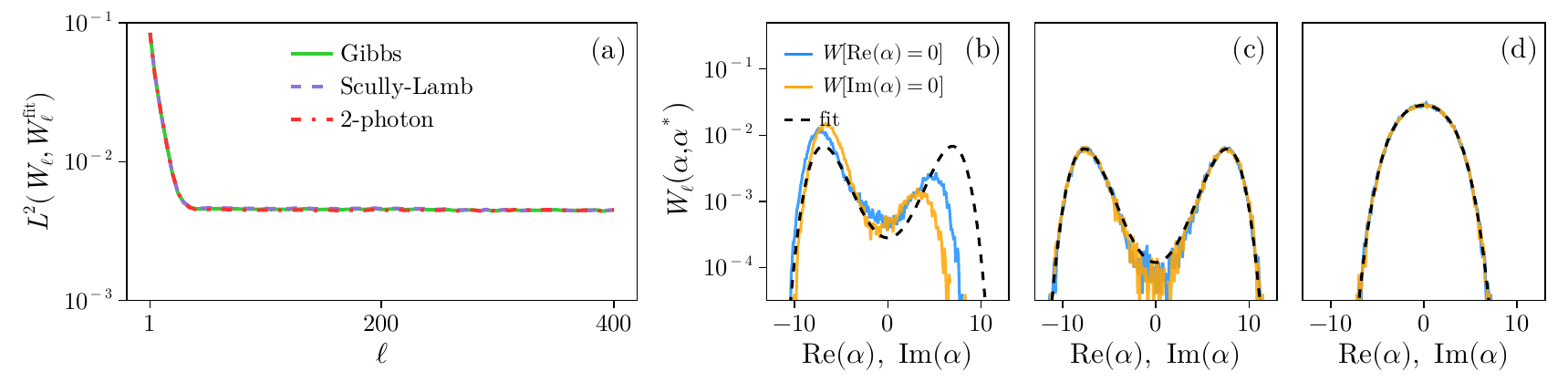}\vspace{0.8em}
\caption{Fit of the $\ell$-th site Wigner function of the driven-dissipative bosonic chain described by Eqs.~\eqref{eqs:hamiltonian} and~\eqref{eqs:lindblad} for a chain with length $L=400$.
In panel $(a)$, we show the $L^2$ norm in Eq.~\eqref{eqs:L2_norm} between the $L$-th site Wigner function and the fitted Wigner function obtained from the three ans\"atze considered in the article: the Gibbs state in Eqs.~\eqref{eq:Gibbs} and \eqref{eqs:SC_hamiltonian} (green line), the Scully-Lamb model in Eq.~\eqref{eqs:impurity_2} (purple-dashed line), and the 2-photon driven-dissipative impurity model in Eq.~\eqref{eqs:impurity_1} (red-dashed line).
Panels (b-e) shows the comparison between two cuts of $W(\alpha, \alpha^*)$ [the one at $\textrm{Re}(\alpha)=0$ (blue curve) and the one at the one at $\textrm{Im}(\alpha)=0$ (orange curve)] and the fitted Wigner function (black dashed line). 
The values of $\ell$ are~(b) $\ell=1$,~(c) $\ell=200$,~(d) $\ell=400$.  
}
\label{fig:wigner_fit2}
\end{figure*}

Given the Wigner function $W_\ell(\alpha, \alpha^*)$ of the $\ell$-th site of the bosonic chain in Eqs.~\eqref{eqs:hamiltonian} and~\eqref{eqs:lindblad} we perform a 2$D$ fit using a Levenberg-Marquardt fitting algorithm. 
The fitting function is the NESS of the local impurity model described by Eqs.~\eqref{eqs:impurity_1} or~\eqref{eqs:impurity_2}.
The fitting parameters are the thermal gain $\gamma^{\uparrow}$, the effective dissipation rate $\gamma^{\downarrow}$ and the nonlinear parameter, either the 2-photon decay rate $\gamma^{\rm s}$ or the saturation $\mathcal{S}$.
Looking at the conditions in Eq.~\eqref{eqs:SC_conditions}, it is important to choose reasonable initial fitting parameters.
We quantify the accuracy of the single-site impurity ans\"atze by computing the $L^2$ norm between the system's Wigner function and the fitted Wigner function, specifically
\begin{equation}\label{eqs:L2_norm}
    L^2(W_\ell, W^{\textrm{fit}}_\ell) = \left[\int_\mathcal{A}\rmd\alpha \rmd\alpha^* \, |W_\ell(\alpha, \alpha^*) - W^{\textrm{fit}}_\ell(\alpha, \alpha^*)|^2\right]^{1/2}.
\end{equation}

In Fig.~\ref{fig:wigner_fit2}, we show the results of the fitting procedure corresponding to Fig.~\ref{fig:pre_thermalization}.
In Fig.~\ref{fig:wigner_fit2} (a), we plot the $L^2$ norm as a function of the index site $\ell$ for the chain with length $L=400$, using both the 2-photon decay and the generalized Scully-Lamb model as single-site ans\"atze.
The results are compared with the $L^2$ norm corresponding to the fit of the thermal ansatz discussed in the following section.
When $\ell\le 50$ the fitting procedure does not give reliable results. In the bulk and in the right tail of the chain, instead, the local Wigner function is completely captured by both the single-site impurity models.
In Figs.~\ref{fig:wigner_fit2} (b-d) we show the comparison between $W_\ell(\alpha, \alpha^*)$ and $W^{\rm fit}_\ell(\alpha, \alpha^*)$.
For $\ell=1$ [c.f. Fig.~\ref{fig:wigner_fit2}(b)], the local Wigner function $W_\ell(\alpha, \alpha^*)$ is not captured by the single-site ansatz.
We conclude that within the left tail of the chain, in proximity of the coherent drive, different cavities remain correlated and thus a single-site description is not sufficient.
For $\ell = 200, 400$ [see Fig.~\ref{fig:wigner_fit2} (c) and~(d)] we see how $W_\ell(\alpha, \alpha^*)$ matches with $W^{\rm fit}_\ell(\alpha, \alpha^*)$.

\subsubsection*{Classical auto-correlation function}

\begin{figure*}[t!]
\includegraphics[width=0.85 \textwidth]{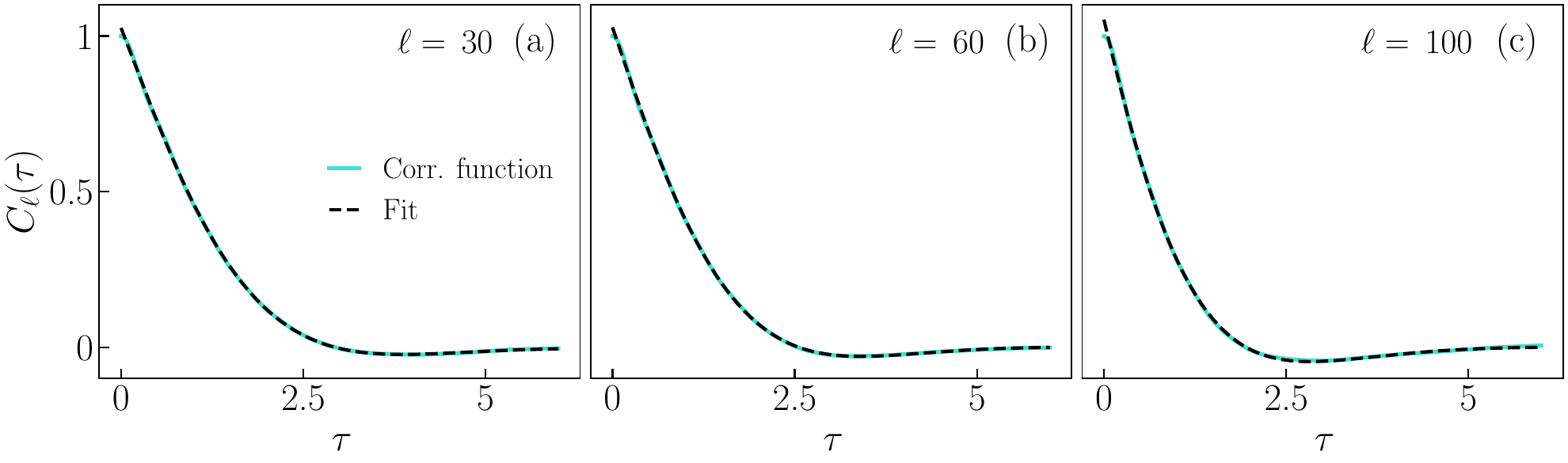}\vspace{0.8em}
\caption{
Classical auto-correlation function in the prethermal phase of a chaotic Bose-Hubbard chain with length $L=200$.
We plot $C_\ell(\tau)$ computed according to Eq.~\eqref{eqs:autocorrelation} for (a) $\ell=30$, (b) $\ell=60$, and (c) $\ell=100$. Results have been obtained from a single long Wigner trajectory in the steady state.
The black dashed line indicates the fit performed according to Eq.~\eqref{eqs:autocorrelation_ansatz}.
The drive strength is fixed to $F=7.5$ and the other parameters are set as in Fig.~\ref{fig:phase_diagram}.
}
\label{fig:autocorrelation}
\end{figure*}

Here, we show that, in the prethermal phase of the chaotic Bose-Hubbard chain, the dephasing rate $\gamma^{\phi}$ is much larger than the 2-photon decay rate $\gamma^{\rm s}$ or the saturation $\mathcal{S}$, hence establishing the absence of long-lived phase coherence typical of lasing states.
As discussed above, $\gamma^\phi$ cannot be extracted from the local steady-state Wigner function $W_\ell(\alpha, \alpha^*)$. 
Here, we extract a rough estimate of $\gamma^\phi$ from two-time correlation functions in the prethermal domain in an $L=200$ chaotic chain by fitting them to a simple ansatz.

We consider the classical steady-state auto-correlation function defined as
\begin{equation}\label{eqs:autocorrelation}
    C_\ell(\tau) := \textrm{Re} \lim_{t, T \to +\infty}\frac{1}{T}\int_{t}^{t+T}\rmd t'\,\,\frac{\langle\hat{a}_\ell^{\dagger}(t'+\tau)\rangle\langle\hat{a}_\ell(t')\rangle}{|\langle\hat{a}_\ell(t')\rangle|^2},
\end{equation}
that can be computed over a single long Wigner trajectory in the steady state, due to the ergodic nature of steady-state trajectories~\cite{beaulieu_observation_2025}.

For the Scully-Lamb model, \textit{i.e.} $\gamma^{\phi} = 3\mathcal{S}/2$, time-correlation functions exhibit a slow oscillatory decay controlled by the rate $\mathcal{S}$.
For the generalized Scully-Lamb model, the much faster decay of time-correlation functions is controlled by $\gamma^\phi \gg \mathcal{S}$~\cite{minganti_liouvillian_2021}.
A similar phenomenology occurs in the 2-photon decay impurity model.
In Fig.~\ref{fig:autocorrelation}, we plot $C_\ell(\tau)$ for a chain of length $L=200$ with a drive amplitude $F=7.5$ and at three sites within the prethermal domain: (a) $\ell=30$, (b) $\ell=60$, and (c) $\ell=100$.
The auto-correlation function exhibits a rapid decay towards zero, with an additional weak oscillation. 
To quantify the decay rate, we fit the initial decay of $C_\ell(\tau)$ with 
\begin{equation}\label{eqs:autocorrelation_ansatz}
    C_{\ell}^{\rm fit}(\tau) = \sum_{j=1}^2A_{\ell, j} \rme^{-\Gamma_{\ell, j}\tau}\cos(\Omega_{\ell, j}\tau).
\end{equation}
Here, $A_{\ell, j}$ are real-valued amplitudes, $\Gamma_{\ell, j} \geq 0$ are decaying rates with magnitudes on the order of $\gamma^\phi$~\cite{minganti_liouvillian_2021}, while the frequencies $\Omega_{\ell, j}$ capture possible oscillations in the early times.
This simple ansatz successfully captures the auto-correlation function in the prethermal phase, as shown in Fig.~\ref{fig:autocorrelation} (black dashed lines).
In the following table, we compare the resulting decay rates $\Gamma_1$ and $\Gamma_2$ with the saturation parameters obtained from the fit of $W_\ell(\alpha, \alpha^*)$.
\begin{center}
\begin{tabular}{|c|c|c|c|c|}
    \hline
    & $\Gamma_1$ & $\Gamma_2$ & $\gamma^{\rm s}$ & $\mathcal{S}$\\
    \hline
    $\ell=30$ & \,\,\, 0.676 \,\,\, & \,\,\, 0.358 \,\,\, & \,\,\, 0.0012 \,\,\, & \,\,\, 0.0011 \,\,\,\\
    \hline
    $\ell=60$ & \,\,\, 0.670 \,\,\, & \,\,\, 0.678 \,\,\, & \,\,\, 0.0010 \,\,\, & \,\,\, 0.0009  \,\,\,\\
    \hline
    $\ell=100$ & \,\,\, 0.954 \,\,\, & \,\,\, 0.954 \,\,\, & \,\,\, 0.0007 \,\,\, & \,\,\, 0.0007 \,\,\,\\
    \hline
\end{tabular}
\end{center}
In the three cases, the exponential decay rate of $C_\ell(\tau)$ is much larger than  $\gamma^{\rm s}$ and $\mathcal{S}$, indicating the presence of a large dephasing rate $\gamma^\phi$, that is included in the 2-photon decay and generalized Scully-Lamb models.
This analysis shows that the prethermal domain is not a laser since the latter would exhibit a significant phase coherence in time.
Note that this distinction with a laser relies on the analysis of dynamical properties of the prethermal domain that go beyond the characterization of its static properties. This motivates our choice to describe the local physics of the chain in terms of a driven-dissipative impurity model and not only the Gibbs state ansatz (see the following section).

\subsection*{Modeling the local state of the chaotic Bose-Hubbard chain with a Gibbs state}\label{sec:thermal_ansatz}

\fil{As explained in the main text, in the chaotic regime the local statics can be captured by a Gibbs state reading
\begin{equation}\label{eqs:thermal_ansatz}
    \hat{\rho}^{\rm eq}_\ell = \frac{\exp[-(\hat h - \mu_\ell \, \hat a^\dagger \hat a)/T_\ell]}{\operatorname{Tr}\left(\exp[-(\hat h - \mu_\ell \, \hat a^\dagger \hat a)/T_\ell]\right)}\,,
\end{equation}
where the impurity Hamiltonian $\hat h$ is given by Eq.~\eqref{eqs:SC_hamiltonian}, $T_\ell$ is the effective temperature and $\mu_\ell$ is the chemical potential at the $\ell$-th site.
By reabsorbing all the onsite quadratic contributions into the effective chemical potential $\mu_\ell \leftarrow \mu_\ell - (\omega_0 - U/2)$, we find that the local Wigner functions can simply be described by the two parameters $T_\ell$ and $\mu_\ell$.
Note that, contrary to the driven-dissipative impurity models, the precise content of $\hat{h}$ is relevant for the fitting procedure to the Gibbs state.
In particular, we fix $U=0.1$ [the value we set in all the numerical simulations of Eq.~\eqref{eqs:hamiltonian} presented in this work].
The $L^2$ norm associated to this fitting procedure is plotted in Fig.~\ref{fig:wigner_fit2} (a). 
The result shows that the Gibbs state in Eq.~\eqref{eqs:thermal_ansatz} is successful in describing the local physics of the chain in the prethermal and thermal domains. In Fig.~\ref{fig:pre_thermalization} (d), we presented the dimensionless quantity $\mu_\ell/T_\ell$ for the three models considered in the paper, namely the 2-photon decay model in Eq.~\eqref{eqs:impurity_1}, the Scully-Lamb model in Eq.~\eqref{eqs:impurity_2}, and the Gibbs state in Eq.~\eqref{eqs:thermal_ansatz}.
All the proposed single-site ans\"atze give qualitatively similar results, thus validating the robustness of our local approach in capturing the features of $W_\ell(\alpha, \alpha^*)$. 
A more sophisticated study of space and time correlation functions, and their characterization with multiple-site impurities, should be addressed in future studies.}

\begin{figure*}[t!]
\includegraphics[width=0.8 \textwidth]{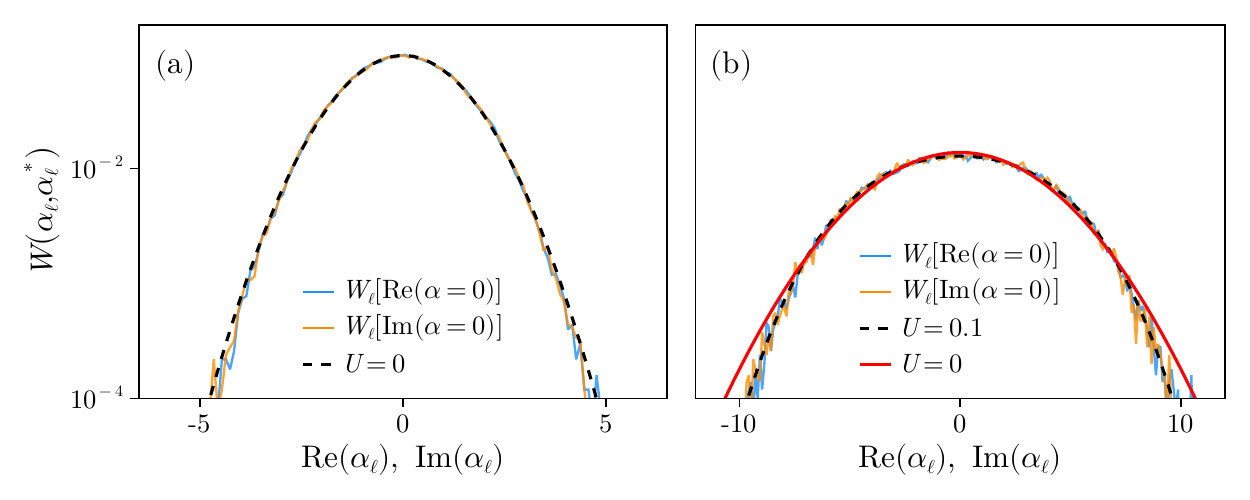}\vspace{0.8em}
\caption{\fil{Comparison between two cuts of $W(\alpha, \alpha^*)$ and the fitted Wigner functions with the Gibbs state in Eq.~\eqref{eqs:thermal_ansatz}.
(a) Comparison between $W_\ell[\textrm{Re}(\alpha=0)]$ (blue curve) and $W_\ell[\textrm{Im}(\alpha=0)]$ (orange curve) and the fitted Wigner function for $\ell=370$ in a $L=400$ chain at low drive amplitude $F=5.5$, by imposing $U=0$ (black-dashed line). 
(b) Comparison between $W_\ell[\textrm{Re}(\alpha=0)]$ (blue curve) and $W_\ell[\textrm{Im}(\alpha=0)]$ (orange curve) and the fitted Wigner function for $\ell=370$ in a $L=400$ chain at intermediate drive amplitude $F=7.5$, by imposing $U=0.1$ (black-dashed line) and $U=0$ (red line) and $U=0.1$ (black dash dotted line). 
The other parameters are set as in Fig.~\ref{fig:phase_diagram}.}
}
\label{fig:gibbs_fit}
\end{figure*}

\fil{Finally, we show that taking into account the Kerr nonlinearity $U$ is important to the characterization of the local state in the chaotic phase at intermediate drive amplitude, even in the thermal domain, while it is not relevant in the chaotic phase at low drive amplitude, where the thermal domain extends across the whole chain and the equipartition theorem describes the local distribution of momenta.
In Fig.~\ref{fig:gibbs_fit} (a), we plot the Wigner function for $\ell=370$ at $\textrm{Im}(\alpha)=0$ (blue curve) and $\textrm{Re}(\alpha)=0$ (orange curve) and $F=5.5$. We perform the fit to the Gibbs state by imposing $U=0$ (black dashed line), which coincides with a Gaussian fit with a single parameter, $\mu_\ell/T_\ell$.
The Gaussian ansatz reproduces the shape of the Wigner function, thus further validating the use of the equipartition theorem in this regime.
In Fig.~\ref{fig:gibbs_fit} (b), we plot the Wigner function for $\ell=370$ at $\textrm{Im}(\alpha)=0$ (blue curve) and $\textrm{Re}(\alpha)=0$ (orange curve) and $F=7.5$. We first perform the the single-parameter Gaussian fit. In this case, the Gaussian Ansatz does not reproduce the local Wigner functions. Instead, when we impose $U=0.1$ (gray dash-dotted line), the full Gibbs state perfectly reproduces the shape of $W_{\ell}(\alpha, \alpha^*)$.
This proves that the contribution of the nonlinearity at intermediate drive amplitudes is present not only in the prethermal domain, where the nonlinearity determines the saturation of the phton number, but also in the thermal domain, where the vacuum dressed by thermal fluctuations accounts for a small but finite amount of nonlinearity.}

\subsection*{Additional details on the RNW regime}\label{sec:classical_RNW}

\begin{figure*}[t!]
\includegraphics[width=1 \textwidth]{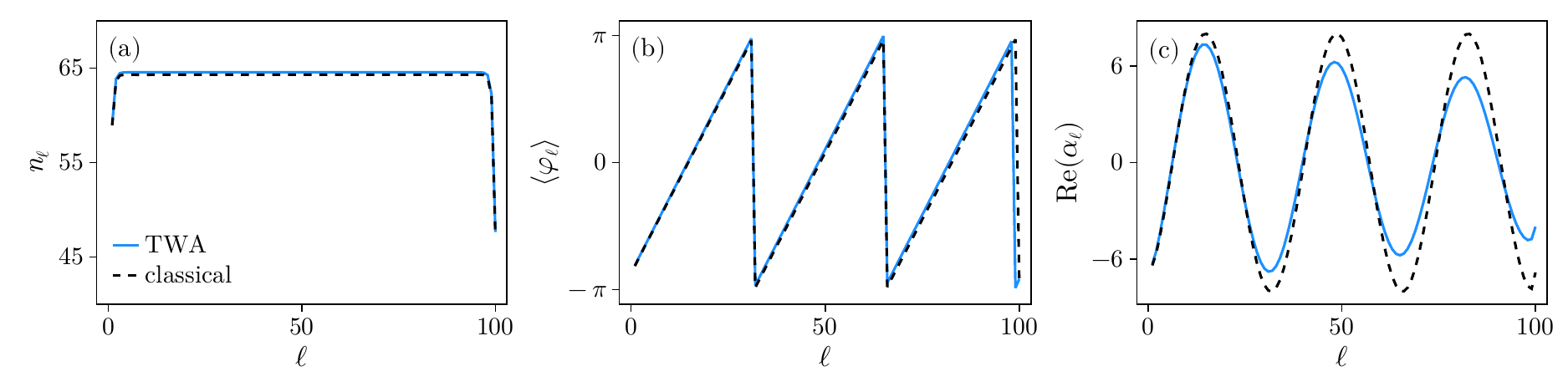}\vspace{0.8em}
\caption{\fil{Additional comparison between the classical and TWA solutions for the RNW regime.
(a) Photon number $n_\ell$ as a function of $\ell$ for a $L=100$ chain in the RNW regime. The blue curve is are the TWA values whereas the black-dashed curve represents the classical Gross-Pitaevskii prediction.
(b) Same as in (a) but for the averaged phase $\langle\varphi_\ell\rangle$.
(c) Same as in (a) but for the real part of the bosonic field, $\textrm{Re}(\alpha_\ell)$.
Results are computed by averaging over $N_{\rm traj} = 5\times10^3$ independent Wigner trajectories and over a time window $\Delta\tau=10^3$ once the steady state is reached.
The drive amplitude is set to $F=12.5$.
The other parameters are set as in Fig.~\ref{fig:phase_diagram}.}
}
\label{fig:rnw_comparison}
\end{figure*}

\fil{In Sec.~\ref{sec:RNW_regime} we studied extensively the RNW regime, highlighting its distinctive quantum features, encoded in the phase fluctuations at the same site $\ell$, measured by means of the circular variance $\Delta\varphi_\ell := 1-|\langle\rme^{\rmi\varphi_\ell}\rangle|$, and at different sites $k$ and $\ell$, measured by means of the first-order coherence function $g^{(1)}_{k\ell}$ in Eq.~\eqref{eqs:g1}. We present here three additional observables, the photon number $n_\ell$, the average phase $\langle\varphi_\ell\rangle$, and the bosonic field $\alpha_\ell$.
The comparison between the classical and TWA solutions is presented in Fig.~\ref{fig:rnw_comparison}.}

\fil{We find that both the photon number and the averaged phase coincide when estimated with the TWA and the classical approach, as shown in Fig.~\ref{fig:rnw_comparison} (a) and (b).
For the bosonic field $\alpha_\ell$, reported in Fig.~\ref{fig:rnw_comparison} (c), we clearly observe the effect of quantum fluctuations, which causes the decay of $\textrm{Re}(\alpha_\ell)$ with $\ell$ [the same pattern can be observed for $\textrm{Im}(\alpha_\ell)$ and $|\alpha_\ell|$].
This distinction can be explained as follows: quantum fluctuations induce an angular spreading of the Wigner function [see Figs.~\ref{fig:RNW} (c-f) in the main text] but keeping a radial localization around the classical value.
As a result, the average $\langle|\alpha_\ell|^2\rangle$ only encodes the information about the radius of $W_\ell(\alpha, \alpha^*)$, which coincides with the classical value, while the phase spreads symmetrically around the classical value, and its average again returns the Gross-Pitaevskii solution.
An increasing $\Delta\varphi_\ell$ leads to the decay of $\alpha_\ell$ because the latter quantity encodes both the radius and the angular spreading of $W_\ell(\alpha, \alpha^*)$, resulting in a decreasing $\textrm{Re}(\alpha_\ell)$ when the average is performed.}

\subsection*{Effects of an intrinsic dissipation rate}

\begin{figure*}[t!]
\includegraphics[width=1\textwidth]{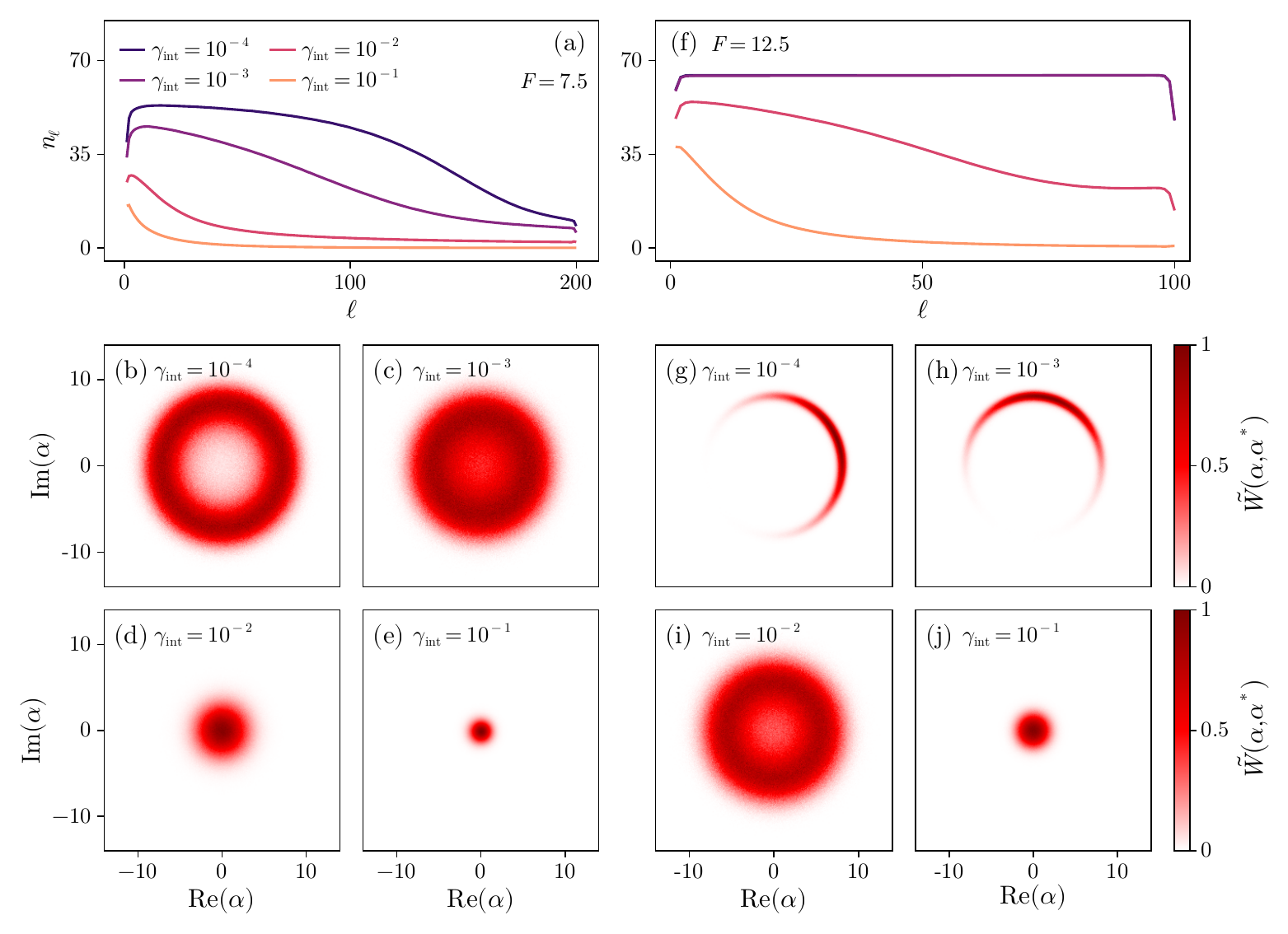}\vspace{0.8em}
\caption{\fil{Effects of an instrinsic dissipation rate $\gamma_{\rm int}$ on the chaotic and the RNW regime.
(a) photon number $n_\ell$ as a function of $\ell$ for a $L=200$ chain in the chaotic regime at $F=7.5$ for various intrinsic dissipation rates $\gamma_{\rm int}=10^{-4}, 10^{-3}, 10^{-2}, 10^{-1}$ (from dark purple to light pink).
(b-e) Local Wigner function $W_\ell\left(\alpha, \alpha^*\right)$ in the chaotic chain at $\ell=50$ for the same values of $\gamma_{\rm int}$ considered in panel (a).
(f) Same as in panel (a) but for a $L=100$ chain in the RNW regime at $F=12.5$.
(g-j) Same but for the regular RNW steady state at $\ell=50$.
The other parameters are set as in Fig.~\ref{fig:phase_diagram}.}
}
\label{fig:Intrinsic_dissipation}
\end{figure*}

\fil{The setup described by Eqs.~\eqref{eqs:lindblad} and \eqref{eqs:hamiltonian} presents a notable idealization: the intrinsic loss of each resonator is assumed to be zero.
In arrays of superconducting nonlinear resonators \cite{FitzpatrickPRX17, FedorovPRL21}, a finite intrinsic dissipation rate $\gamma_{\rm int}\ll\gamma$ is always present. 
Therefore, Eq.~\eqref{eqs:lindblad} generalizes to
\begin{align}\label{eqs:lindblad_intrinsic_dissipation}
    \frac{\partial\hat{\rho}}{\partial t}  &= -\rmi[\hat{H}, \hat{\rho}] +
       \mathcal{D}[\hat L_1] \hat \rho +    \mathcal{D}[\hat L_L]  \hat \rho + \sum_{\ell=1}^L\mathcal{D}[\hat L_\ell^{\rm int}]\hat{\rho},
\end{align}
where $\hat L^{\rm int}_\ell = \sqrt{\gamma_{\rm int}}\hat{a}_\ell$.
An important question concerns the robustness of the classical and quantum photonic regimes studied in the main text with respect to spurious intrinsic dissipations.
This study is relevant for the experimental realization of the driven-dissipative architecture and the observation of the phenomena we describe.
Here, we focus on the effects of a finite $\gamma_{\rm int}$ on the chaotic prethermal domain in a $L=200$ chain at $F=7.5$, and on the regular RNW regime in a $L=100$ chain at $F=12.5$.}

\fil{In Fig.~\ref{fig:Intrinsic_dissipation} (a) we plot the photon number $n_\ell$ as a function of the site index $\ell$ in the chaotic phase.
For small dissipation rates equal to $\gamma_{\rm int} = 10^{-4}$ the photon number profile resembles the one presented in the main text [see Fig.~\ref{fig:correlations} (a)].
For larger dissipation rates equal to $\gamma_{\rm int} = 10^{-3}$ the prethermal phase visibly shrinks.
For even larger dissipation rates up to $\gamma_{\rm int} = 10^{-1}$, the prethermal regime disappears completely and most of the resonator end in a vacuum-like state with around zero photons.
This discussion is supported by the plots of the local Wigner functions $W_\ell\left(\alpha, \alpha^*\right)$ reported in Figs.~\ref{fig:Intrinsic_dissipation} (b-e) for $\ell=50$.
Similar considerations hold for the regular RNW regime.
In Fig.~\ref{fig:Intrinsic_dissipation} (f) we plot the photon number $n_\ell$ as a function of the site index $\ell$.
For small dissipation rates equal to $\gamma_{\rm int} = 10^{-4}-10^{-3}$ we observe flat photon number profile of the RNW regime [see Fig.~\ref{fig:regular_array} (a) and Fig.~\ref{fig:rnw_comparison} (a)].
For larger intrinsic dissipation rates, like $\gamma_{\rm int} = 10^{-2}$, the photon number acquires a gradient with $\ell$ and the RNW regime is destroyed. 
For even larger intrinsic dissipation rates up to $\gamma_{\rm int} = 10^{-1}$, the bulk of the chain goes into a vacuum-like states with few or zero photons in each resonator.
This discussion is supported by the plots of the local Wigner functions $W_\ell\left(\alpha, \alpha^*\right)$ reported in Figs.~\ref{fig:Intrinsic_dissipation} (g-j) for $\ell=50$.}

\fil{We conclude that an intrinsic dissipation rate, reasonably, threatens the out-of-equilibrium regimes we studied in the main text.
Too large values of $\gamma_{\rm int}$ prevent photon transport across the chain and most of the system is in a trivial vacuum-like state.
On the other hand, both the prethermal and the RNW regimes are still present with a small but finite $\gamma_{\rm int}$.}

\end{document}